\documentclass[10pt,letterpaper]{article}
\usepackage[text={5.5in, 8.25in},centering]{geometry}
\usepackage{fancyhdr}
\usepackage{amssymb}
\usepackage{amsfonts}
\usepackage{amsmath}
\usepackage{amsthm}

\usepackage[dvipsnames]{xcolor}
\usepackage{listings}
\usepackage{matlab-prettifier}
\usepackage{graphicx}
\usepackage{placeins}
\usepackage[skip=6pt]{caption}
\usepackage{epsfig,subfigure}
\usepackage{mathrsfs}
\usepackage{multicol,multirow}
\usepackage{placeins}
\usepackage{setspace}
\usepackage{appendix}
\usepackage{scalerel} 
\usepackage{ulem}

\usepackage{caption}
\usepackage{url}
\usepackage{todonotes}
\usepackage{xfrac}
\usepackage[alphabetic]{amsrefs}
\usepackage[colorlinks=true, linkcolor=blue, citecolor=black]{hyperref}%
\usepackage[capitalize]{cleveref}
\newcommand\myurl[1]{\url{#1}}

\BibSpec{webpage}{%
  +{}{\PrintAuthors} {author}
  +{,}{ \textit} {title}
  +{}{ \parenthesize} {date}
  +{,}{ \myurl} {myurl}
  +{,}{ } {note}
  +{.}{ } {transition}
}

\theoremstyle{remark}

\newtheorem{remark}{Remark}

\crefname{definition}{Definition}{Definitions}
\crefname{lemma}{Lemma}{Lemmas}
\crefname{theorem}{Theorem}{Theorems}
\crefname{proposition}{Proposition}{Propositions}
\crefname{figure}{Figure}{Figures}

\Crefname{definition}{definition}{definitions}
\Crefname{lemma}{lemma}{lemmas}
\Crefname{theorem}{theorem}{theorems}
\Crefname{proposition}{proposition}{propositions}
\Crefname{figure}{figure}{figures}

\usepackage{relsize}
\usepackage{bm} 
\providecommand{\mathbold}[1]{\bm{\mathsf{#1}}}
\newcommand{\R}{\mathbb{R}}

\newcommand{\vct}[1]{\bm{#1}}

\newcommand{\mtx}[1]{\mathbold{#1}}

\usepackage{stmaryrd}

\newcommand\scalemath[2]{\scalebox{#1}{\mbox{\ensuremath{\displaystyle #2}}}}

\newcommand{\erowblock}{\mathit{r}}
\newcommand{\ecolblock}{\mathit{c}}
\newcommand{\sblock}{\mathit{s}}

\newcommand{\sktext}{\mathrm{sk}}
\newcommand{\sk}[1]{
    {#1}^{\sktext}
}
\newcommand{\pre}[1]{
    {#1}^{\mathrm{pre}}
}
\newcommand{\perm}[1]{
    {#1}^{\mathrm{perm}}
}

\usepackage{lmodern}

\DeclareMathOperator{\rank}{rank}

\usepackage{soul}

\newcommand{\maxm}[1]{{\color{blue}\sf{[Max: #1]}}}

\usepackage{makecell} 

\usepackage{xspace}

\newcommand{\RandLAPACK}{\textsf{RandLAPACK}\xspace}
\newcommand{\RBLAS}{\textsf{RandBLAS}\xspace}
\newcommand{\RandBLAS}{\RBLAS}

\newcommand{\code}[1]{\texttt{#1}}

\newcommand{\leadslicespace}{\hspace{0.125em}}
\newcommand{\trailslicespace}{\hspace{0.025em}}
\newcommand{\fslice}{\leadslicespace{}{:}\trailslicespace{}}  
\newcommand{\lslice}[1]{{0}{:}{#1}}  
\newcommand{\tslice}[1]{{#1}{:}}  
\newcommand{\trans}{\text{\tiny{T}}}

\newcommand{\ignore}[1]{}

\usepackage{algorithm}
\usepackage{algorithmicx}   
\usepackage[noend]{algpseudocode}  
\algrenewcommand\textproc{\texttt}
\algdef{SE}[SUBALG]{Indent}{EndIndent}{}{\algorithmicend\ }
\algtext*{Indent}
\algtext*{EndIndent}
\algdef{SE}[DOWHILE]{Do}{doWhile}{\algorithmicdo}[1]{\algorithmicwhile\ #1}%
\newcommand{\codecomment}[1]{\textcolor{gray}{\texttt{//} #1}}
\usepackage[capitalize]{cleveref}
\usepackage{overpic}
\usepackage{stackengine}
\usepackage{tikz}

\ExplSyntaxOn

\NewDocumentCommand{\showvariable}{sm}
 {
  \IfBooleanTF { #1 }
   {
    \exp_args:Nc \agava_showvar:N { #2 }
   }
   {
    \agava_showvar:N #2
   }
 }

\cs_new:Nn \agava_showvar:N
 {
  \texttt{\token_to_str:N #1\unskip} ~
  \bool_case:n
   {
    { \token_if_macro_p:N #1 } { #1 ~ (macro) }
    { \token_if_chardef_p:N #1 } { \int_eval:n { #1 } ~ (chardef) }
    { \token_if_mathchardef_p:N #1 } { \int_eval:n { #1 } ~ (mathchardef) }
    { \token_if_dim_register_p:N #1 } { \dim_eval:n { #1 } ~ (dimension) }
    { \token_if_skip_register_p:N #1 } { \skip_eval:n { #1 } ~ (skip) }
    { \token_if_primitive_p:N #1 } { \the#1 ~ (primitive) }
   }
 }

\ExplSyntaxOff

\newcommand{\footremember}[2]{%
    \footnote{#2}
    \newcounter{#1}
    \setcounter{#1}{\value{footnote}}%
}
\newcommand{\footrecall}[1]{%
    \footnotemark[\value{#1}]%
} 

\newcommand\blfootnote[1]{%
  \begingroup
  \renewcommand\thefootnote{}\footnote{#1}%
  \addtocounter{footnote}{-1}%
  \endgroup
}

\usepackage{relsize}

\title{

Anatomy of High-Performance Column-Pivoted QR Decomposition
}
\author{
    Maksim Melnichenko\footremember{ICL}{Innovative Computing Laboratory, University of Tennessee, Knoxville}
    \and 
    Riley Murray\footremember{Sandia}{Sandia National Laboratories}
    \and
    William Killian\footremember{NVIDIA}{NVIDIA}
    \and
    James Demmel\footremember{UCB}{University of California Berkeley}
    \and
    Michael W. Mahoney\footremember{ICSI}{International Computer Science Institute (ICSI)} \footremember{LBNL}{Lawrence Berkeley National Laboratory} \footrecall{UCB}
    \and
    Piotr Luszczek\footremember{MIT}{MIT Lincoln Lab} $\:$ \footrecall{ICL}
    \and
    Mark Gates\footrecall{ICL}
}

\begin{document}

\maketitle


\begin{abstract}
We introduce an algorithmic framework for performing QR factorization with column pivoting (QRCP) on general matrices. 
The framework enables the design of practical QRCP algorithms through user-controlled choices for the core subroutines. 
We provide a comprehensive overview of how to navigate these choices on modern hardware platforms, offering detailed descriptions of alternative methods for both CPUs and GPUs.
The practical QRCP algorithms developed within this framework are implemented as part of the open-source RandLAPACK library. 
Our empirical evaluation demonstrates that, on a dual AMD EPYC 9734 system, the proposed method achieves performance improvements of up to two orders of magnitude over LAPACK's standard QRCP routine and greatly surpasses the performance of the current state-of-the-art randomized QRCP algorithm \cite{MOHvdG:2017:QR}. 
Additionally, on an NVIDIA H100 GPU, our method attains approximately $65\%$ of the performance of cuSOLVER's unpivoted QR factorization.
 
\blfootnote{$\!^1$
     Send correspondence to the first author, at mmelnic1@vols.utk.edu.
    }
\end{abstract}

\section{Introduction}
\label{sec:intro}

Randomized Numerical Linear Algebra (RandNLA) is a relatively young branch of the field of numerical linear algebra (NLA).
It leverages randomization as a computational resource to
develop algorithms for computing solutions to classical linear algebra problems, with performance superior to deterministic NLA schemes.
Some of these algorithms reliably compute \textit{approximate solutions} to a given problem.
Others, under some suitable assumptions, provide the \textit{exact
solution} by generating a full matrix factorization.
Algorithms for full matrix factorizations are the essential part of the \textit{decompositional approach} to matrix computations, which revolutionized the field of computational science ~\cite{814652}.

Despite the ongoing widespread influence of RandNLA, it has yet to have a major practical impact on how we compute the ``classical''
\textit{full} matrix decompositions, e.g., Cholesky, LU, QR, Schur,
eigenvalue decomposition, and the full SVD~\cite{Higham:blog:big6}.
This is largely due to the fact that, in modern implementations of such decompositions, all of the computations contributing to the
leading-order terms in algorithms' complexity are cast in terms of Level~3 Basic Linear Algebra Subprograms (BLAS) operations.
Such operations are ideally suited to achieve high performance on modern
hardware, and hence it is notoriously difficult to
improve upon the performance of schemes that predominantly rely on Level~3 BLAS.
There is, however, a classical matrix decomposition, the widely adapted implementation for which does \emph{not} predominantly rely on Level~3 BLAS: the QR decomposition with column pivoting (QRCP).
The fundamental problem
with the classical approach to QR with column pivoting (Householder
QRCP) is that only half of the computations can be cast in terms of
Level~3 BLAS operations~\cite{LAWN114}.
Consequently, the standard LAPACK function for QRCP, called \code{GEQP3}, remains very slow due to memory bandwidth constraints.

In recent work~\cite{MBM2024}, we introduced a novel QRCP algorithm called ``Cholesky QR with Randomization and Pivoting for
Tall matrices'' (CQRRPT).
CQRRPT \textit{carefully} uses techniques from RandNLA to deliver acceleration over the alternative methods for QRCP.
These include specialized communication-avoiding
algorithms~\cite{FK2020}, and, in certain cases, standard
LAPACK unpivoted QR, called \code{GEQRF}, together with \code{LATSQR}, which is tailored specifically for tall matrices \cite{demmel2012communication}.
Furthermore, CQRRPT can be implemented to achieve numerical stability unconditionally of the properties of its input data.
This highlights not only its reliability, but it also advertises the method as an appealing tool for \textit{orthogonalization}.
The main limitation of this scheme, however, hides in its
name: CQRRPT is only applicable to rectangular matrices,
where the number of rows is much larger than the number of columns.
This limitation of CQRRPT disqualifies it from candidacy for a
definitive ``solution'' to the problem of designing a
QRCP algorithm.

This paper introduces an algorithmic framework, referred to as \textit{BQRRP},\footnote{pronounced ``bee-crip''} which stands for ``Blocked QR with Randomization and Pivoting.'' 
BQRRP serves as a natural evolution of CQRRPT,\footnote{pronounced ``see-cript''} encapsulating much of CQRRPT's core features, while resolving its main aforementioned limitation.
 
This paper does not dwell too much on the theoretical properties of the algorithmic framework we propose.
Rather, it concentrates on providing many details on the practical aspects of what goes into a high-performance QRCP implementation. 
As such, while our work is particularly beneficial for numerical software engineers, it is accessible to a broader audience who may find value in exploring it at their own pace.

\subsection{Existing work and our contribution}
\label{subsec:contribution}

Naturally, efforts to revise the standard approach to QRCP have emerged
in the past, among which are the works of Bischof and
Quintana-Ortí~\cite{BQ1998_Alg,BQ1998_Code}, as well as the independent works of Martinsson~\cite{Martinsson:2015:QR} and Duersch and
Gu~\cite{DG:2017:QR}, with subsequent extensions by Martinsson
\textit{et al.}~\cite{MOHvdG:2017:QR} and also by Xiao, Gu, and
Langou~\cite{XGL:2017:RandQRCP}. 
Of particular interest are randomized algorithms described in \cite{DG:2017:QR} and \cite{MOHvdG:2017:QR}.
Both these papers, in addition to the derivation of the pseudocode schemes, present software with practical QRCP implementations.

\paragraph{The early randomized Householder methods.} In \cite{DG:2017:QR}, the authors show several benchmarks of prospective QRCP methods (written in
Fortran~$90$),
compared against both the standard pivoted and unpivoted QR algorithms, provided by the contemporary (though unspecified) version of
Intel Math Kernel Library (MKL).
The \code{RQRCP} method, described in the
pseudocode \cite[Algorithm 4]{DG:2017:QR} exhibits an
order-of-magnitude speedup over the standard
pivoted QR and achieves asymptotically
up to $60\%$ of the performance of the standard unpivoted QR, when applied to full-rank square matrices.
Their work also presents a \code{TRQRCP}
scheme (\cite[Algorithm 6]{DG:2017:QR}), which proves faster than \code{RQRCP} in low-rank cases.

In the other work~\cite{MOHvdG:2017:QR},
the authors present a randomized algorithm called \code{HQRRP} that
offers an order-of-magnitude speedup over \code{GEQP3} and that achieves
up to $80$\% of the performance of the
standard unpivoted QR, \code{GEQRF}.
In the benchmarks shown therein, the standard pivoted and unpivoted QR functions were taken from LAPACK version $3.4.0$, and the implementations were linked to BLAS from MKL version $11.2.3$.
The performance results were reported for
HQRRP that was implemented with libflame version $11104$ \cite{FVE09}.
In addition to the libflame-based implementation, the authors presented an LAPACK-compatible version of HQRRP.\footnote{Available at \url{https://github.com/flame/hqrrp}}
The emphasis on the practicality and portability of HQRRP can be seen
throughout the \cite{MOHvdG:2017:QR} paper,
suggesting that, at its time, this algorithm qualified for an ideal replacement of \code{GEQP3} in LAPACK.

\begin{figure}[htb!]
  \centering
  \hspace*{-0.8cm}
  \begin{tikzpicture}
    \node[inner sep=0pt] at (0,0) {\includegraphics[width=1.0\linewidth]{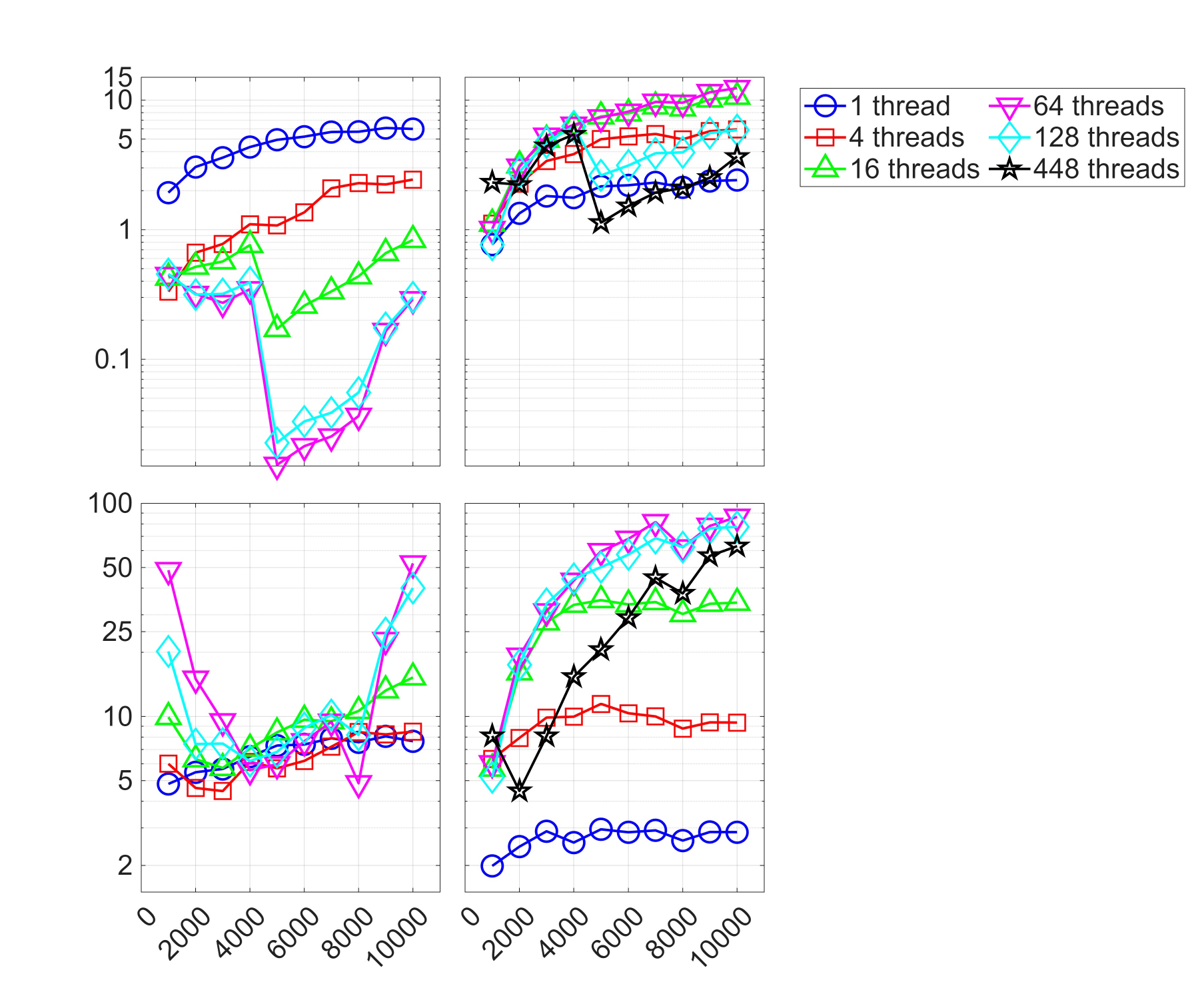}};

    \node[anchor=north west, text width=5.3cm, align=justify] at (1.9, 3.5) {%
      \caption{\footnotesize{Speedup of \code{HQRRP} \cite{MOHvdG:2017:QR} (first row) and standard LAPACK unpivoted QR, \code{GEQRF} (second row), over the standard LAPACK QRCP, \code{GEQP3}, attained on matrices of sizes between $1{,}000$-by-$1{,}000$ to
      $10{,}000$-by-$10{,}000$, when varying the number of OpenMP threads used.
      In all experiments, \code{HQRRP} block size was set to $128$, as suggested in \cite[Sec. 4.1]{MOHvdG:2017:QR}.
      Results captured on machines, described in \cref{table:cpu_config}.
      On an Intel system, the performance of \code{HQRRP} relative to that of \code{GEQP3} stagnates with the increase of the number of OpenMP threads used;
      on an AMD system, \code{HQRRP} is able to achieve performance above what was reported in \cite[Fig. 1]{MOHvdG:2017:QR}.
      On both systems, the speedup of \code{GEQRF} over \code{GEQP3} greatly exceeds the speedup observed in \cite[Fig. 5]{MOHvdG:2017:QR}.
      We further discuss the observed results in \cref{app:hqrrp_performance}.}}
      \label{fig:hqrrp_plot_remake}
    };

    \node[anchor=north west, font=\bfseries] at (-0.33\linewidth, 5.5) {Intel CPU};
    \node[anchor=north west, font=\bfseries] at (-0.08\linewidth, 5.5) {AMD CPU};

    \node[anchor=north west, rotate=90, font=\bfseries] at (-0.46\linewidth, 0.04\linewidth) {\textbf{Speedup \code{HQRRP}/\code{GEQP3}}};
    \node[anchor=north west, rotate=90, font=\bfseries] at (-0.46\linewidth, -0.3\linewidth) {\textbf{Speedup \code{GEQRF}/\code{GEQP3}}};

    \node at (-0.27\linewidth, -5.6) {$\mathbf{m=n}$};
    \node at (0.005\linewidth, -5.6) {$\mathbf{m=n}$};
  \end{tikzpicture}
\end{figure}
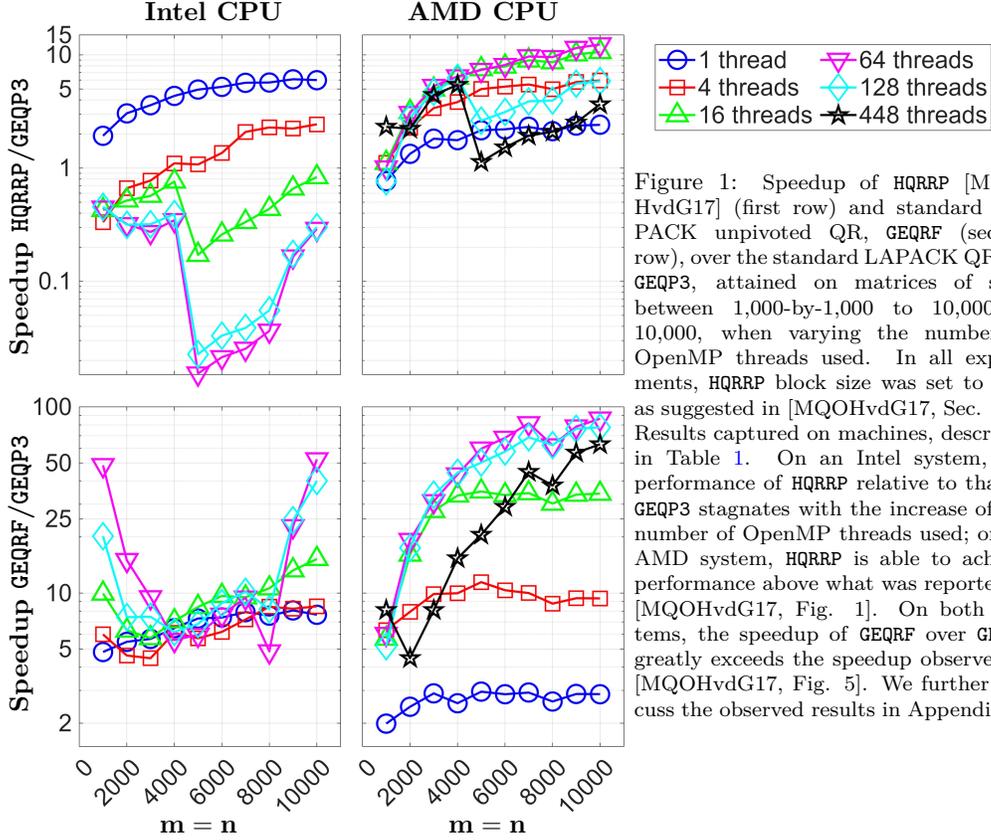

\paragraph{Our motivation.}
The development of these algorithms, however, was not the final chapter in the quest for a high-performance QRCP method.
As with its predecessors, neither \code{HQRRP} nor
any of the variants of \code{RQRCP}
made it into LAPACK. 
Many years of hardware development have
passed, and benchmarking on modern machines shows that the gap in
performance between the standard pivoted and unpivoted QR factorization
algorithms has grown from roughly $10\times$ to near $100\times$.
Simultaneously, while the LAPACK-compatible version of HQRRP proved to be easily portable to modern libraries, its performance did not increase commensurately. 
See~\cref{fig:hqrrp_plot_remake}.

With that, following
the steps of our predecessors,
we accepted the challenge of formulating a modern approach for QR with column pivoting that would match the contemporary performance of the algorithms
using Level~3 BLAS functionality.


\paragraph{Modern randomized QRCP.}
Our newest installment in the QRCP family of algorithms is the \textit{BQRRP} \textit{algorithmic
framework}.
In this framework, one implements a \code{BQRRP}
instantiation by selecting a set of subroutines, most suitable for a system on which \code{BQRRP} is to be used.
Our main contributions are:
(a) a detailed description of the most suitable practical choices for such subroutines on two modern systems; and (b) CPU and GPU implementations of \code{BQRRP}, as part of an open-source \RandLAPACK{} library, that allows users to configure the algorithm in a way that is most suitable for their needs.
This manuscript describes both CPU and GPU versions of \code{BQRRP}.
As we show in \cref{sec:performance}, an implementation of
\code{BQRRP\_CPU} not only outperforms \code{HQRRP}, but also it is up
to two orders of magnitude faster than
the standard pivoted QR (\code{GEQP3}) implementation available in MKL 2025.
\code{BQRRP\_GPU} exhibits excellent throughput and shows reasonable performance relative to an unpivoted QR (\code{GEQRF}) implementation
available in NVIDIA's
cuSOLVER 11.6.1.9.
This makes \code{BQRRP\_GPU} a strong candidate for adoption as a standard general QRCP method by GPU linear algebra library vendors.

The dominant cost of \code{BQRRP\_CPU} comes from
a routine that applies an implicitly-stored
orthonormal matrix to a portion of the input data in a loop.
On a GPU, in addition to this bottleneck, there is the added complication in the form of the cost of permuting the columns of a portion of the input matrix at every iteration of the main loop. 
This cost can be mitigated by employing advanced pivoting strategies that allow for additional levels of parallelism, such as the ``parallel pivots'' approach. 
We discuss these in \cref{sec:bqrrp_gpu}.
The rest of the major operations in \code{BQRRP} are all highly efficient, reaching the Level~3 BLAS performance.

\code{BQRRP\_CPU} is designed to be storage-efficient, applying all of its subsequent operations in
place (modulo comparatively tiny workspace buffers). Furthermore, the
output format of both \code{BQRRP\_CPU}
and \code{BQRRP\_GPU} is identical to that of \code{GEQP3}. 

Whether our novel approach remains relevant in the long run is an open question. However, even in the worst-case scenario, we emphasize that the value of our work extends beyond delivering a modern high-performance algorithm for QRCP. It also lies in establishing a solid foundation for analyzing the performance and implementation details of future QRCP methods, as well as exposing the underlying structure of such algorithms, which can be invaluable for future developers if the scourge of slow QRCP returns.

\subsection{Outline of the manuscript}
\label{subsec:outline}

The rest of \cref{sec:intro} is dedicated to describing the notation (\cref{subsec:def_and_notations}), as well as the setup of the experiments (\cref{subsec:exp_setup}) used throughout the paper.
\cref{sec:introduce_BQRRP} then starts by formally introducing the generalized view of the BQRRP algorithmic framework in \cref{alg:BQRRP}.
Our formalism emphasizes that the \code{BQRRP} is defined by the choices of its core subroutines.
The choices for such subroutines are presented in \crefrange{subsec:qrcp_wide}{subsec:col_perm}.
The subroutine details there are aided
by a series of pseudocode algorithms and CPU performance plots.
\cref{sec:bqrrp_practical} presents a step-by-step guide to how \code{BQRRP\_CPU} is constructed in \RandLAPACK{}, with particular attention to its in-place storage feature.
The discussion is aided by a detailed storage visualization in \cref{fig:bqrrp_storage}.
\cref{sec:bqrrp_gpu} discusses how constructing a GPU version of \code{BQRRP} differs from constructing the previously-described CPU version, making references to \cref{alg:BQRRP} and \crefrange{subsec:qrcp_wide}{subsec:col_perm}.
\cref{sec:performance_profiling} shows details on how each subroutine that comprises the CPU and GPU
versions of \code{BQRRP} affects the overall performance of the algorithm, presenting runtime breakdown plots.
This is done to help identify specific performance bottlenecks inside the algorithm.
\cref{sec:piv_qual} provides empirical investigations of pivot quality. 
It shows how easy-to-compute metrics of pivot quality compare when running LAPACK's default function
\code{GEQP3} versus when running the versions of \code{BQRRP}, defined previously in \cref{sec:bqrrp_practical}.
The results are shown for input matrices that are notoriously difficult to be processed by QRCP.
\cref{sec:performance} provides performance experiments with both GPU and CPU implementations of \code{BQRRP} in \RandLAPACK{}.
Concluding remarks are given in \cref{sec:conclusion}.

\subsection{Definitions and notation}
\label{subsec:def_and_notations}

\paragraph{Preliminaries.}
Matrices appear in boldface sans-serif capital letters.
The transpose of a matrix $\mtx{X}$ is given by $\mtx{X}^{\trans}$.
Numerical vectors appear as boldface lowercase letters, while index vectors appear as uppercase letters.
The identity matrix is denoted by $\mtx{I}$.
We enumerate components of matrices and vectors with indices starting from zero, rather than starting from one.
We extract the leading $k$ columns of $\mtx{X}$ by writing $\mtx{X}(\fslice{},\lslice{k})$.
This range excludes column $k$,
while its trailing $n-k$ columns are extracted by writing $\mtx{X}(\fslice{},\tslice{k}{n})$.
The $(i,j)^{\text{th}}$ entry of $\mtx{X}$ is $\mtx{X}(i,j)$.
Similar conventions apply to extracting the rows of a matrix or components of a vector.

The matrix we ultimately aim to decompose is denoted by $\mtx{M}$ and has
dimensions $m$-by-$n$,
with $m$ and $n$ not related in any particular way.
We use the letters $i$ and $j$ as zero-based
integer indices between 0 and $\min\{m, n\}-1$.
In some contexts, $i$ denotes the iteration of an algorithm loop.
The symbol $b$ is used to refer to the block size parameter used in a given algorithm; and $b$ is expected to be small relative to the matrix size, i.e., $b \ll \min\{m, n\}$.
The symbol $\ell$ denotes the estimated rank of a given matrix.
The symbol $k$ denotes the block rank, or the rank of the
given submatrix, $k \leq b$.

Given an iteration $i \in \lslice{(\lceil n/b \rceil-1)}$ of an arbitrary algorithm that parses the given matrix in increments of block size $b$, we use $\sblock = i\times b$ to denote the start of the current row/column block range;
$\erowblock = \min\{m, (i+1)b\}$ denotes the (exclusive) end of the current row block range; and $\ecolblock = \min\{n, (i+1)b\}$ denotes the (exclusive) end of the current column block range.

Throughout this paper, we aim to clearly identify
the purpose of each function name, which are presented in \code{typewriter-style} font. 

We often refer to linear algebraic functions using their conventional BLAS and LAPACK names in uppercase.
The explanation of the standard naming conventions can be found at \url{https://www.netlib.org/lapack/lug/node24.html}.
Whenever we mention a given LAPACK function name, we avoid using the letter that denotes the precision, assuming that all computations are performed in double precision (for example, we use ``\code{GEQP3}'' instead of ``\code{DGEQP3}'').
We use \code{BQRRP} (\code{fixed-width} font) when referring to \textit{any} practical algorithm developed from the BQRRP (standard font) algorithmic framework.
We use \code{BQRRP\_CPU} and \code{BQRRP\_GPU} to specifically denote the respective CPU and GPU versions of \code{BQRRP}.

\paragraph{\code{BQRRP} intended output format.}
A \code{BQRRP} algorithm intends to provide output in a format that is identical to the standard LAPACK QRCP routine, \code{GEQP3}.
Assuming that a matrix $\mtx{M} \in \R^{m \times n}$ is passed as
input into \code{GEQP3}, the
output format is described as follows:
\begin{itemize}
\item Vector $\vct{\tau}$ of length $\min\{m, n\}$ that holds the scalar factors of the elementary reflectors.

\item Modified in-place $\mtx{M}$ with its above-diagonal portion
occupied by an explicit upper-trapezoidal $\mtx{R}$ of size $\min\{m, n\} \times n$.
The below-diagonal portion of output $\mtx{M}$ stores Householder vectors $\vct{v}_{i}$ for $i = \lslice{\min\{m, n\}}$,
which, together with the vector $\vct{\tau}$, compactly
represents the orthonormal matrix $\mtx{Q}$ as a product of $\min\{m, n\}$ elementary reflectors:
    \begin{align*}
            & \mtx{Q} = \mtx{H}_{1}\mtx{H}_{2}\dots\mtx{H}_{\min\{m, n\}} \\
            & \mtx{H}_{i} = \mtx{I} - \tau \vct{v}_{i} \vct{v}_{i}^\trans  .
    \end{align*}

\item Permutation vector $\vct{J}$ of size $n$ such that
if $\vct{J}(j)=i+1$, then the $j^{\mathrm{th}}$ column of $\mtx{M}(:,\vct{J})$ was the $i^{\mathrm{th}}$ column of $\mtx{M}$. Note that in LAPACK convention, the permutation vector $\vct{J}$ stores entries in a \textit{one-based} index format used by Fortran and MATLAB (hence the presence of ``$+1$'' term in the $\vct{J}(j)=i+1$ expression).
\end{itemize}

We refer back to the above output format throughout the paper when talking about the ``intended'' output format for \code{BQRRP}. 
The described format of the $\mtx{Q}$ factor is referred to as the \textit{implicit economical} storage format.
By contrast, some of the algorithms described in this manuscript use an \textit{explicit economical} storage format, where $\mtx{Q}$ is comprised of $\min\{m, n\}$ explicitly-defined orthogonal column vectors. 

\subsection{Experiments setup}\label{subsec:exp_setup}

Each of the further sections of this manuscript is interlaced with the practical performance experiments.
For that reason, we present upfront the details of the hardware and software setup of our experiments.

\paragraph{Our software.}
 Versions of \code{BQRRP} algorithm discussed in this work, as well as their subcomponent functions, are implemented as part of an open-source C\texttt{++} library called \RandLAPACK{}.
Together with its counterpart, \RandBLAS, \RandLAPACK{} provides a platform for developing, testing, and benchmarking high-performance RandNLA algorithms.
This software was produced as part of the BALLISTIC project~\cite{BALLISTIC}, and it is actively developed by the authors of this paper.
It is important to note that \RandBLAS{} and \RandLAPACK{} rely on BLAS\texttt{++} and LAPACK\texttt{++}~\cite{gates2022portable}
for the purpose of obtaining basic linear algebra
functions; BLAS\texttt{++} and LAPACK\texttt{++} themselves serve as wrappers around the low-level vendor-optimized BLAS and LAPACK packages.
As such, many vendor-optimized libraries (Intel's MKL, AMD's AOCL, Apple Accelerate, etc.) can be used to supply our software with basic linear
algebra capabilities.
Furthermore, \RandBLAS{} relies on the \code{Random123}\footnote{Available at \url{https://github.com/DEShawResearch/random123}.} library for the purpose of exporting random number generators.
For a detailed description of \RandBLAS, visit: 
\begin{quote}
\url{https://randblas.readthedocs.io/en/latest/}.
\end{quote}
In contrast to \RandBLAS, \RandLAPACK's scope still continues to change and expand.
Because of this, we have yet to define explicit project documentation.

All experiments in this manuscript were run using the following version of our software:
\begin{quote}
    \url{https://github.com/BallisticLA/RandLAPACK/releases/tag/BQRRP-benchmark}.
\end{quote}
The code for \code{BQRRP} can be found in \code{/RandLAPACK/drivers/rl\_bqrrp*}.
The code for constructing and dispatching the experiments can be found in \code{/benchmark/bench\_BQRRP/*}, as well as \code{/test/drivers/bench\_bqrrp*}.

An automated script for re-running all of the experiments shown in this paper, as well as the MATLAB plotting scripts for replicating our plots, can be found in: 
\begin{quote}
    \url{https://github.com/BallisticLA/BQRRP_benchmarking}.
\end{quote}

\paragraph{Algorithm performance metric.}
Throughout this paper, we measure the algorithm performance via the \textit{canonical} FLOP rate.
This metric is obtained by dividing the FLOP count of a \textit{standard} algorithm for a given matrix size by the wall clock time required to run
a comparable algorithm under consideration. 
For example, for comparing canonical FLOP rates of various versions of QR and QRCP factorizations, we use the flop count of the standard LAPACK unpivoted QR called \code{GEQRF} \cite[Page 121]{LAWN41:1994}
and divide it by the wall clock time of any given QR or QRCP algorithm being compared.
The canonical FLOP rate gives a clear way to compare algorithm performance by using a standard FLOP count divided by the algorithm’s actual runtime. 
Unlike the raw FLOP rate, which gives credit to unnecessary work, the canonical FLOP rate sets a fair comparison by holding all algorithms to the same amount of work. 
While runtime alone could be used for comparisons, the
FLOP rates are useful because they connect to familiar performance standards, such as the peak FLOP rates advertised for the hardware or standard algorithms like \code{GEMM} for matrix-matrix multiply.
FLOP rates also help reveal scalability issues across problem sizes: if an algorithm’s FLOP rate drops a lot as the problem size grows, it may
reveal an inefficient method or implementation.
Canonical FLOP rates, therefore, give a balanced measure that considers both runtime and standard performance benchmarks.

\paragraph{Numerical properties of test matrices.}
All test matrices were generated using \RandBLAS. For the performance experiments, each matrix entry is independently sampled from the standard normal distribution. 
However, it should be noted that such matrices are appropriate for our
performance tests as being representative of ``typical'' user inputs, and they have
predictable numerical and performance properties. For the numerically
challenging matrices, see \cref{sec:piv_qual}. The average-case
performance, on which we focus extensively here, is much better observed
with our primary choice of matrix elements drawn from i.i.d. distribution, giving a representative behavior in pivot column
interchanges.

\paragraph{Sizes of test matrices.}
We focus our experiments on large matrices where the performance gap between Level-2 and Level-3 BLAS based algorithms is readily apparent.
Simultaneously, we want to ensure that the experiments we set up terminate in \textit{reasonable} time.
Hence, we do not conduct runs on input matrices of the absolute largest sizes that can fit on our systems. 
Once we finish analyzing the established versions of the \code{BQRRP} algorithm for the larger matrix sizes, we present the investigation of the performance of various methods when applied to input matrices of a wide range of sizes, using various numbers of OpenMP threads, in \cref{sec:performance}. 

There are some known performance concerns when it comes to selecting the specific sizes of the input matrices. Specifically,
matrices with column sizes being powers of 2 experience increased
demand for the main memory bandwidth based on how the modern cache memory works: elements from nearby columns are mapped to the exact cache
line because power-of-2 memory addresses have all lower bits set to 0.
Note that all modern libraries use a packed storage format for an efficient in-cache matrix multiply kernel, but they still have to read and write the cached data from/to the main memory, causing false sharing of cache lines. Due to a much-simplified structure of caches, GPUs do not experience this problem, and thus, we needed to compromise on the matrix size
selection for comparable results between the types of computing devices.
Because of this, we chose to run our experiments on matrices with dimensions that are powers of two and multiples of ten.
Additionally, in our main performance experiments with various QR and QRCP algorithms, we focus on square input matrices. This choice provides a meaningful baseline for performance comparisons, while ensuring a balanced computational workload and even distribution of work across threads.
Nonetheless, we do present the performance results captured on both tall and wide matrices in \cref{app:more_aspect_ratios}.

\paragraph{Hardware and software configuration.}
All of our tests used double-precision arithmetic, all code was compiled with the \code{-O3} flag.
The performance of each algorithm is determined by selecting its best execution time from five consecutive runs. 
Instead of running each algorithm individually five times, we execute the entire set of compared algorithms in sequence, repeating this set five times to capture comparatively consistent performance.

The detailed configurations of the machines that we used for conducting the CPU experiments are shown in \cref{table:cpu_config}.

\FloatBarrier
\begin{table}[htp!]
\small\def\arraystretch{1.3}
\centering
\begin{tabular}{|cc|c|c|}

\hline
\multicolumn{2}{|c|}{\textbf{}}                                                                                                                                            & \textbf{Intel Xeon 8462Y+ (2$\times$)} &\textbf{AMD EPYC 9734 (2$\times$)}   \\ \hline
\multicolumn{2}{|c|}{\textbf{Cores per socket}}                                                                                                                            & 32                                  &112   \\ \hline
\multicolumn{2}{|c|}{\textbf{Total threads}}                                                                                                                            & 128                                  &448   \\ \hline
\multicolumn{1}{|c|}{\multirow{2}{*}{\textbf{Clock Speed}}}                                                 & \textbf{\begin{tabular}[c]{@{}c@{}}Base\end{tabular}}         & 2.80 GHz                            &2.2 GHz   \\ \cline{2-4} 
\multicolumn{1}{|c|}{}                                                                                      & \textbf{\begin{tabular}[c]{@{}c@{}}Boost\end{tabular}}        & 4.10 GHz                            &3.0 GHz   \\ \hline
\multicolumn{1}{|c|}{\multirow{3}{*}{\textbf{\begin{tabular}[c]{@{}c@{}}Cache sizes \\ per socket\end{tabular}}}}         & \textbf{L1}                                        & 80 KB                             &64 KB   \\ \cline{2-4} 
\multicolumn{1}{|c|}{}                                                                                      & \textbf{L2}                                                   & 2 MB                               & 1 MB   \\ \cline{2-4} 
\multicolumn{1}{|c|}{}                                                                                      & \textbf{L3}                                                   & 60 MB                            &256 MB   \\ \hline
\multicolumn{2}{|c|}{\textbf{RAM}} &\multicolumn{2}{c|}{DDR5 1TB}   \\ \hline
\multicolumn{2}{|c|}{\textbf{\begin{tabular}[c]{@{}c@{}} FP64 Peak Performance\end{tabular}}}                                                                                        & 5.4 TeraFLOPs                                  & 6.5 TeraFLOPs    \\ \hline
\multicolumn{2}{|c|}{\textbf{BLAS \& LAPACK}} &\multicolumn{2}{c|}{MKL 2025.0}   \\ \hline
\multicolumn{2}{|c|}{\textbf{Compiler}} &\multicolumn{2}{c|}{GCC 13.2.0}   \\ \hline
\multicolumn{2}{|c|}{\textbf{CMake}} &\multicolumn{2}{c|}{3.31.4}   \\ \hline
\multicolumn{2}{|c|}{\textbf{OS}}  &\multicolumn{2}{c|}{Red Hat Enterprise Linux 8.9}   \\ \hline
\end{tabular}
\caption{\small Key details of the hardware and software configuration of the platforms where CPU testing was performed. 
Note that we are using MKL on both Intel and AMD systems, since it performs better than AOCL 5.0 on AMD hardware. 
This observation is discussed in \cref{app:subsec:hqrrp_remake_alternative}.}
\label{table:cpu_config}
\end{table} 
\FloatBarrier

When running benchmarks on CPUs described in \cref{table:cpu_config}, we set the number of OpenMP threads (with \code{OMP\_NUM\_THREADS} environment variable) to the maximum value of threads available, unless specified otherwise.
While using the maximum number of available threads may not always be the optimal approach to achieve peak performance, we advocate for harnessing all available computational resources.
Both tested Intel and AMD systems featured dual-socket configurations. We launched tests with the \code{numactl --interleave=all} command when running our experiments to balance the memory usage across the two memory controllers.

The system setup used for the GPU experiments can be found in \cref{table:gpu_config}.
Since the GPU listed in \cref{table:gpu_config} has ``only'' $80$ GB of memory, we are unable to run double-precision experiments with the input matrices of certain sizes on this GPU.

\FloatBarrier
\begin{table}[htp!]
\small\def\arraystretch{1.3}
\centering
\begin{tabular}{|cc|c|}

\hline
\multicolumn{2}{|c|}{\textbf{}}                                                                                                                                            & \textbf{NVIDIA Hopper (H100)}     \\ \hline
\multicolumn{2}{|c|}{\textbf{CUDA cores}}                                                                                                                                     & $16{,}896$                                       \\ \hline
\multicolumn{1}{|c|}{\multirow{3}{*}{\textbf{\begin{tabular}[c]{@{}c@{}}Memory\end{tabular}}}}         & \textbf{Type}                                                   & HBM3                                 \\ \cline{2-3} 
\multicolumn{1}{|c|}{}                                                                                      & \textbf{Size}                                                   & 80 GB                                   \\ \cline{2-3} 
\multicolumn{1}{|c|}{}                                                                                      & \textbf{Bandwidth}                                                   & 3.35 TB/s                                \\ \hline
\multicolumn{2}{|c|}{\textbf{BLAS \& LAPACK}}                                                                                                                                     &  cuBLAS 12.4.5.8 \& cuSOLVER  11.6.1.9                                     \\ \hline
\multicolumn{2}{|c|}{\textbf{\begin{tabular}[c]{@{}c@{}} FP64 Peak Performance\end{tabular}}}                                                                                     & 60 TeraFLOPs                                    \\ \hline
\multicolumn{2}{|c|}{ \textbf{CUDA NVCC}}                                                                                                                                     & 12.4.1                                     \\ \hline
\multicolumn{2}{|c|}{\textbf{CMake}}                                                                                                                                     & 3.27                                     \\ \hline
\end{tabular}
\caption{\small Key details of the hardware and software configuration of the platform where GPU testing was performed.}
\label{table:gpu_config}
\end{table} 

\section{The framework}
\label{sec:introduce_BQRRP}

As stated in \cref{subsec:contribution}, the intended format for a \code{BQRRP} algorithm is to be in line with that of standard LAPACK QRCP, \code{GEQP3}.
\cref{alg:BQRRP} gives pseudocode toward this end.
For simplicity of presentation, it uses a \textit{simplified} representation of orthogonal factors from
QR factorizations, stating that such factors represent a \textit{square} matrix using some number of elementary reflectors (implicit economical storage format).
The pseudocode is valid in a setting where variables are passed by value rather than by reference, and where functions can return nontrivial datastructures.
Details on how a \code{BQRRP} algorithm can be properly implemented in-place and when working with raw pointers are deferred to \cref{sec:bqrrp_practical}.

The pseudocode in \cref{alg:BQRRP} has two required inputs: the matrix $\mtx{M}$ and an integer block size $b$. 
It relies on five essential
helper functions inside its main loop.
These helper functions are applied to various submatrices whose bounds are computed at 
step \ref{bqrrp:block_partitions}.

\begin{itemize}

\item \code{qrcp\_wide} in step \ref{bqrrp:qrcp} represents
a column-pivoted QR factorization method, suitable for wide matrices.

\item \code{tri\_rank} in
step \ref{bqrrp:rank_est} computes some notion of numerical rank of an input triangular matrix.

\item \code{col\_perm} in
steps \ref{bqrrp:permute_r}, \ref{bqrrp:permute_m}, \ref{bqrrp:update_j}
is responsible for permuting the columns of a given matrix in accordance with the indices stored in a given vector.

\item \code{qr\_tall} in step \ref{bqrrp:qr_tall} performs a tall unpivoted QR factorization.

\item \code{apply\_trans\_q} in steps \ref{bqrrp:apply_q_1} and \ref{bqrrp:apply_q_2} applies the transpose $\mtx{Q}$-factor output by \code{qr\_tall} to a given matrix.
\end{itemize}

\noindent \crefrange{subsec:qrcp_wide}{subsec:col_perm} describe possibilities of how each of these functions might be implemented. 
We recommend specific implementations targeting the CPU architectures from \cref{table:cpu_config}.

\begin{algorithm}[htb!]
\small \caption{Blocked QR with Randomization and Pivoting}
\label{alg:BQRRP}
\begin{algorithmic}[1]

\vspace{0.15em}
\Statex \textbf{Required input:} An $m$-by-$n$ matrix $\mtx{M}$; and an integer block size parameter $b$.

\vspace{0.45em}

\Statex \textbf{Optional input:} A scalar $\gamma$ that sets the size of the sketch relative to $b$ ($\gamma \geq 1$).

\vspace{0.45em}

\Statex \textbf{Output:} Numerical rank $\ell$, orthonormal $m \times m$ matrix $\mtx{Q}$ based on $\ell$ elementary reflectors, upper-trapezoidal $\ell \times n$ matrix $\mtx{R}$, and a column permutation vector $\vct{J}$ of length $n$.

\vspace{0.35em}

\setstretch{1.25}
\Function{bqrrp}{$\mtx{M}, b, \gamma$} \label{bqrrp:input}
\label{bqrrp:input}
    \State Set $d = \lceil \gamma \cdot b \rceil$ and sample a
    $d$-by-$n$ sketching operator $\mtx{S}$ from a Gaussian distribution\label{bqrrp:sample}
    \State Allocate empty $\mtx{Q}$, $\mtx{R}$; $\vct{J} = 1:(n+1)$ \label{bqrrp:alloc}
    \State Sketch $\sk{\mtx{M}} = \mtx{S}\mtx{M}$\label{bqrrp:sketching}
    \For{$i = 0:\lceil n/b \rceil$} \label{bqrrp:loop}
        \State $\sblock = i \cdot b$ is the start of the current row/col block;
        \Statex\hspace{2.6em} $\ecolblock = \min\{n, (i+1)b\}$ is the (exclusive) end
            column of the column block;
        \Statex \hspace{2.6em} $\erowblock = \min\{m, (i+1)b\}$ is the (exclusive) end row of the row block; \label{bqrrp:block_partitions}

        \State Decompose $[\sim, \sk{\mtx{R}}, \sk{\vct{J}}] = \code{qrcp\_wide}(\sk{\mtx{M}}(:, \tslice{s}))$\label{bqrrp:qrcp}
        \Statex ~~~~~~~\, \codecomment{$\sk{\vct{J}}$ is a permutation vector of length $n-s$}
        \State Determine $k = \code{tri\_rank}(\sk{\mtx{R}})$ \label{bqrrp:rank_est}
         \Statex \hspace{2.5em} \codecomment{ $k \leq \min\{b, n-\sblock, m-\sblock\}$}
        \State Permute $\mtx{R}(\lslice{\sblock}, \tslice{\sblock}) = \code{col\_perm}(\mtx{R}(\lslice{\sblock}, \tslice{\sblock}), \sk{\vct{J}})$ \label{bqrrp:permute_r}
        \Statex ~~~~~~~\, \codecomment{ The rectangular portion of the \textbf{computed} rows is permuted
        (no-op at $i = 0$)}
        \State Permute $\mtx{M}(\tslice{\sblock}, \tslice{\sblock}) = \code{col\_perm}(\mtx{M}(\tslice{\sblock}, \tslice{\sblock}), \sk{\vct{J}})$ \label{bqrrp:permute_m}
        \State Update $\vct{J} = \code{col\_perm}(\vct{J}(\tslice{\sblock}), \sk{\vct{J}})$  \label{bqrrp:update_j}
        \State Decompose $[\mtx{Q}^{\mathrm{curr}}, \mtx{R}_{11}] = \texttt{qr\_tall}(\mtx{M}(\tslice{\sblock}, \tslice{\sblock}\ecolblock), k)$\label{bqrrp:qr_tall}
        \Statex ~~~~~~~\, \codecomment{ $\mtx{Q}^{\mathrm{curr}}$ uses $k$ reflectors, implicitly represents $m - \sblock$ columns and $\mtx{R}_{11}$ is $k$-by-$b$}

        \State Update $\mtx{R}(\tslice{\sblock}(\sblock + k), \tslice{\sblock}\ecolblock) = \mtx{R}_{11}$ \label{bqrrp:update_R11}
        \If{$k \neq \min\{b, (n-\sblock), (m-\sblock)\}$} \label{bqrrp:iter_check}
            \State Perform $[\sim, \mtx{R}_{12}] = \code{apply\_trans\_q}(\mtx{Q}^{\mathrm{curr}}, \mtx{M}(\tslice{\sblock}k, \tslice{\ecolblock}))$ \label{bqrrp:apply_q_1}
            \Statex ~~~~~~~~~~~~\, \codecomment{ Output  $\mtx{R}_{12}$ is $k \times (n - \ecolblock)$}
        \Else
            \State Perform $[\mtx{M}^{\mathrm{to\_update}}, \mtx{R}_{12}] = \code{apply\_trans\_q}(\mtx{Q}^{\mathrm{curr}}, \mtx{M}(\tslice{\sblock}, \tslice{\ecolblock}))$ \label{bqrrp:apply_q_2}
            \Statex ~~~~~~~~~~~~\, \codecomment{Output $\mtx{M}^{\mathrm{to\_update}}$ is $(m - \erowblock)$-by-$(n - \ecolblock)$, $\mtx{R}_{12}$ is $b$-by-$(n - \ecolblock)$}
            \State Update $\mtx{M}(\tslice{\erowblock}, \tslice{\ecolblock}) = \mtx{M}^{\mathrm{to\_update}}$ \label{bqrrp:update_M}
        \EndIf
        \State Update $\mtx{R}(\tslice{\sblock}(\sblock+k), \tslice{\ecolblock}) = \mtx{R}_{12}$ \label{bqrrp:update_r12}
        \State Update $\mtx{Q}(\tslice{\sblock}\erowblock, \tslice{\sblock}(\sblock+k)) = \mtx{Q}^{\mathrm{curr}}$ \label{bqrrp:update_q}
        \If{$i = n/b$ or $k \neq \min\{b, (n-\sblock), (m-\sblock)\}$}
        \label{bqrrp:termination}
            \State $\ell = \sblock + k$ \label{bqrrp:update_rank}
            \State \textbf{break}
        \EndIf
        \State Update $\sk{\mtx{M}}(\fslice{}, \tslice{\ecolblock}) = \begin{bmatrix} \sk{\mtx{R}}_{12} - \sk{\mtx{R}}_{11}(\mtx{R}_{11})^{-1}\mtx{R}_{12} \\ \ \sk{\mtx{R}}_{22} \end{bmatrix}$ \label{bqrrp:update_sample}
    \EndFor
    \State \textbf{return} $\ell$, $\mtx{Q}$, $\mtx{R}$, $\vct{J}$ \label{bqrrp:output}
    \EndFunction
\end{algorithmic}
\end{algorithm}


While \code{BQRRP} is a randomized algorithm, it uses randomness only once.
Before entering the main loop, it generates a matrix whose entries are independent mean-zero variance-one Gaussian random variables.\footnote{The unit-variance requirement is not essential, so long as all entries are sampled from the same mean-zero Gaussian distribution.}
This matrix is applied $\mtx{M}$ to obtain a smaller matrix called the \textit{sketch}.
The number of rows in the sketch is $d = \lceil \gamma b \rceil$, where $\gamma$ is called the \textit{sampling factor}.
The natural default value of $\gamma$ depends on the implementation of \code{qcrp\_wide} (in our recommended implementation, the only reasonable value is $\gamma = 1$).
As \code{BQRRP} iterates, the sketch is updated deterministically with a method proposed by Duersch and Gu~\cite[Section 4]{DG:2017:QR}.

\begin{remark}
    In our prior work for QRCP of tall matrices, \cite{MBM2024}, we used a fast sparse operator to prevent sketching from becoming a computational bottleneck.
    In the context of \code{BQRRP}, Gaussian sketching has no risk of being a bottleneck operation.
    This is because \code{BQRRP} sketches in the \textit{sampling regime} rather than the \textit{embedding regime}, to borrow terms from  \cite[Section 2.2]{RandLAPACK_Book}.
\end{remark}

Note that \crefrange{subsec:qrcp_wide}{subsec:col_perm} do not dwell too much on the edge cases of \cref{alg:BQRRP} ($m$, $n$ not divisible by $b$, small square inputs in the described pseudocodes, etc.), instead describing the default formulation of the subroutines that go into \cref{alg:BQRRP} and their performance on the most relevant inputs.

\FloatBarrier
\subsection{A practical wide QRCP selection}
\label{subsec:qrcp_wide}

At iteration $i \in \lslice{(\lceil n/b \rceil-1)}$, step \ref{bqrrp:qrcp} in \cref{alg:BQRRP} uses a \code{qrcp\_wide} function, applied to a wide (for most $i$) sketched matrix $\sk{\mtx{M}} \in \R^{d \times (n-ib)}$.
One could implement this by calling the standard LAPACK QRCP function, \code{GEQP3}. 
In our high-performance implementation of \code{BQRRP}, we employ
an approach that uses the LU factorization with partial pivoting,
\code{GETRF}, to retrieve the pivot vector $\sk{\vct{J}}$ and then unpivoted QR, \code{GEQRF}, to find the matrix $\sk{\mtx{R}}$ needed in the sample updating step (step \ref{bqrrp:update_sample}).
\cref{alg:qrcp_practical} shows how such a method is implemented. 

\begin{algorithm}[htb]
\small \setstretch{1.2}
\caption{ : Practical wide QRCP }
\label{alg:qrcp_practical}
\begin{algorithmic}[1]
\Statex \textbf{Input:} a submatrix $\sk{\mtx{M}} \in \mathbb{R}^{d \times (n-ib)}$, where $d = \lceil \gamma b \rceil \geq b$.
\setstretch{1.25}
\State \textbf{function} $\code{qrcp\_practical}(\sk{\mtx{M}})$
\vspace{-4pt}
\Indent
    \State Allocate $\mtx{M}^{\mathrm{sk\_trans}} = \code{transpose}(\sk{\mtx{M}})$ \label{qrcp:transpose}
    \State Compute $[\sim, \sim,\vct{J}^\mathrm{lu}] = \code{lu}(\mtx{M}^{\mathrm{sk\_trans}})$
    \Statex ~~~\, \codecomment{ Done via standard row-pivoted LU factorization, \code{GETRF}}
    \State Convert $\vct{J}^\mathrm{qr} = \code{piv\_transform}(\vct{J}^\mathrm{lu})$\label{qrcp:piv_transform}
    \State Permute $\sk{\mtx{M}} = \code{col\_perm(}\sk{\mtx{M}}, \vct{J}^\mathrm{qr})$ \label{wide_qrcp:permute}
    \State Compute $[\sk{\mtx{Q}}, \sk{\mtx{R}}] = \code{qr}
    (\sk{\mtx{M}})$ \label{wide_qrcp:compute}
    \Statex ~~~\, \codecomment{ Done via standard unpivoted QR factorization, \code{GEQRF}}
    \State \textbf{return} $\sk{\mtx{Q}}$, $\sk{\mtx{R}}$, $\vct{J}^\mathrm{qr}$
\EndIndent    
\end{algorithmic}
\end{algorithm}

Step \ref{qrcp:transpose} in \cref{alg:qrcp_practical} does not use an in-place transpose on purpose.
Since the matrix $\sk{\mtx{M}}$ that is to be used by the
\code{qrcp\_wide} function in step \ref{alg:qrcp_practical} of \cref{alg:BQRRP} is marginally smaller than $\mtx{M}$, allocating a $d \times n$ buffer for $\mtx{M}^{\mathrm{sk\_trans}}$ is more efficient
than performing an additional in-place transpose to restore $\sk{\mtx{M}}$ at the end of the algorithm.

\begin{remark}
    Note that a portion of $\sk{\mtx{R}}$ can be used in step \ref{bqrrp:qr_tall} for preconditioning the next panel of $\mtx{M}$
    (see details in \cref{subsec:qr_tall}).
\end{remark}

\paragraph{Permutation formats.}
Step \ref{qrcp:piv_transform} in \cref{alg:qrcp_practical} is crucial, since the format in which a pivoted LU represents the permutation vector $\vct{J}$ is different from the pivoted QR format, and hence the format conversion procedure is required. 
In the context of pivoted LU factorization, row $i$ of the input matrix was interchanged with row $\vct{J}^\mathrm{lu}(i)$.
The format conversion 
consists of first creating a vector $\vct{J}^\mathrm{qr}$ of length $n$ with entries from $1$ to $n$ and then serially swapping elements in it according to the entries in $\vct{J}^\mathrm{lu}$. 
Simply put, for element at index $i \in \lslice{(n-1)}$, element at $\vct{J}^\mathrm{qr}(i)$ is to swap positions with the element at $\vct{J}^\mathrm{qr}(\vct{J}^\mathrm{lu}(i) - 1)$. 

\begin{remark}
  When implementing the pivot translation procedure in practice, it is important to remember that pivot vectors in LAPACK store entries in a one-based index format used by Fortran and MATLAB (as stated previously in \cref{subsec:def_and_notations}).
\end{remark}

\paragraph{Wide QRCP and sketching.}
When \cref{alg:qrcp_practical} is in use in \code{BQRRP}, the first $b$ components of $\vct{J}^{\text{qr}}$ are the same for $\gamma = 1$ and $\gamma > 1$.
Therefore the rank-revealing properties of \code{BQRRP} using \cref{alg:qrcp_practical} for \code{qrcp\_wide} cannot be improved by using $\gamma > 1$.

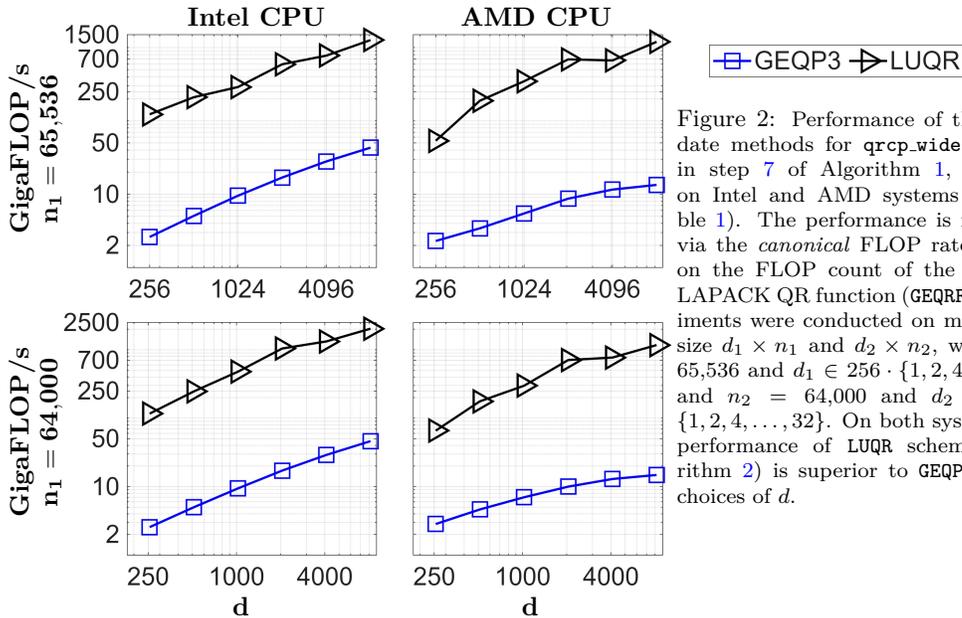
\begin{figure}[htb!]
  \centering
  \hspace*{0.0cm}
  \begin{tikzpicture}
    \node[inner sep=0pt] at (0,0) {\includegraphics[width=1.0\linewidth]{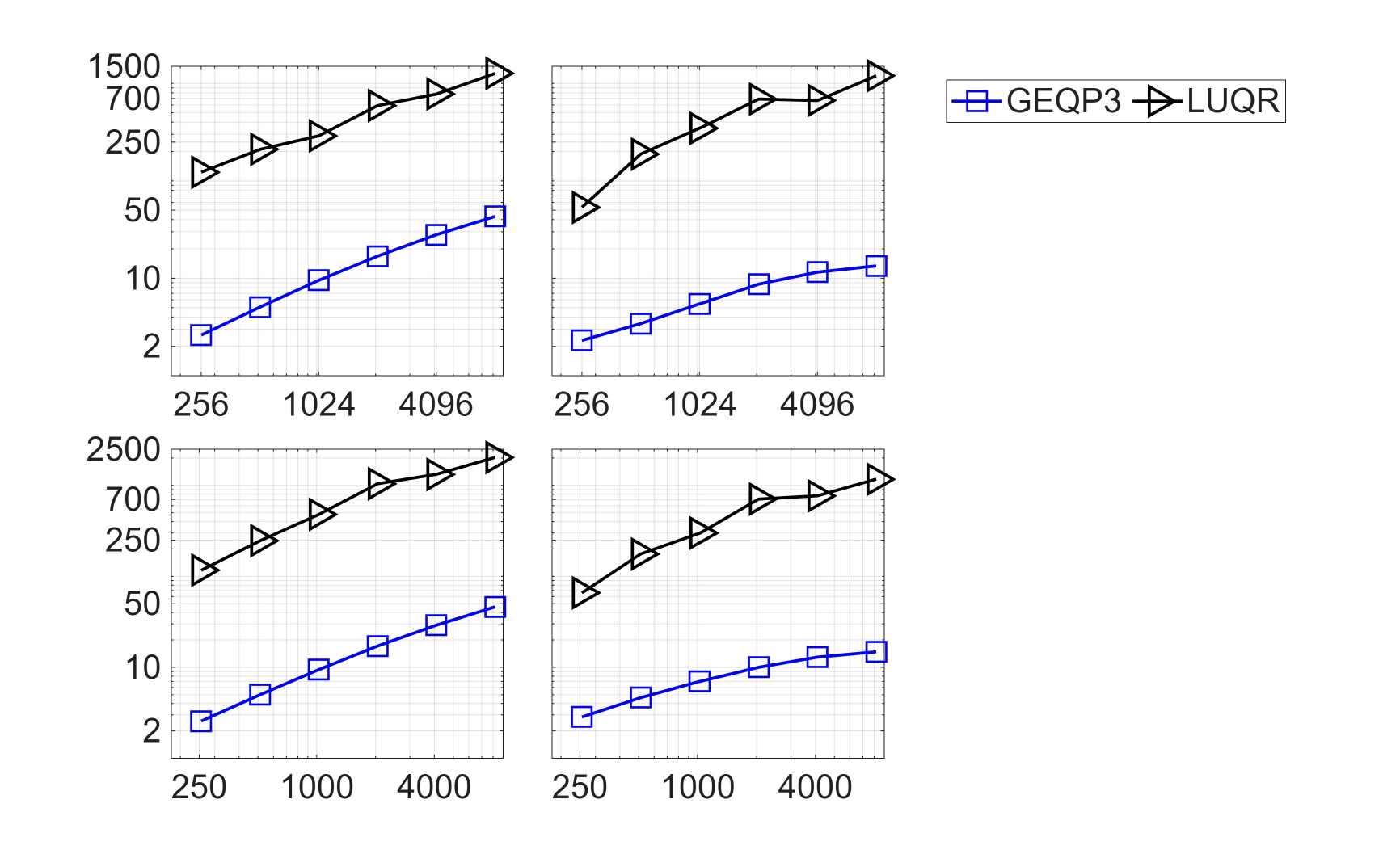}};

    \node[anchor=north west, text width=5.0cm, align=justify] at (1.9, 3) {%
      \caption{\footnotesize{Performance of the candidate methods for \code{qrcp\_wide} function in step \ref{bqrrp:qrcp} of \cref{alg:BQRRP}, captured on Intel and AMD systems (see \cref{table:cpu_config}). 
    The performance is measured via the \textit{canonical} FLOP rate, relying on the FLOP count of the standard LAPACK QR function (\code{GEQRF}).
    Experiments were conducted on matrices of size $d_1 \times n_1$ and $d_2 \times n_2$, with $n_1 = 65{,}536$ and $d_1 \in 256 \cdot \{1, 2, 4, \dots, 32\}$, and $n_2 = 64{,}000$ and $d_2 \in 250 \cdot \{1, 2, 4, \dots, 32\}$.
    On both systems, the performance of \code{{LUQR}} scheme (\cref{alg:qrcp_practical}) is superior to \code{GEQP3}, for all choices of $d$. 
    }}
      \label{fig:wide_qrcp}
    };

    \node[anchor=north west, font=\bfseries] at (-0.33\linewidth, 4.1) {Intel CPU};
    \node[anchor=north west, font=\bfseries] at (-0.07\linewidth, 4.1) {AMD CPU};

    \node[anchor=north west, rotate=90, font=\bfseries] at (-0.50\linewidth, 0.06\linewidth) {\textbf{GigaFLOP/s}};
    \node[anchor=north west, rotate=90, font=\bfseries] at (-0.47\linewidth, 0.07\linewidth) {\textbf{$\mathbf{n_1 = 65{,}536}$}};

    \node[anchor=north west, rotate=90, font=\bfseries] at (-0.50\linewidth, -0.21\linewidth) {\textbf{GigaFLOP/s}};
    \node[anchor=north west, rotate=90, font=\bfseries] at (-0.47\linewidth, -0.20\linewidth) {\textbf{$\mathbf{n_1 = 64{,}000}$}};

    \node at (-0.27\linewidth, -4) {$\mathbf{d}$};
    \node at (0.005\linewidth, -4) {$\mathbf{d}$};
  \end{tikzpicture}
\end{figure}

\paragraph{Candidates' performance.}
Observe a practical comparison of the performance of the two approaches in \cref{fig:wide_qrcp}.
The performance superiority of \cref{alg:qrcp_practical} to the standard QRCP approach makes it the best option to be used in a \code{BQRRP} implementation.

\FloatBarrier
\subsection{Numerical rank selection}
\label{subsec:rank_est}

In order to understand the role of \code{tri\_rank} at step \ref{bqrrp:rank_est}, it is useful to pretend that all computations in \code{BQRRP} and its subroutines are performed in exact arithmetic.
If the current block is not full-rank ($k \neq  \min\{b, (n-\sblock), (m-\sblock)\}$), then the current iteration of \code{BQRRP} will be the final one.
To see why this is reasonable, suppose \code{tri\_rank} returns the exact rank of its input matrix.
One can show that, conditional on an event which occurs with probability 1, \code{BQRRP} returns $\ell = \rank(\mtx{M})$ and a full decomposition of $\mtx{M}$.
If the update at Step \ref{bqrrp:update_sample} were still performed, $\sk{\mtx{M}}(\fslice{},\tslice{c})$ would be the zero matrix.

Of course, it is unrealistic to assume black-box access to a method to compute the exact rank of a floating-point matrix.
This raises the question of whether, setting aside rounding errors, we could ensure a full decomposition of $\mtx{M}$ if we allowed for overestimation of rank.
The answer is that \textit{we can};
it is valid for \code{tri\_rank} to simply return the dimension $p$ of the input $p \times p$ triangular matrix.
This choice can be used with minor modifications to the rest of \code{BQRRP} and would ensure that \code{BQRRP} mimics \code{GEQP3} as closely as possible.
It would be sufficient to use a different updating formula at step \ref{bqrrp:update_sample} that remains well-posed if  $\mtx{R}_{11}$ is singular; two such methods are described in \cite[\S 4]{DG:2017:QR}.
It would also be sufficient to keep the ill-posed update, while ensuring that \code{qrcp\_wide} returns $\sk{\vct{J}} = (1,2,\ldots,n-s+1)$ when $\sk{\mtx{M}}$ contains infs or NaNs.
Since $\sk{\vct{J}}$ only affects the pivot decisions used in $\mtx{M}$, a stable Householder QR method can safely be applied in \code{qr\_tall}, without being impacted by $\sk{\mtx{M}}$ becoming ill-formed.

The \code{BQRRP} implementation in RandLAPACK uses a naive rank estimator that is described in our prior work on preconditioned column-pivoted Cholesky QR~\cite{MBM2024}.
This strategy is only needed for implementations of \code{qr\_tall} based on Cholesky QR.
More sophisticated strategies would be needed if \code{BQRRP} were intended as a drop-in replacement for LAPACK 3.12's \code{GEQP3RK} function for truncated QRCP of low-rank matrices.
Such strategies are beyond the scope of this manuscript.
In particular, all performance experiments we conduct are on matrices of full numerical rank.

\subsection{Tall QR selection}
\label{subsec:qr_tall}

Step \ref{bqrrp:qr_tall} in \cref{alg:BQRRP} is concerned with performing unpivoted QR on a (generally) tall submatrix of the input matrix that has been permuted via a permutation vector, computed according to the description in \cref{subsec:qrcp_wide}.
Thus, any QR factorization method that can handle tall matrices is suitable here. The resulting $\mtx{Q}$-factor must follow the format outlined in \cref{subsec:def_and_notations}.
If the selected QR method produces $\mtx{Q}$ in a different format, it should be converted to the required representation.
Additionally, at iteration $i \in \lslice{(\lceil n/b \rceil-1)}$, the tall QR is to be performed on a matrix of size
$(m-ib)$-by-$b$ (except possibly in the last iteration if $n$ is not evenly divisible by $b$), regardless of whether the block has full rank $k$ (estimated with the procedure described in \cref{subsec:rank_est}).
Despite the fact that $b$ reflectors would be computed in that case, we would use only $k$ of them outside of this step.

\paragraph{Available methods.} 
LAPACK offers several algorithms suitable for tall QR.
The natural choice in this context is \code{LATSQR} \cite{lapack_latsqr}:
a specialized Householder QR factorization for tall matrices.
Alternatively, one could use a standard Householder QR factorization, \code{GEQRF} \cite{lapack_geqrf}. 
As the third option, \code{GEQRT} \cite{lapack_geqrt}, is a blocked algorithm based on compact WY representation~\cite{Bischof:1987:WRP:37316.37324}.
Finally, LAPACK's \code{GEQR} calls either \code{LATSQR} or \code{GEQRT}, depending on the size of the input matrix and the output from LAPACK's
\code{ILAENV} inquiry function~\cite{lapack_ilaenv}.
In addition to listing the standard LAPACK algorithms, we consider the use of \textit{Cholesky QR} in this context, following our approach from
prior work~\cite{MBM2024}. 
Background on Cholesky QR is given in \cref{app:cholqr}.

\begin{remark} In principle, one could use a version of the Gram-Schmidt method in order to obtain an \textit{explicit} version of the $\mtx{Q}$-factor. 
While this may be a viable option some settings, it would increase space requirements by $m^2$ words and it would not meet our output format requirements.
\end{remark}

\paragraph{Cholesky QR dependencies.} Although Cholesky QR can be safe to use in the context of \code{BQRRP}, its use produces the following complication: steps \ref{bqrrp:apply_q_1} and \ref{bqrrp:apply_q_2} require access to the fully-formed $m$-by-$m$ representation of the orthonormal factor computed at the current iteration, while Cholesky QR only outputs an \textit{explicit} economical representation  $\mtx{Q}^{\mathrm{chol}} \in \R^{m \times k}$.
We can acquire the full representation by using a specialized LAPACK routine \code{ORHR\_COL}.
This routine transforms an explicit economical $\mtx{Q}^{\mathrm{chol}}$ output by Cholesky QR into an implicit representation via \textit{Householder reconstruction}. 
The description of the basic approach for performing the Householder reconstruction can be found in \cite[Algorithms 5, 6]{BD2015}.
An advanced recursive implementation of \code{ORHR\_COL} is described in its respective Netlib LAPACK documentation \cite{lapack_orhr_col}; see also \cite{F1997}.

\cref{alg:cholqr_deps} shows
a practical implementation
of Cholesky QR and its dependencies, offering a method that can be used in step \ref{bqrrp:qr_tall}.

\FloatBarrier

\begin{algorithm}[htb]
\small \setstretch{1.2}
\caption{ : Cholesky QR + dependencies in the context of \code{qr\_tall}}
\label{alg:cholqr_deps}
\begin{algorithmic}[1]
\Statex \textbf{Input:} Iteration $i \in \lslice{(\lceil n/b \rceil-1)}$ (from \cref{alg:BQRRP}); matrix $\perm{\mtx{M}} \in \R^{(m-ib) \times b}$, where $b \ll m$; upper-triangular $\sk{\mtx{R}} \in \R^{d \times (n-ib)}$ (output from step \ref{bqrrp:qrcp} in \cref{alg:BQRRP}), where $d = \lceil \gamma b \rceil \geq \gamma b \geq b$ when $\gamma \geq 1$; block rank $k \leq b$
\setstretch{1.25}
\State \textbf{function} $\code{cholqr\_deps}(\perm{\mtx{M}}, \sk{\mtx{R}}, k)$
\vspace{-4pt}
\Indent
    \State Truncate and precondition $\pre{\mtx{M}} = \perm{\mtx{M}}(\fslice{}, \lslice{k})(\sk{\mtx{R}})^{-1}$ \label{cholqr:precond}
    \State Decompose $[\mtx{Q}^{\mathrm{chol}}, \mtx{R}^{\mathrm{chol}}] = \texttt{cholqr}(\pre{\mtx{M}})$\label{cholqr:cholqr}
    \Statex ~~~~~~~\, \codecomment{ $\mtx{Q}^{\mathrm{chol}}$ is explicit $(m-ib)$-by-$k$; $\mtx{R}^{\mathrm{chol}}$ is $k$-by-$k$}
    \State Reconstruct $[\mtx{Q}^{\mathrm{curr}}, \vct{D}] = \code{householder\_reconstruct}(\mtx{Q}^{\mathrm{chol}})$ \label{cholqr:orhr_col}
    \Statex ~~~~~~~\, \codecomment{ Using \code{ORHR\_COL}; $\mtx{Q}^{\mathrm{curr}}$ uses $k$ reflectors, represents $m-ib$;  $\vct{D}$ is a sign vector}
    \State Compute $\mtx{R}_{11} = \mtx{R}^{\mathrm{chol}}\code{diag}(\vct{D})\sk{\mtx{R}}(\lslice{k}, \lslice{b})$\label{cholqr:undo_precond}
    \Statex ~~~~~~~\, \codecomment{ Undoing the preconditioning; $\mtx{R}_{11}$ is $k \times b$}
    \State \textbf{return} $\mtx{Q}^{\mathrm{curr}}$, $\mtx{R}_{11}$
\EndIndent    
\end{algorithmic}
\end{algorithm}

\begin{figure}[htb!]
  \centering
  \hspace*{-0.2cm}
  \begin{tikzpicture}
    \node[inner sep=0pt] at (0,0) {\includegraphics[width=1.0\linewidth]{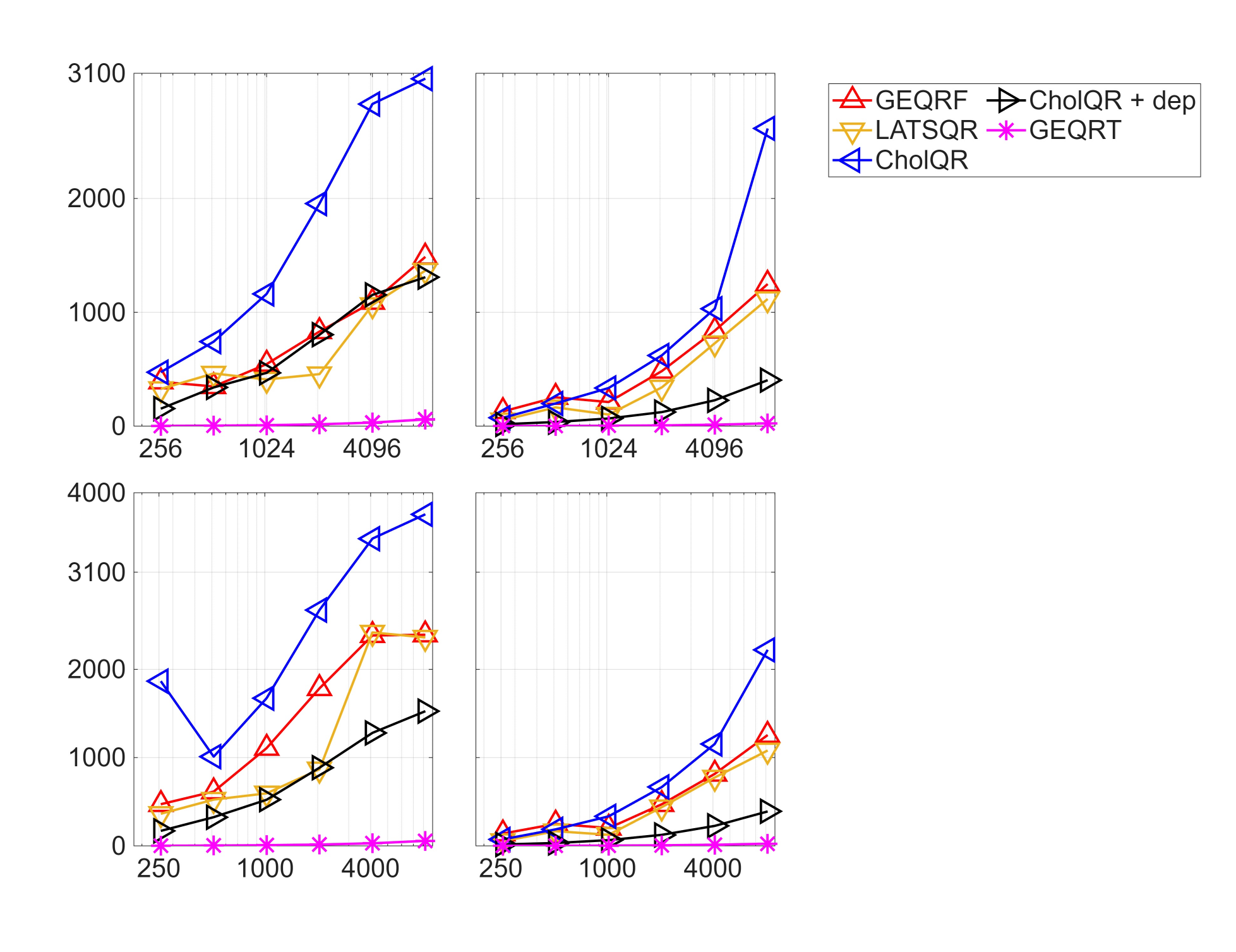}};

    \node[anchor=north west, text width=4.9cm, align=justify] at (1.9, 3.5) {%
    \caption{\footnotesize{Performance of the candidate methods for \code{qr\_tall} function in step \ref{bqrrp:qr_tall} of \cref{alg:BQRRP}, captured on Intel and AMD systems (see \cref{table:cpu_config}). 
    The performance is measured via \textit{canonical} FLOP rate, using the FLOP count of LAPACK's \code{GEQRF}, on matrices of sizes $m_1 \times k_1$ and $m_2 \times k_2$, with $m_1 = 65{,}536$ and $k_1 \in 256 \cdot \{1, 2, 4, \dots, 32\}$, and $m_2 = 64{,}000$ and $k_2 \in {250\cdot\{1, 2, 4, \dots, 32\}}$.
    Across all systems, we observe that Cholesky QR outperforms all alternative methods; however, the version of Cholesky QR with all of its dependencies (\cref{alg:cholqr_deps}) is only competitive on the Intel system with the input data of dimensions that are powers of two.
    Otherwise, on both Intel and AMD systems, \code{GEQRF} shows the best performance out of the practical alternatives.
    }}
    \label{fig:qr_tall}
    };

    \node[anchor=north west, font=\bfseries] at (-0.35\linewidth, 5.0) {Intel CPU};
    \node[anchor=north west, font=\bfseries] at (-0.08\linewidth, 5.0) {AMD CPU};

    \node[anchor=north west, rotate=90, font=\bfseries] at (-0.52\linewidth, 0.1\linewidth) {\textbf{GigaFLOP/s}};
    \node[anchor=north west, rotate=90, font=\bfseries] at (-0.49\linewidth, 0.11\linewidth) {\textbf{$\mathbf{m_1 = 65{,}536}$}};

    \node[anchor=north west, rotate=90, font=\bfseries] at (-0.52\linewidth, -0.23\linewidth) {\textbf{GigaFLOP/s}};
    \node[anchor=north west, rotate=90, font=\bfseries] at (-0.49\linewidth, -0.22\linewidth) {\textbf{$\mathbf{m_1 = 64{,}000}$}};

    \node at (-0.27\linewidth, -5) {$\mathbf{k}$};
    \node at (0.005\linewidth, -5) {$\mathbf{k}$};
  \end{tikzpicture}
\end{figure}
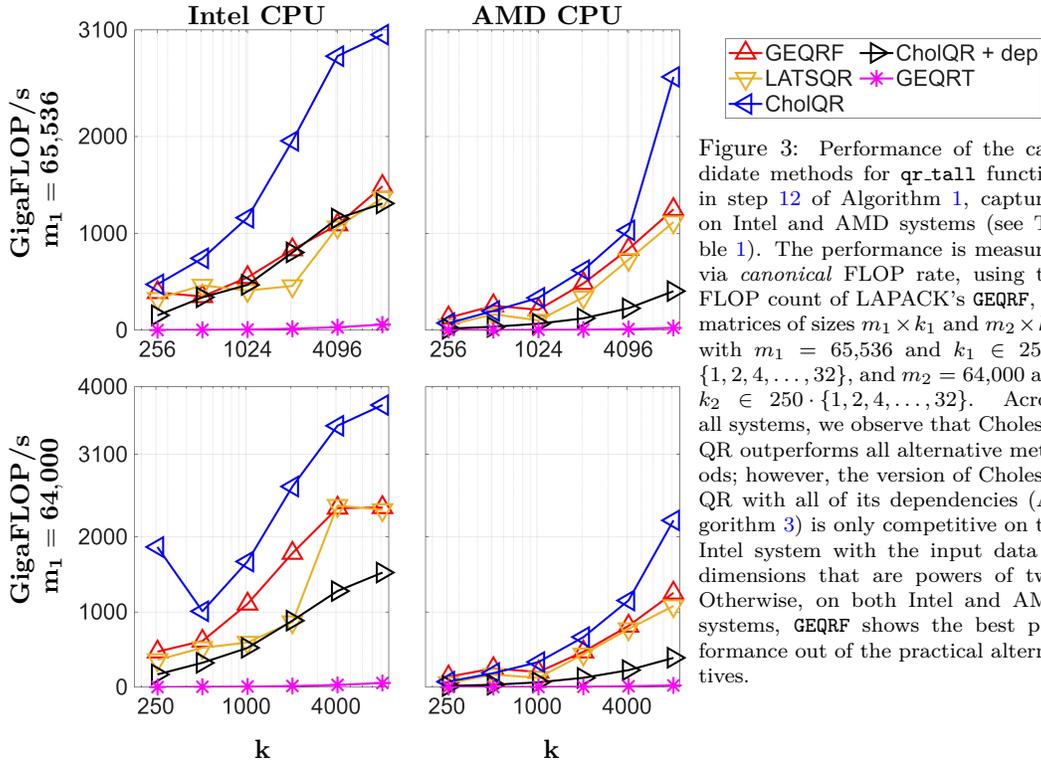

\paragraph{Candidates' performance.} 
\cref{fig:qr_tall} illustrates how the use of preconditioning and Householder restoration affect the performance of Cholesky QR, relative to alternative methods for tall QR factorization.
\cref{fig:qr_tall} shows that \code{GEQRF} is the best alternative method for tall QR factorization across both systems. 
Additionally, favoring the use of Cholesky QR on the Intel system for the input matrices of select sizes can be a reasonable option.

\begin{remark}
    In theory, the performance of \code{GEQRT} should be comparable to that of \code{GEQRF}, as the primary distinction between the two is that \code{GEQRT} provides access to the $\mtx{T}$ matrix, which encodes the sequence of upper triangular block reflectors. However, as shown in \cref{fig:qr_tall}, \code{GEQRT} performs significantly worse than all other tall QR implementations. Varying the internal block size had no noticeable impact on performance. We suspect this is because \code{GEQRT} is not as heavily optimized in MKL 2025.0 as other QR routines.
\end{remark}

\paragraph{Output $\mtx{Q}$ representation}
Only \code{GEQRF} outputs the representation of $\mtx{Q}$ that precisely matches the one we are after (described in \cref{subsec:def_and_notations}). 
The format of $\mtx{Q}$ produced by \code{ORHR\_COL} differs from the intended one in the following way: instead of generating a vector with scalar factors of the elementary reflectors $\vct{\tau}$, it produces an upper-trapezoidal matrix $\mtx{T}$ that represents a sequence of upper triangular block reflectors stored compactly. Each block in $\mtx{T}$ is of size $n_b \times b$
(except possibly in the last iteration if $n$ is not evenly divisible by $b$),
with $n_b$ being the block size parameter.
The scalar factors of the elementary reflectors, originally intended for $\vct{\tau}$, are stored along the diagonals of the block matrix $\mtx{T}$.

For details of the representations used by \code{GEQR}, \code{LATSQR}, and \code{GEQRT},
we refer the readers to the official LAPACK documentation.

\subsection{Selection of methods for applying the Q-factor}
\label{subsec:updating_rules}

In \code{BQRRP}, steps \ref{bqrrp:apply_q_1} and \ref{bqrrp:apply_q_2} represent an application of a transpose of a full representation of the current iterations's $\mtx{Q}$ factor, $(\mtx{Q}^{\mathrm{curr}})^{\trans}$ (obtained as described in \cref{subsec:qr_tall}) to a pivoted trailing portion of the current iteration's matrix $\mtx{M}$.
Whether line \ref{bqrrp:apply_q_1} or \ref{bqrrp:apply_q_2} is executed
depends on whether the current block rank, estimated as outlined in \cref{subsec:rank_est}, equals the number of columns in the current block.
The distinction between these lines lies in what is updated during the current iteration of the main \code{BQRRP} loop. 
One approach updates only a portion of the output 
$\mtx{R}$-factor (in this case, the current iteration is the final one). The other approach updates both the 
$\mtx{R}$-factor and the trailing portion of the input matrix, enabling the algorithm to proceed with further iterations.
In an actual implementation (i.e., one where the decomposition is performed in place), no logical branching is needed to distinguish between lines \ref{bqrrp:apply_q_1}  and \ref{bqrrp:apply_q_2}.
The difference is only a matter of how many rows of $\mtx{M}$ are involved in the computation.

\paragraph{Available methods.}
There are two main methods for applying $(\mtx{Q}^{\mathrm{curr}})^{\trans}$ to a given matrix; the choice of the right method relates to the decision made in the tall QR step.

The first available method is LAPACK's \code{ORMQR} function~\cite{lapack_ormqr},
which takes reflectors produced by \code{GEQRF} or \code{GEQP3}.
Recall that \cref{subsec:qr_tall} states that the \code{GEQP3}-compatible representation of $\mtx{Q}$ must be obtained regardless of the algorithm used in the tall QR step.
Therefore \code{ORMQR} is always an option after obtaining the \code{GEQP3}-compatible representation.
The cost of \code{ORMQR} is $4nmk - 2nk^2 + 3nk$ FLOPs \cite[Page 122]{LAWN41:1994}.

The second option to use in this context is the \code{GEMQRT} function~\cite{lapack_gemqrt}.
The input format of \code{GEMQRT} matches the output format of \code{ORHR\_COL}.
Since Cholesky QR must use \code{ORHR\_COL}, this implies pairing Cholesky QR with \code{GEMQRT} can be
an efficient choice.
Alternatively, one could pair \code{GEMQRT} with any tall QR method, first ensuring that the representation of the output $\mtx{Q}$-factor is made compatible with the input representation of \code{GEMQRT}.

One more alternative to the methods described above is avoiding the trailing update altogether, as described in \cite[Algorithm 5.1]{DG:2017:QR}.
This approach is, however, only applicable to low-rank data, and in an implementation, it would further increase the complexity of a rather nontrivial \code{BQRRP} scheme.

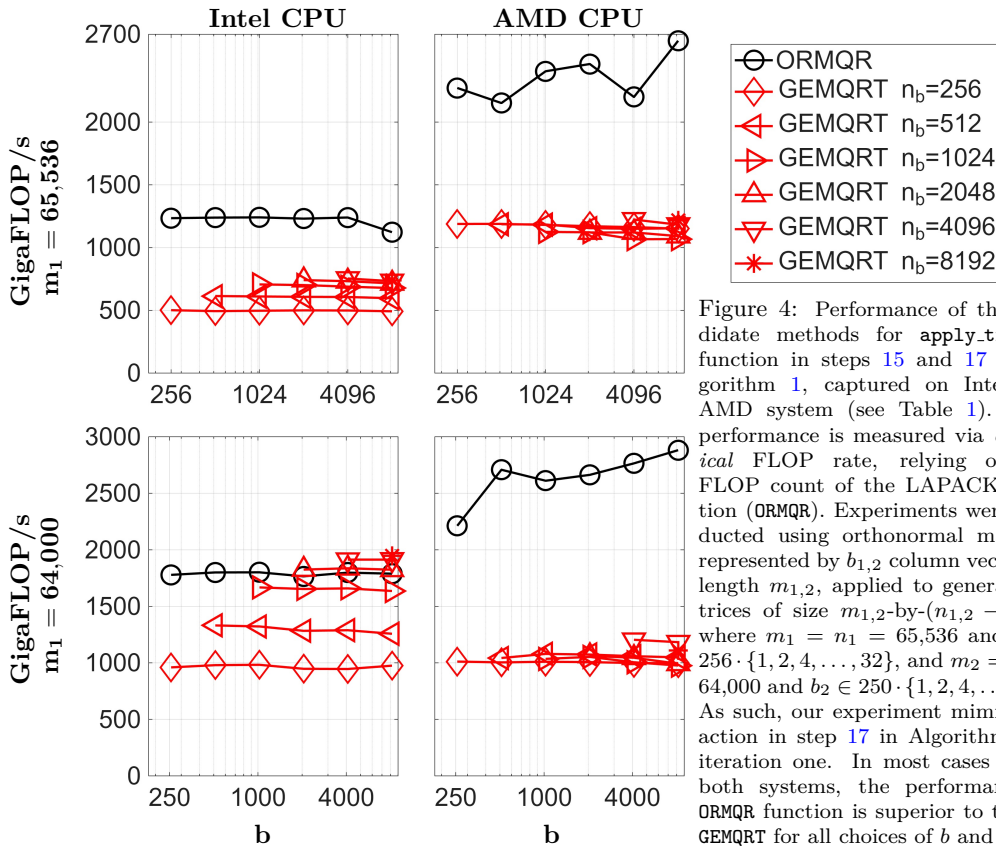
\begin{figure}[htb!]
  \centering
  \hspace*{-0.2cm}
  \begin{tikzpicture}
    \node[inner sep=0pt] at (0,0) {\includegraphics[width=1.0\linewidth]{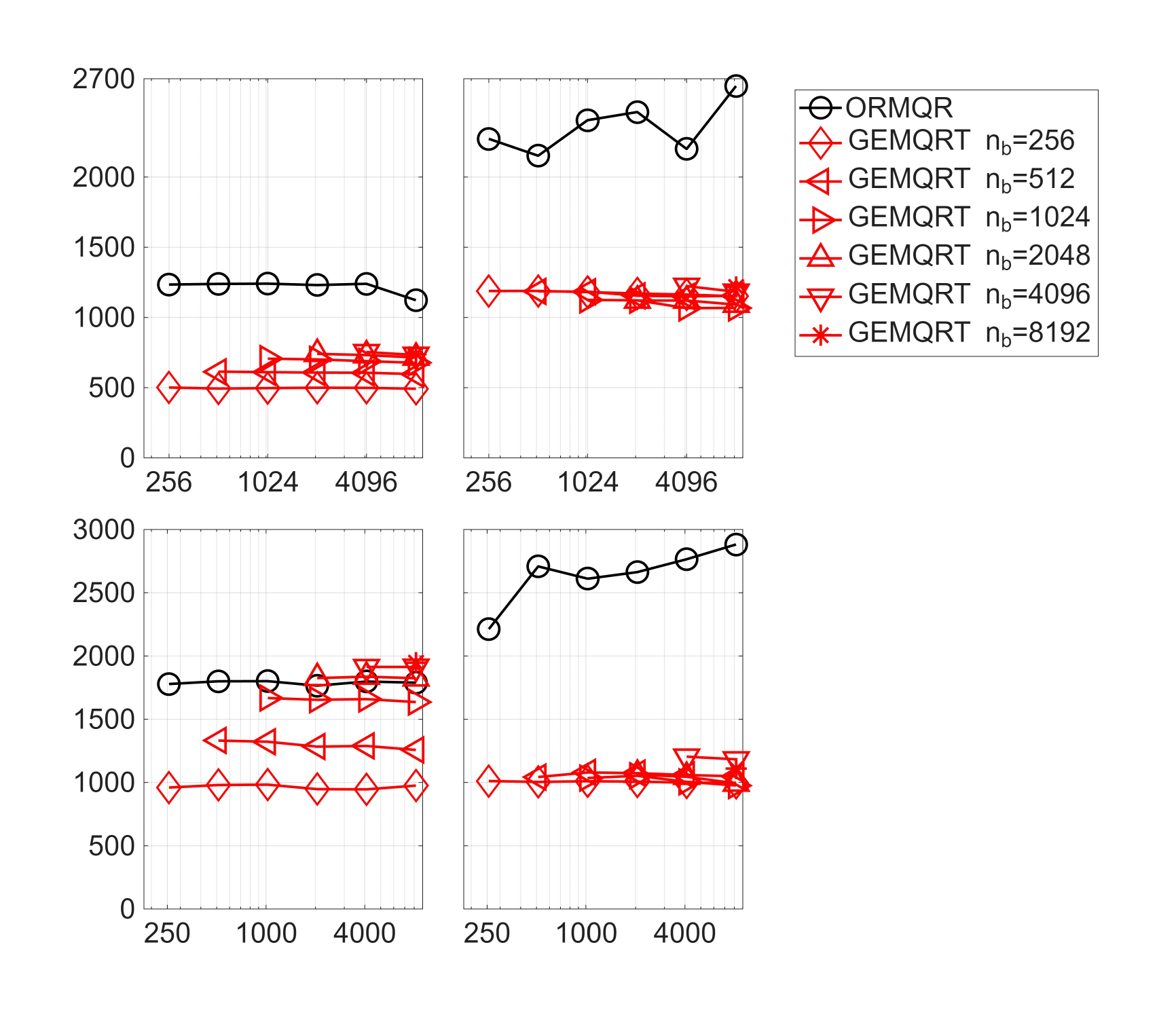}};

    \node[anchor=north west, text width=4.8cm, align=justify] at (1.9, 2) {%
    \caption{\footnotesize{Performance of the candidate methods for \code{apply\_trans\_q} function in steps \ref{bqrrp:apply_q_1} and \ref{bqrrp:apply_q_2} of \cref{alg:BQRRP}, captured on Intel and AMD system (see \cref{table:cpu_config}). 
    The performance is measured via \textit{canonical} FLOP rate, relying on the FLOP count of the LAPACK function (\code{ORMQR}).
    Experiments were conducted using orthonormal matrices represented by $b_{1,2}$ column vectors of length $m_{1,2}$, applied to general matrices of size $m_{1,2}$-by-$(n_{1,2}-b_{1,2})$, where $m_1 = n_1 = 65{,}536$ and $b_1 \in 256 \cdot \{1, 2, 4, \dots, 32\}$, and $m_2 = n_2 = 64{,}000$ and $b_2 \in 250 \cdot \{1, 2, 4, \dots, 32\}$.
    As such, our experiment mimics the action in step \ref{bqrrp:apply_q_2} in \cref{alg:BQRRP} at iteration one.
    In most cases across both systems, the performance of \code{{ORMQR}} function is superior to that of \code{GEMQRT} for all choices of $b$ and $n_b$.
    }}
    \label{fig:apply_Q}
    };

    \node[anchor=north west, font=\bfseries] at (-0.33\linewidth, 5.6) {Intel CPU};
    \node[anchor=north west, font=\bfseries] at (-0.06\linewidth, 5.6) {AMD CPU};

    \node[anchor=north west, rotate=90, font=\bfseries] at (-0.52\linewidth, 0.1\linewidth) {\textbf{GigaFLOP/s}};
    \node[anchor=north west, rotate=90, font=\bfseries] at (-0.49\linewidth, 0.11\linewidth) {\textbf{$\mathbf{m_1 = 65{,}536}$}};

    \node[anchor=north west, rotate=90, font=\bfseries] at (-0.52\linewidth, -0.26\linewidth) {\textbf{GigaFLOP/s}};
    \node[anchor=north west, rotate=90, font=\bfseries] at (-0.49\linewidth, -0.25\linewidth) {\textbf{$\mathbf{m_1 = 64{,}000}$}};

    \node at (-0.27\linewidth, -5.5) {$\mathbf{b}$};
    \node at (0.005\linewidth, -5.5) {$\mathbf{b}$};
  \end{tikzpicture}
\end{figure}

\paragraph{Candidates' performance.}
The relative performance of the alternative ways of applying an orthonormal factor $(\mtx{Q}^{\mathrm{curr}})^{\trans}$ to an arbitrary matrix of size $m$-by-$(n-b)$ is shown in \cref{fig:apply_Q}.
The performance superiority of \code{ORMQR} function to the alternatives in the majority of the explored cases makes it the best option to be used in a \code{BQRRP} implementation.

\begin{remark}
    Similar to the underperformance of \code{GEQRT}, which we attributed to limited optimization in MKL 2025.0 \code{GEMQRT} also exhibits subpar performance compared to \code{ORMQR}, as shown in \cref{fig:apply_Q}, despite the expectation that their performance should be comparable.
\end{remark}

\subsection{Selection of implementation for column permutation}
\label{subsec:col_perm}

A \code{BQRRP} algorithm requires a conceptually trivial, yet crucially important kernel:
a function for permuting columns of a given matrix in accordance with the pivot vector output from  \code{qrcp\_wide} at step \ref{bqrrp:qrcp}.
This kernel
is used in \cref{alg:BQRRP} at step \ref{bqrrp:permute_r} for permuting the columns in the rectangular portion of the $\mtx{R}$-factor, at step \ref{bqrrp:permute_m} for permuting the columns of a submatrix
of the input matrix, and at step \ref{bqrrp:update_j} for updating the permutation vector.

In the context of pivoted QR factorization that generates
a pivot vector $\vct{J}^\mathrm{qr}$ indicating that
if $\vct{J}^\mathrm{qr}(j)=i+1$, then the $j^{\mathrm{th}}$ column of $\mtx{M}(:,\vct{J}^\mathrm{qr})$ was the $i^{\mathrm{th}}$ column of $\mtx{M}$.
This representation of the pivot vector can be referred to as ``permutation format.''
These are stored with one-based indexing for consistency with LAPACK.
The approach for permuting the columns of a given matrix in accordance with $\vct{J}^\mathrm{qr}$ is described in \cref{alg:col_perm_seq}.

\begin{algorithm}[htb]
\small \setstretch{1.2}
\caption{ : Sequential approach to column permutation}
\label{alg:col_perm_seq}
\begin{algorithmic}[1]
\Statex \textbf{Input:} A matrix $\mtx{M} \in \mathbb{R}^{m \times n}$, a pivot vector $\vct{J}^\mathrm{qr}$ of length $n$ produced by a black-box \code{qrcp} function.
\setstretch{1.25}
\State \textbf{function} $\code{col\_perm\_sequential}(\mtx{M}, \vct{J}^\mathrm{qr})$
\vspace{-4pt}
\Indent
    \For{$i = 0:n$}
    \State $j = \vct{J}^\mathrm{qr}(i)-1$
    \State \code{swap}($\mtx{M}(:, i), \mtx{M}(:, j)$)
    \Statex ~~~~~~~\, \codecomment{ Swap entirety of two columns in $\mtx{M}$}
    \State $\mathtt{idx} = \code{find}(\vct{J}^\mathrm{qr}, i+1)$
    \Statex ~~~~~~~\, \codecomment{ Find the index of an element with value $i$}
    \State $\vct{J}^\mathrm{qr}(\mathtt{idx}) = j+1$ \label{perm:update_J}
    \EndFor
    \State \textbf{return} $\mtx{M}$
\EndIndent    
\end{algorithmic}
\end{algorithm}


There are two important things to note about \cref{alg:col_perm_seq}. 
First, as seen from step \ref{perm:update_J}, the pivot vector $\vct{J}^\mathrm{qr}$ is updated at every iteration of the main loop.
In the context of a \code{BQRRP} algorithm, we want to preserve $\vct{J}^\mathrm{qr}$ after column permutation is done, and hence copying the pivot vector is required. 
Second, the idea behind how the permutations are performed implies that the same column can be moved several times.
This prevents us from parallelizing the main loop in this algorithm (hence the keyword ``\code{sequential}'' in its name).

In principle, the sequential nature of \cref{alg:col_perm_seq} could cause a performance bottleneck in a \code{BQRRP} algorithm. 
However, we do not anticipate this happening when running \code{BQRRP} on a CPU, as it is considered latency tolerant hardware
with a very low level of parallelism required to saturate the available
memory bandwidth. The column permutation is inherently data intensive
operation with no floating-point computation involved, and thus its main
hardware bottleneck is the maximum rate at which the matrix elements can be
transferred between their main memory locations. Historically, CPUs
continue to suffer from decreasing bandwidth per compute cores, and thus
very few of them are needed to saturate the local memory controller in a
socket. Specialized vector load and store instructions inside the
\code{swap} function (and other Level 1 BLAS) allow one to take advantage
of all memory channels, even with the majority of the cores remaining
idle.
The remaining issue may stem from the sequential nature
of the loop in \cref{alg:col_perm_seq}. 
However, modern CPUs feature latency-hiding mechanisms like out-of-order execution, register renaming across an extended shadow register file,
branch prediction, and data prefetching. These hardware resources allow
for multiple iterations of the loop to be unrolled at runtime onto the
CPU units, waiting for the moment when the \code{swap} operation
finishes. If the non-temporal data moving instructions are used for
swapping, then there is no interference between all levels of cache memory and the matrix column elements that have to be moved through
only a few vector registers. In summary, the utilization of the CPU
components is evenly distributed, and the only bottleneck is the main
memory bandwidth during column swapping.

\FloatBarrier
\section{Practical implementation of \code{BQRRP} and storage management}
\label{sec:bqrrp_practical}


The ideas from \cref{sec:introduce_BQRRP} can be consolidated in the form of the two visions for \code{BQRRP} algorithms:
one relying on Cholesky QR (represented by \cref{alg:cholqr_deps} and $\mtx{Q}$ matrix storage format conversion);
and another relying on Householder QR in step \ref{bqrrp:qr_tall}.
In both cases, \code{ORMQR} is the function of choice in steps \ref{bqrrp:apply_q_1} and \ref{bqrrp:apply_q_2}, due to its reliable performance and straightforward input format requirements.
Both approaches would rely on \cref{alg:qrcp_practical} due to its unarguably superior efficiency compared to \code{GEQP3}.

From the storage standpoint, both approaches preserve the central feature of \code{BQRRP\_CPU}, namely
the fact that it can be implemented in a way that all the major computations take place in the space occupied by the given input matrix (disregarding the use of some relatively small workspace buffers).
To explain how this can be achieved, this section provides a step-by-step breakdown of how each operation in the two versions of \code{BQRRP\_CPU} is implemented and how the data in each operation is stored.

\begin{figure}[htb!]
    \centering
    \begin{overpic}[width=1.0\textwidth]{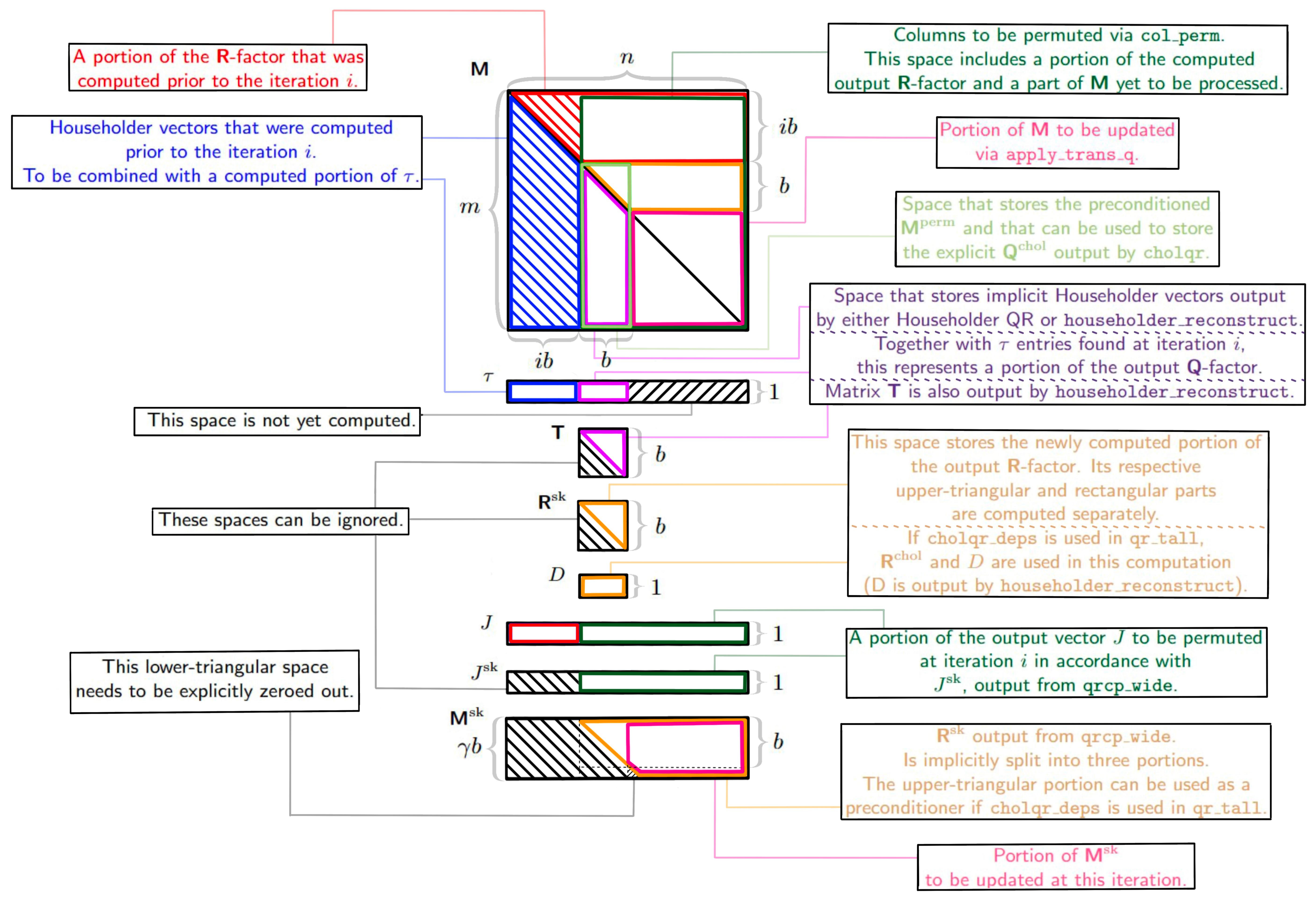}
    \end{overpic}
\captionof{figure}{\small A visualization of how all the major components of \code{BQRRP} are stored at iteration $i \in \lslice{(\lceil n/b \rceil-1)}$ of its main loop.
Refer to pseudocode \cref{alg:BQRRP} for the parameter and buffer names.
\code{BQRRP} would require a maximum of $dm + 2dn + 2b^2 + 4n + b$ additional words of memory for the internal workspace buffers.
This includes buffers for $\mtx{M}^{\mathrm{sk\_trans}}$ (size $n \times d$, $d = \gamma b$), $\vct{D}$ (size $b$) and $\vct{J}^{\mathrm{lu}}$ (of length $n$), not depicted in the figure.
These storage costs are modest, if not negligible, considering the relative sizes of parameters used in practical settings: $b \leq d \ll n$.
}
\label{fig:bqrrp_storage}
\end{figure}

The rest of this section is split into subsections that correspond to specific steps (or collections of steps) in \cref{alg:BQRRP}.
We make references to the previous sections (\crefrange{subsec:qrcp_wide}{subsec:updating_rules}) and the steps in \cref{alg:BQRRP}.
We emphasize that our description matches the way \code{BQRRP} is implemented as part of \RandLAPACK{}.
The aforementioned two versions of \code{BQRRP\_CPU} are implicitly defined through the choices that the user makes when configuring the algorithm.
To aid the upcoming detailed description,
we present in \cref{fig:bqrrp_storage} the visualization of how all the major components of \code{BQRRP} are stored at iteration
$i$.

\subsection{Input and output specification (steps \ref{bqrrp:input}, \ref{bqrrp:output})}

On input, \code{BQRRP} has access to the following: a matrix $\mtx{M} \in \R^{m \times n}$, with no specific requirements on how $m$ and $n$ relate, two empty buffers of length $n$ to store the permutation vector $\vct{J}$ and a vector with scalar factors of the elementary reflectors $\vct{\tau}$, an integer block size parameter $b \ll n$, and a scalar $\gamma$ that sets the sketch size relative to $b$ (if using LU-based QRCP on the sketch $\mtx{M}^{\text{sk}}$, $\gamma = 1.0$). 
Additionally, \code{BQRRP} receives instructions on which internal subroutine decisions to take; in this case, the choice is to use Householder QR or Cholesky QR in step \ref{bqrrp:qr_tall}.
If Cholesky QR is chosen, the user can supply an optional integer parameter $n_b$, defining the number of column block reflectors to be used in \code{ORHR\_COL}.

On output, \code{BQRRP} would return the following: the inferred numerical rank $\ell$; the $\mtx{Q}$-factor represented by $\ell$ Householder reflectors that are stored implicitly and occupy the portion of the space below the diagonal (that initially contained the input matrix $\mtx{M}$) together with a vector $\vct{\tau}$ of length $\ell$; an upper-trapezoidal $\mtx{R}$ factor sized $\ell \times n$ and stored in the upper-triangular portion of $\mtx{M}$'s space; and a permutation vector $\vct{J}$ of length $n$.
    
\subsection{Initial sketching details (steps \ref{bqrrp:sample}-\ref{bqrrp:sketching})}

At step \ref{bqrrp:sample} of \cref{alg:BQRRP}, a sketching operator $\mtx{S} \in \R^{d \times m}$ (where $d = \lceil \gamma b \rceil$) is constructed with independent mean-zero variance-one Gaussian random variables as its entries.\footnote{Variance $1/d$ could also be used. This would have the effect of making the expected squared norm of each column in $\mtx{S}\mtx{M}$ match the squared norm of the corresponding column in $\mtx{M}$.} 
The generation of this sketching operator is governed by \RandBLAS{}; the memory required to store $\mtx{S}$ will be automatically allocated upon the operator construction and deallocated when the operator application function returns. 
In our in-place implementation of \code{BQRRP}, step \ref{bqrrp:alloc} is, naturally, not explicitly performed.
 At step \ref{bqrrp:sketching}, a sketch resulting from applying an operator $\mtx{S}$ to the input $\mtx{M}$ is stored in a $d \times n$ buffer $\sk{\mtx{M}}$. 
A portion of this space will further be used to store an updated sketch at every iteration of \code{BQRRP}'s main loop.

\subsection{Block partitions in \code{BQRRP} (steps \ref{bqrrp:loop}, \ref{bqrrp:block_partitions})}
\label{subsec:block_partitions}
In Sections \ref{subsec:practical_sketch_permutation}-\ref{subsec:practical_sample_update}, we refer to the current iteration of the \cref{alg:BQRRP} main loop as $i \in \lslice{(\lceil n/b \rceil-1)}$.
We use $\sblock = ib$ to denote the start of the current row/column block range, $\erowblock = \min\{m, (i+1)b\}$ to denote the (exclusive) end of the current row block range, and $\ecolblock = \min\{n, (i+1)b\}$ to denote the (exclusive) end of the current column block range (the same notation used in step \ref{bqrrp:block_partitions} of \cref{alg:BQRRP}).
This ensures that the final iteration is well-defined even if neither $m$ or $n$ are evenly divisible by $b$.

\subsection{Processing the sketch and column permutation (steps \ref{bqrrp:qrcp}-\ref{bqrrp:update_j})}
\label{subsec:practical_sketch_permutation}
        
\paragraph{QRCP on the sketch.} At step \ref{bqrrp:qrcp}, performing QRCP on the sketch as described in \cref{alg:qrcp_practical} requires an $n \times d$ buffer for transposing $\sk{\mtx{M}}$ (the full buffer size is only needed at $i = 0$).
The important data computed in this step at the iteration $i$ is stored as follows:
the upper-trapezoidal $\sk{\mtx{R}} \in \R^{d \times (n - \sblock)}$ is stored at $\sk{\mtx{M}}(:, \tslice{\sblock})$; and a vector $\sk{\vct{J}} \in \R^{(n - \sblock)}$ is stored in a buffer of length $n$.
In our further elaboration, we partition $\sk{\mtx{R}}$ into three components: upper-triangular $\sk{\mtx{R}}_{11} = \sk{\mtx{M}}(\lslice{b}, \tslice{\sblock}\ecolblock)$; rectangular $\sk{\mtx{R}}_{12} = \sk{\mtx{M}}(\lslice{b}, \tslice{\ecolblock})$; and an upper-trapezoidal $\sk{\mtx{R}}_{22} = \sk{\mtx{M}}(\tslice{b}, \tslice{\ecolblock})$.

Our implementation of \code{BQRRP} allows for alternatively performing QRCP via \code{GEQP3}, in which case a buffer for transposing $\sk{\mtx{M}}$ is not needed.

\paragraph{Block rank computation.} In step \ref{bqrrp:rank_est}, if the block rank $k$ (computed as described in \cref{subsec:rank_est}) is not equal to $\min\{b, (n-\sblock), (m - \sblock)\}$
(when the current block is not full rank), then the given iteration would be the final one, as the termination criteria at step \ref{bqrrp:termination} states.
In that case, the trailing portion of the matrix $\mtx{M}$ is not updated at step \ref{bqrrp:update_M}.

If \code{BQRRP} is set to use Cholesky QR in step \ref{bqrrp:qr_tall}, then a rank estimate is employed to ensure no infinite or not-a-number values appear when the preconditioning is performed.

\paragraph{Combining column permutations.} Since the intended output format of \code{BQRRP} matches that of \code{GEQP3}, the output $\mtx{R}$ factor will be stored in the upper-triangular portion of $\mtx{M}$'s memory space.
At iteration $i \geq 1$, step \ref{bqrrp:permute_r} permutes the trailing columns in the \textit{computed rectangular} portion of the $\mtx{R}$-factor, implying that $\mtx{M}(\lslice{\sblock}, \tslice{\sblock})$ is to be permuted in accordance with $\sk{\vct{J}}$.
Meanwhile, step \ref{bqrrp:permute_m} requires $\mtx{M}(\tslice{\sblock}, \tslice{\sblock})$ to be permuted by the same vector.
Therefore, steps \ref{bqrrp:permute_r} and \ref{bqrrp:permute_m} can be combined into permuting $\mtx{M}(\fslice{}, \tslice{\sblock})$ at every iteration of the \code{BQRRP}'s main loop.

\begin{remark}
    If \cref{alg:col_perm_seq} is used for permuting columns, a copy of $\sk{\vct{J}}$ is required.
\end{remark}

After the columns of $\mtx{M}(\fslice{}, \tslice{\sblock})$ have been permuted, we may perform an early termination check by verifying whether the first column in $\mtx{M}(\fslice{}, \tslice{\sblock})$ consists of all zeros. 
In that case, \code{BQRRP} terminates immediately. 
If at iteration $i$ we have $k \neq \min\{b, n-\sblock, m-\sblock\}$, then \code{BQRRP} should avoid permuting $\mtx{M}(\tslice{\sblock + k}, \tslice{\ecolblock})$, since this trailing portion of the input matrix would not be used (i.e., the trailing update to $\mtx{M}$ would be avoided in step \ref{bqrrp:apply_q_1}).
This was not shown in \cref{alg:BQRRP} for simplicity of presentation.

\paragraph{Updating the permutation vector.} By step \ref{bqrrp:update_j}, the pivot vector $\sk{\vct{J}}$, computed at the current iteration, has been used for all the necessary internal operations.
It can now be incorporated into the vector $\vct{J}$ that is to be output by \code{BQRRP}.
When $i$ is $0$, this is done by copying $\sk{\vct{J}}$ into $\vct{J}$; at any subsequent iteration, the trailing portion of $\vct{J}$, $\vct{J}(\tslice{\sblock})$, will be permuted in accordance with $\sk{\vct{J}}$ via (a vector version of) \cref{alg:col_perm_seq}.
Since $\sk{\vct{J}}$ is not needed after this step, we do not have to create a copy of it in \cref{alg:col_perm_seq}.

\subsection{Panel QR factorizations details (steps \ref{bqrrp:qr_tall}, \ref{bqrrp:update_R11})}
        
        The internal details of step \ref{bqrrp:qr_tall} depend on the user's choice of tall QR function.
        As said before, in our implementation of \code{BQRRP}, we allow users to choose between \code{GEQRF} and the Cholesky QR-based approach.

        \paragraph{Householder QR on a panel.} When using \code{GEQRF} in step \ref{bqrrp:qr_tall}, we perform the factorization on all columns of $\perm{\mtx{M}} = \mtx{M}(\tslice{\sblock}, \tslice{\sblock}\ecolblock)$, regardless of whether $k$, the current column block rank, is equal to $k_{\mathrm{max}}=\min\{b, (n-\sblock), (m-\sblock)\}$.
        This is because we want $\mtx{R}_{11}$ to have $(\ecolblock - \sblock) = \min\{b, (n - \sblock)\}$ columns.
        Despite the fact that $(\ecolblock - \sblock)$ reflectors are computed in that case, we use only $k$ of them \textit{outside} of this step.
        Using \code{GEQRF} on $\perm{\mtx{M}}$ in this step requires no additional storage.
        This function produces $\mtx{Q}^{\mathrm{curr}}$, represented by $(\ecolblock - \sblock)$ reflectors of length $(m - \sblock)$ (stored in the portion of $\mtx{M}(\tslice{\sblock}, \tslice{\sblock}\ecolblock)$ below the diagonal) and the vector of scalar factors of the elementary reflectors (stored in $\vct{\tau}(\tslice{\sblock}\ecolblock)$).
        The explicit $\mtx{R}_{11}$ of size $(\erowblock - \sblock) \times (\ecolblock - \sblock)$ is stored right above the $\mtx{Q}^{\mathrm{curr}}$, in the upper-triangular portion of $\mtx{M}(\tslice{\sblock}\erowblock, \tslice{\sblock}\ecolblock)$.
        Note that in this case, step \ref{bqrrp:update_R11} is performed implicitly, since $\mtx{R}_{11}$ will be stored where it needs to be on output from \code{GEQRF}.

        \paragraph{Cholesky QR on a panel.} Using the Cholesky QR approach in step \ref{bqrrp:qr_tall} is comprised of four parts: preconditioning; performing Cholesky QR on the preconditioned matrix; implicit $\mtx{Q}^{\mathrm{curr}}$ reconstruction; and computing $\mtx{R}_{11}$. All of these are done in accordance with \cref{alg:cholqr_deps}.

        The preconditioning is to be done on a portion of the matrix $\perm{\mtx{M}}$. We perform $\pre{\mtx{M}} = \perm{\mtx{M}}(\fslice{}, \lslice{k}) (\sk{\mtx{R}}_{11}(\lslice{k}, \lslice{k}))^{-1}$, using $\sk{\mtx{R}}_{11}$ stored in $\sk{\mtx{M}}(\lslice{b}, \tslice{\sblock}\ecolblock)$. This step requires no workspace buffers, and the matrix $\pre{\mtx{M}} \in \R^{(m-\sblock) \times k}$ is stored at $\mtx{M}(\tslice{\sblock}, \tslice{\sblock}(\sblock+k))$ after the preconditioning.
        The next step is to perform Cholesky QR on~$\pre{\mtx{M}}$.
        
        As explained in \cref{subsec:qr_tall}, Cholesky QR is comprised of: computing the Gram matrix (\code{SYRK}), performing the Cholesky factorization (\code{POTRF}), and forming the explicit economical version of the orthonormal factor (\code{TRSM}).
        These functions require (at most) a single $b \times b$ workspace buffer to store the Gram matrix and the output upper-triangular $\mtx{R}^{\mathrm{chol}}$ factor, both of size $k \times k$, $k \leq \min\{b, (n-\sblock), (m-\sblock)\}$.
        The explicit orthonormal factor $\mtx{Q}^{\mathrm{chol}} \in \R^{(m - \sblock) \times k}$ is stored at $\mtx{M}(\tslice{\sblock}, \tslice{\sblock}(\sblock+k))$, in place of~$\pre{\mtx{M}}$.
        
        \code{ORHR\_COL} used for the Householder reconstruction returns an implicit representation of $\mtx{Q}^{\mathrm{curr}}$ in the form of $k$ reflectors and an upper-trapezoidal matrix $\mtx{T}$ of size $n_b \times k$.
        The $k$ reflectors fit exactly in place of the portion of $\mtx{Q}^{\mathrm{chol}}$ below the diagonal of $\mtx{M}(\tslice{\sblock}, \tslice{\sblock}(\sblock+k))$, while $\mtx{T}$ requires an additional buffer of size $n_b \times b$.
        Furthermore, \code{ORHR\_COL} produces a sign vector $\vct{D}$ of length $k$ that requires a storage buffer of length $b$.
    
        Computing $\mtx{R}_{11}$ requires first applying the entries from the sign vector $\vct{D}$ to the columns of $\mtx{R}^{\mathrm{chol}}$ (this is done so that $\mtx{R}^{\mathrm{chol}}$ is in line with $\mtx{Q}^{\mathrm{curr}}$, computed by \code{ORHR\_COL}).
        The next step amounts to undoing the preconditioning on $\mtx{R}^{\mathrm{chol}}$ via multiplying it by an upper-triangular $\sk{\mtx{R}}_{11}(\lslice{k}, \lslice{(\ecolblock-\sblock)})$, stored at $\sk{\mtx{M}}(\lslice{k}, \tslice{\sblock}\ecolblock)$.
        The product is temporarily stored in place of $\mtx{R}^{\mathrm{chol}}$ and then is copied into the upper-triangular portion of $\mtx{M}(\tslice{\sblock}(\sblock+k), \tslice{\sblock}\ecolblock)$, placing it right above the implicitly-stored Householder reflectors (step \ref{bqrrp:update_R11}). 

\begin{remark}
        This additional copy is currently unavoidable because no standard BLAS function for multiplying two triangular matrices exists (although one could implement \code{trtrmm}).
\end{remark}

        For the output format of \code{BQRRP} to match that described in \cref{subsec:def_and_notations}, we would need to copy the entries from the block diagonals of $\mtx{T}$ into $\vct{\tau}(\tslice{\sblock}(\sblock+k))$.

\subsection{Applying transposed Q and updating factors (steps \ref{bqrrp:iter_check}-\ref{bqrrp:update_q})}

        Step \ref{bqrrp:apply_q_1} or \ref{bqrrp:apply_q_2} perform operations described in \cref{subsec:updating_rules}.
        Depending on whether the estimated block rank $k$ matches the column block size $b$, we apply a transpose of $\mtx{Q}^{\mathrm{curr}}$ (from the Householder representation in $\mtx{M}(\tslice{\sblock}, \tslice{\sblock}(\sblock+k))$ and $\vct{\tau}(\tslice{\sblock}(\sblock+k))$), to either the first $k$ or to all $(m - \sblock)$ rows of $\mtx{M}(\tslice{\sblock}, \tslice{\ecolblock})$.

        Regardless of whether step \ref{bqrrp:apply_q_1} or \ref{bqrrp:apply_q_2} is executed, the first $k$ rows of $\mtx{M}(\tslice{\sblock}(\sblock+k), \tslice{\ecolblock})$ represents $\mtx{R}_{12}$ (this implicitly fulfills step \ref{bqrrp:update_r12}).
        If step \ref{bqrrp:apply_q_2} is executed, then the remaining rows $\mtx{M}(\tslice{\erowblock}, \tslice{\ecolblock})$ represent the ``working submatrix'' of the matrix $\mtx{M}$ at the next iteration (this implicitly fulfills step \ref{bqrrp:update_M}).
        Otherwise, the trailing $\mtx{M}$ update is avoided.

         In \code{BQRRP}, step \ref{bqrrp:update_q} is performed implicitly, as the matrix $\mtx{Q}$ is constructed from the Householder vectors, stored below the main diagonal of $\mtx{M}$ and the scalar factors of the elementary reflectors stored in $\vct{\tau}$.

\subsection{Algorithm termination and sample update (steps \ref{bqrrp:termination}-\ref{bqrrp:update_sample})}
\label{subsec:practical_sample_update}

        \paragraph{Termination criteria.} The termination criteria check (step \ref{bqrrp:termination}) simply amounts to verifying whether the maximum number of iterations has been exceeded ($i$ reached $\lceil n/b \rceil-1$) or whether the block rank $k$ does not match the column block size $\min\{b, n-\sblock, m-\sblock\}$.
        Before the termination, the rank parameter $\ell$ is updated to $\sblock + k$ (step \ref{bqrrp:update_rank}).

        \paragraph{Sample update.} If the maximum number of iterations has not been reached at this point, the sketch $\sk{\mtx{M}}$ is updated at step \ref{bqrrp:update_sample}.
        This step first computes the expression $\sk{\mtx{R}}_{11}(\mtx{R}_{11})^{-1}$, where $\sk{\mtx{R}}_{11}$ is stored in the upper-triangular portion of $\sk{\mtx{M}}(\lslice{b}, \tslice{\sblock}\ecolblock)$ and $\mtx{R}_{11}$ is stored in the upper-triangular portion of $\mtx{M}(\tslice{\sblock}\erowblock, \tslice{\sblock}\ecolblock)$.
        The result is written into the space of $\sk{\mtx{R}}_{11}$, at $\sk{\mtx{M}}(\lslice{b}, \tslice{\sblock}\ecolblock)$; 
        we explicitly zero out the entries in this space below the diagonal.
        Next, the full expression $\sk{\mtx{R}}_{12} - \sk{\mtx{R}}_{11}(\mtx{R}_{11})^{-1}\mtx{R}_{12}$ is computed.
        The matrix $\sk{\mtx{R}}_{12}$ is stored in $\sk{\mtx{M}}(\lslice{b}, \tslice{\ecolblock})$ and $\mtx{R}_{12}$ is stored at $\mtx{M}(\tslice{\sblock}\erowblock, \tslice{\ecolblock})$.
        The result of this expression is placed into the space of $\sk{\mtx{R}}_{12}$, at $\sk{\mtx{M}}(\lslice{b}, \tslice{\ecolblock})$.
        The upper-trapezoidal matrix $\sk{\mtx{R}}_{22}$ is already stored at its intended location, in $\sk{\mtx{M}}(\tslice{b}(d - b), \tslice{\ecolblock})$, but we need to make sure that the entries are zeroed out below the diagonal.
        After the sketch update is completed, the ``working submatrix'' of $\sk{\mtx{M}}$ is implicitly located at $\sk{\mtx{M}}(\fslice{}, \tslice{\ecolblock})$.


\section{Consideration specific to implementations targeting GPU accelerators}
\label{sec:bqrrp_gpu}


\subsection{Limited LAPACK functionality} \label{subsec:bqrrp_gpu_blas_lapack}
Implementing a GPU version of the \code{BQRRP} algorithm involves several challenges, primarily due to the limited availability of GPU variants of most LAPACK and BLAS-level functions. Specifically, NVIDIA cuSOLVER lacks support for the full range of LAPACK functions. 
This issue, however, is rather understandable, as the latest version of LAPACK 3.12.0 includes over $2000$ functions, many of which are not widely used.
Nevertheless, the lack of a wide range of LAPACK functions limits our choices for the internal subroutines in a \code{BQRRP\_GPU}.

For the purpose of constructing \code{BQRRP\_GPU}, the most notable function that cuSOLVER does not offer is \code{ORHR\_COL}, which we require when using Cholesky QR in the \code{qr\_tall} subroutine.
In order to provide some investigation of Cholesky QR methods for \code{qr\_tall} we developed a basic CUDA implementation of \cite[Algorithm $5$]{BD2015}.
We note that dramatically improved performance could be achieved by the Cholesky QR approach if cuSOLVER offered a recursive LU-based implementation of \code{ORHR\_COL} similar to the one used by LAPACK.
It is also worth noting that there is no standard QRCP algorithm offered in cuSOLVER.
Therefore our \code{BQRRP\_GPU} implementation is forced to use the LU-based \cref{alg:qrcp_practical} for \code{qrcp\_wide}.


\subsection{Column permutation}
\label{subsec:bqrrp_gpu_col_perm}
A major difference between the CPU and GPU versions of \code{BQRRP} is the importance of a high-performance implementation of a function for permuting columns of a given matrix on the GPU in accordance with a pivot vector.

\paragraph{Hardware considerations.}
Recall that \cref{alg:col_perm_seq} illustrated a sequential approach to permuting columns. 
Despite this approach being sequential, we do not anticipate it having a major negative effect on the overall performance of \code{BQRRP\_CPU}, as described in \cref{subsec:col_perm}.
By contrast, on a GPU, there are very few hardware resources
devoted to dealing with inherently sequential instruction streams or even tolerating the high latency of the main memory transactions. 
The initially presented sequential loop for pivoting is
anathema to the parallel GPU hardware.

\paragraph{The alternative approach.}
The widely-used alternative solution is the ``parallel pivoting'' strategy, shown in \cref{alg:col_perm_par}.
By contrast to \cref{alg:col_perm_seq}, \cref{alg:col_perm_par} does not swap columns within the memory space of a single matrix.
Instead, it copies columns from one space into another at the new location from the permutation vector, which allows for parallelizing the main loop, significantly increasing the performance of this column permutation approach.
The performance gain offered by \cref{alg:col_perm_par} comes at the cost of increased space usage, as it requires a copy of the input $\mtx{M}$.

\begin{algorithm}[htb]
\small \setstretch{1.2}
\caption{ : Column-parallel approach to column permutation}
\label{alg:col_perm_par}
\begin{algorithmic}[1]
\Statex \textbf{Input:} A matrix $\mtx{M} \in \mathbb{R}^{m \times n}$, a pivot vector $\vct{J}^\mathrm{qr}$ output by a black-box \code{qrcp} function
\setstretch{1.25}
\State \textbf{function} $\code{col\_perm\_parallel}(\mtx{M}, \vct{J}^\mathrm{qr})$
\vspace{-4pt}
\Indent
    \State $\mtx{M}^{\mathrm{cpy}} = \code{copy}(\mtx{M})$
    \For{$i = 0:n$} \label{perm_par:loop}
    \State $j = \vct{J}^\mathrm{qr}(i)$
    \State $\mtx{M}(:, i) = \mtx{M}^{\mathrm{cpy}}(:, j)$
    \EndFor
\EndIndent 
\State \textbf{return} $\mtx{M}$
\end{algorithmic}
\end{algorithm}

\paragraph{Managing data copies in \cref{alg:col_perm_par}.}
In order to minimize the number of explicit copies performed as part of using \cref{alg:col_perm_par} in practice, one could swap the pointers that point to the memory spaces of $\mtx{M}$ and $\mtx{M}^{\mathrm{cpy}}$ before performing column permutation. 
This approach is safe as long as either the \emph{entire} matrix $\mtx{M}$ is permuted in \cref{alg:col_perm_par}, or the non-permuted part of $\mtx{M}$ is not used outside of \cref{alg:col_perm_par}.
As stated in \cref{subsec:col_perm}, the column permutation is used
in steps \ref{bqrrp:permute_r}, \ref{bqrrp:permute_m}, and \ref{bqrrp:update_j} in \cref{alg:BQRRP} (where steps  \ref{bqrrp:permute_r} and \ref{bqrrp:permute_m} can be merged) and in step \ref{wide_qrcp:permute} of \cref{alg:qrcp_practical}.
As such, we will be applying permutations to portions of matrices $\mtx{M}^{\mathrm{sk}}$ and $\mtx{M}$, as well as a vector $\sk{\vct{J}}$ at every iteration of \code{BQRRP\_GPU}.

\begin{remark}
    To avoid overcomplicating the notation, the description below does not account for cases where $n$ is not evenly divisible by $b$.
\end{remark}

Since the full $\mtx{M}^{\mathrm{sk}}$ is not used outside of \code{BQRRP\_GPU}, the pointer-swapping strategy can be used without any memory safety concerns.
This is, however, not the case with $\mtx{M}$ and $\sk{\vct{J}}$, as both of their entire memory spaces will have to store correct data on output from the \code{BQRRP\_GPU}.
To illustrate the complication with the pointer-swapping strategy, suppose the main loop in \code{BQRRP\_GPU} is executing starting with $i = 0$, where index $0$ is considered ``even.''
Then, at the end of the $0^{th}$ iteration, the space of $\mtx{M}^{\mathrm{cpy}}$, will contain the correct data to be a part of the output of \code{BQRRP\_GPU} since the pointer swapping took place.
At iteration $1$, the space $\mtx{M}^{\mathrm{cpy}}$ will contain the first $b$ properly-computed columns, the rest of the correct
entries will be contained in $\mtx{M}(:, \tslice{b})$.
Continuing with that logic, we conclude that the space $\mtx{M}^{\mathrm{cpy}}$ will contain the ``correct'' entries in $\mtx{M}^{\mathrm{cpy}}(:, \tslice{ib}(i+1)b)$ for all even $i \leq n/b$, and the space $\mtx{M}$ will contain the correct entries for all odd $i$ in $\mtx{M}(:, \tslice{ib}(i+1)b)$ when \code{BQRRP\_GPU} terminates.
If \code{BQRRP\_GPU} terminated at an even iteration, the pointers to $\mtx{M}$ and $\mtx{M}^{\mathrm{cpy}}$ will need to be swapped back around.
Regardless of the final parity status of the loop's index,
the ``correct'' columns will need to be copied from $\mtx{M}^{\mathrm{cpy}}$ to $\mtx{M}$.
Note that if \code{BQRRP\_GPU} was to terminate early at an even iteration, the trailing entries in range $(i+1)b$ to $m$
in $\mtx{M}^{\mathrm{cpy}}$ will need to be copied into $\mtx{M}$.

As per the vector $\vct{J}$, since it is not permuted at iteration $0$, the space of
$\vct{J}^{\mathrm{cpy}}$ would contain the ``correct'' entries in
$\vct{J}^{\mathrm{cpy}}(\tslice{ib}(i+1)b)$ for all odd $i \leq n/b$,
while the space $\vct{J}$ will contain the correct entries for all even $i$ in $\vct{J}(\tslice{ib}(i+1)b)$ when \code{BQRRP\_GPU} terminates.
Otherwise, the vector $\vct{J}$ is to be processed similarly to the matrix $\mtx{M}$. 

\subsection{The view of a practical \code{BQRRP\_GPU}}
Considering the constraints of designing a GPU version of \code{BQRRP} described in Sections \ref{subsec:bqrrp_gpu_blas_lapack} and \ref{subsec:bqrrp_gpu_col_perm}, we adjust our view of the \code{BQRRP\_GPU} algorithms from the two implicit CPU versions of \code{BQRRP} described in the beginning of \cref{sec:bqrrp_practical}.

All in all, we stick to the vision of \code{BQRRP\_GPU} being represented by two implicit algorithms (defined by the user choices during algorithm configuration),
where one version relies on Cholesky QR and its dependencies, and the other version relies on Householder QR in step \ref{bqrrp:qr_tall}.
Cholesky QR is paired with the \textit{sequential} approach to \code{ORHR\_COL}, as we decided that implementing a high-performance GPU version of \code{ORHR\_COL} is beyond the scope of this work.
In steps \ref{bqrrp:apply_q_1} and \ref{bqrrp:apply_q_2}, we use cuSOLVER's \code{ORMQR} function.
Furthermore, we decided to sacrifice storage for performance by using \cref{alg:col_perm_par} (and the pointer-swapping logic) for column permutation in our implementation of \code{BQRRP\_GPU}.

At the time of writing, the current version 1.0.1 of RandBLAS does not offer GPU support.
Because of that, steps \ref{bqrrp:sample} and \ref{bqrrp:sketching} are performed outside of \code{BQRRP\_GPU}, and $\mtx{M}^{\mathrm{sk}}$ is provided as an input into the algorithm.

\section{Performance profiling of major computational kernels}
\label{sec:performance_profiling}

Having established how the two views of \code{BQRRP\_CPU} and \code{BQRRP\_GPU} are constructed, in \cref{sec:bqrrp_practical,sec:bqrrp_gpu}, it is important to analyze how different parts of these algorithms affect the overall algorithm performance, in order to identify the bottlenecks for future optimization.
In this section, we omit depicting the results for matrices with dimensions that are multiples of ten, as their performance profiling plots are nearly identical to those for matrices with dimensions that are powers of two.

\begin{figure}[htb!]
    \centering
    \begin{overpic}[width=0.82\textwidth]{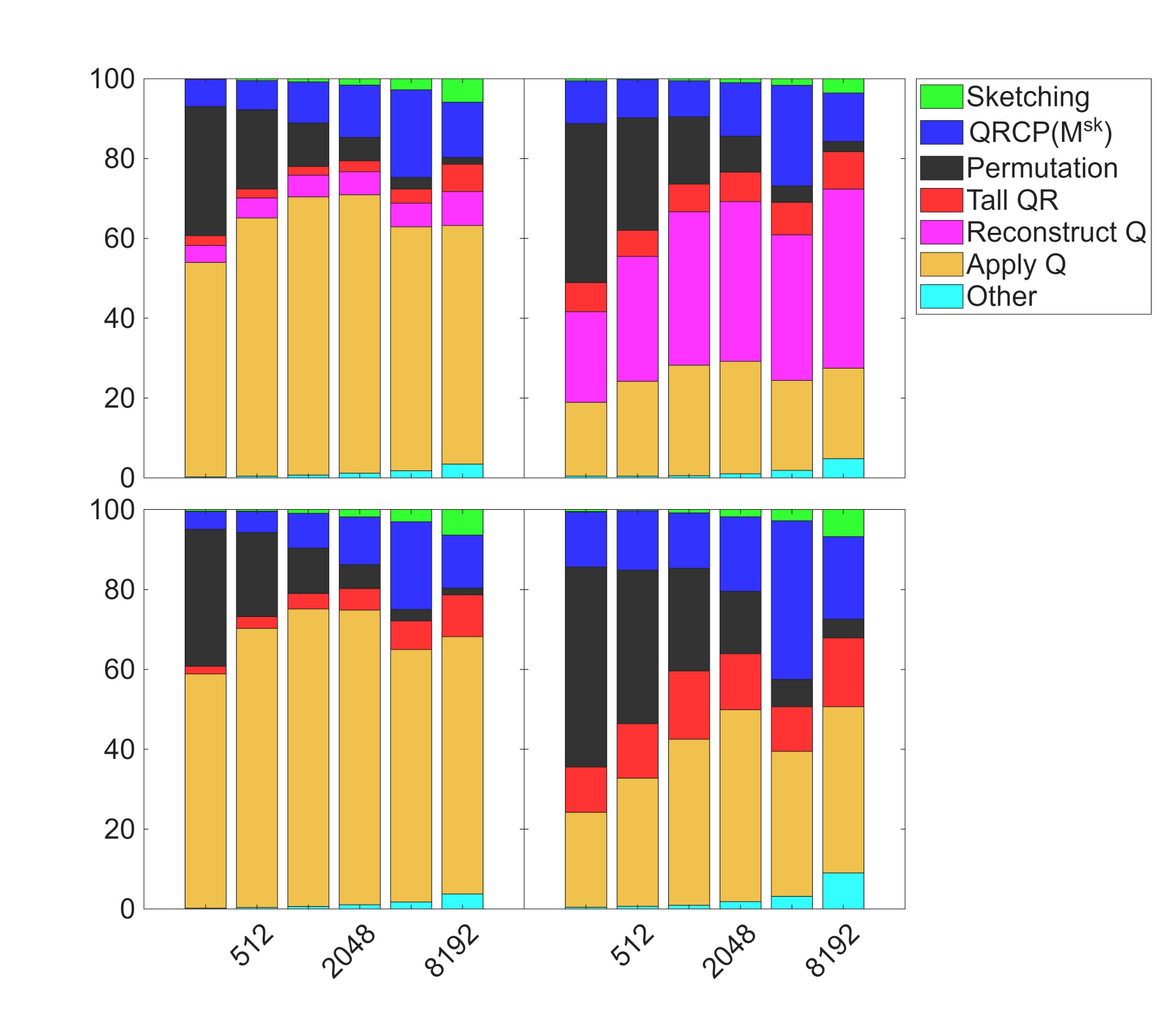}
    \put (20, 82) {\textbf{Intel CPU}}
    \put (52, 82) {\textbf{AMD CPU}}
    \put (1.5, 61) {\rotatebox[origin=c]{90}{\code{BQRRP\_CQR\_CPU}}}
    \put (4, 62) {\rotatebox[origin=c]{90}{\textbf{runtime (\%)}}}
    \put (1.5, 25) {\rotatebox[origin=c]{90}{\code{BQRRP\_HQR\_CPU}}}
    \put (4, 26) {\rotatebox[origin=c]{90}{\textbf{runtime (\%)}}}
    \put (60, 0) {\textbf{b}}
    \put (28, 0) {\textbf{b}}
    \end{overpic}
\captionof{figure}{\footnotesize Percentages of \code{BQRRP\_CPU} runtime, occupied by its respective subroutines. 
  The top row represents \code{BQRRP\_CPU} with Cholesky QR on a panel, and separately shows the percentage of runtime occupied by the preconditioned Cholesky QR and Householder restoration. The bottom row represents \code{BQRRP\_CPU} with Householder QR on a panel.
  The results are captured on an Intel CPU (left) and an AMD CPU (right) (see \cref{table:cpu_config}).
  Observe that in all plots, \code{apply\_trans\_q} (\cref{subsec:updating_rules}) is among the most costly subroutines. 
}
\label{fig:cpu_runtime_breakdown}
\end{figure}
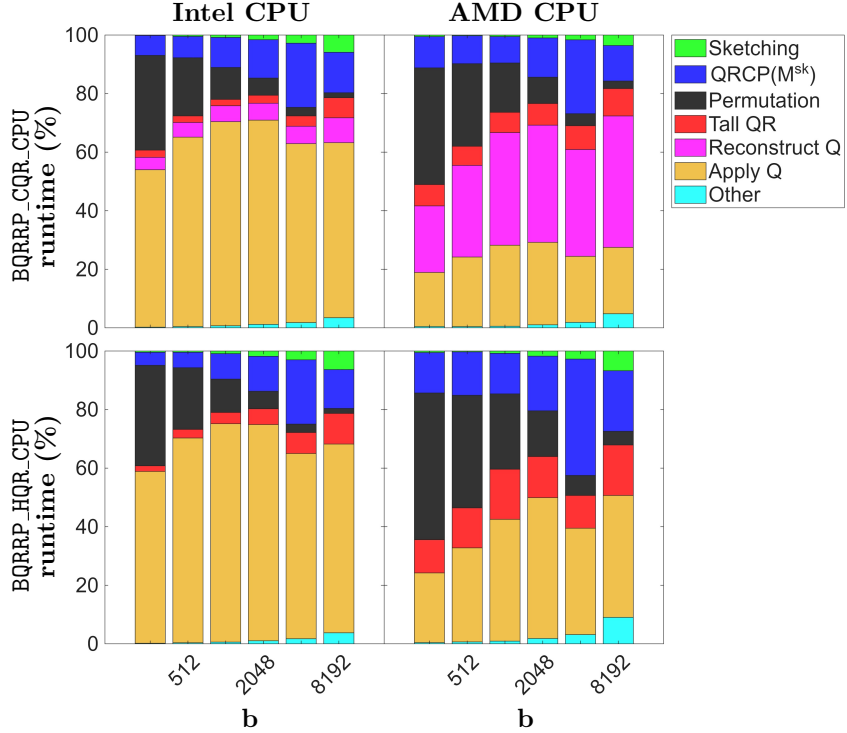

\paragraph{CPU performance breakdown.} 
We first present in \cref{fig:cpu_runtime_breakdown} the algorithm runtime breakdown results for the two CPU versions of \code{BQRRP}.
We depict the percentage of runtime that is occupied by a given
component kernel of the two versions of \code{BQRRP\_CPU} on the $y$-axis.
We use square test matrices with $65{,}536$ rows and columns and the block size parameter $b$ varying as powers of two from $256$ to $2048$ ($x$-axis). 

An immediate observation from \cref{fig:cpu_runtime_breakdown} is that \code{apply\_trans\_q} function is the major bottleneck in the three out of four \code{BQRRP} formations that we consider, despite the fact that we used used the best available function for this step (per our results in \cref{subsec:updating_rules}).
We also observe that \code{col\_perm\_sequential} and \code{ORHR\_COL} are proportionally much slower on the AMD system than the Intel system.
It is plausible that the slowdown in the former function is more visible in the AMD system given its up to 448 threads available.
Moreover, the only parallelism within \code{col\_perm\_sequential} stems from the BLAS \code{SWAP} function. Since MKL is used on the AMD system (as noted in \cref{subsec:exp_setup}), the underlying BLAS functions may not be fully optimized for AMD hardware, which could result in suboptimal performance.
We do not have an explanation for the slow performance of \code{ORHR\_COL} on the AMD system.

\begin{remark}
    The results in \cref{fig:cpu_runtime_breakdown} raise the question: could one have predicted this plot based solely on the operation counts of the core subroutines in \code{BQRRP}? 
    The short answer is \textit{no}, as evidenced by the expense of column permutation, which requires a total of $O(mn)$ operations across the entire algorithm.
    Furthermore, since \code{BQRRP} is a blocked algorithm, the operation count for each subroutine varies across iterations; adding these operation counts would lead to complex expressions that obscure more than they clarify.
\end{remark}

\FloatBarrier

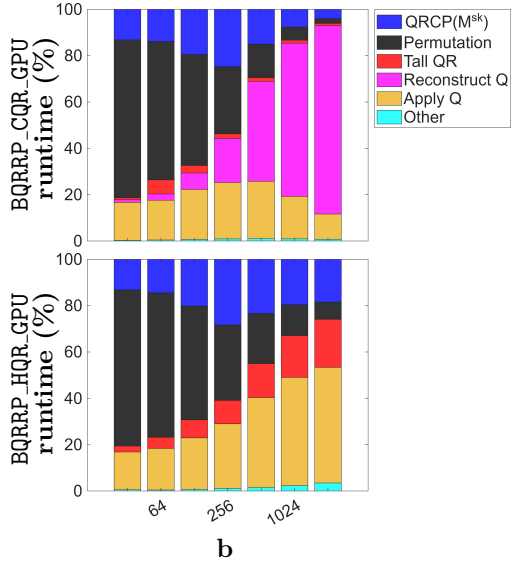
\begin{figure}[htp]
\minipage{0.6\textwidth}
\centering
    \begin{overpic}[width=0.8\textwidth]{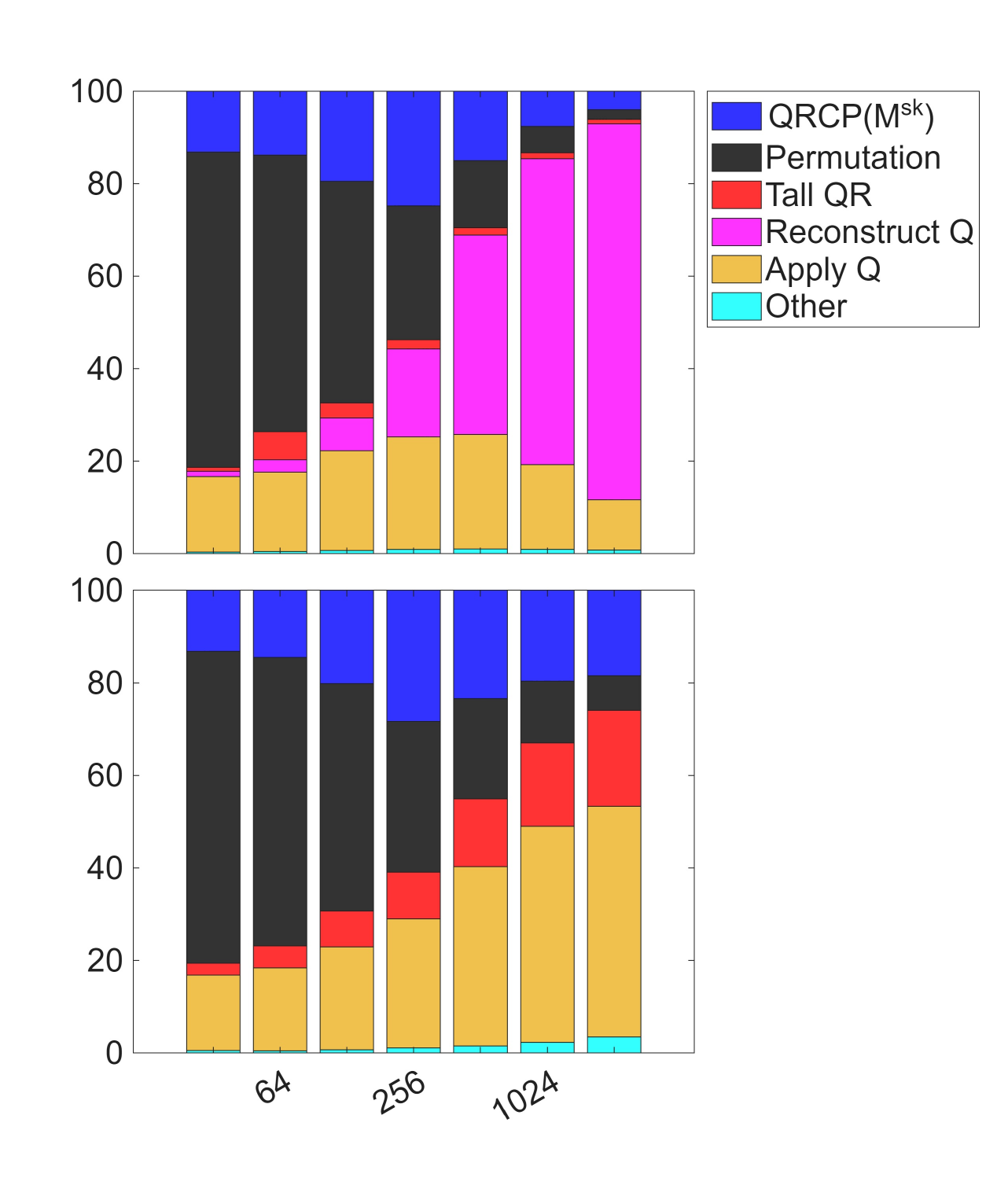}
    \put (-1.5, 72) {\rotatebox[origin=c]{90}{\code{BQRRP\_CQR\_GPU}}}
    \put (2, 73) {\rotatebox[origin=c]{90}{\textbf{runtime (\%)}}}
    \put (-1.5, 29) {\rotatebox[origin=c]{90}{\code{BQRRP\_HQR\_GPU}}}
    \put (2, 30) {\rotatebox[origin=c]{90}{\textbf{runtime (\%)}}}
    \put (33, 0) {\textbf{b}}
    \end{overpic}
  \endminipage
  \minipage{0.4\textwidth}
  \captionof{figure}{\footnotesize Percentages of the \code{BQRRP\_GPU} runtime, occupied by its respective subroutines. 
  The top row represents \code{BQRRP\_GPU} with Cholesky QR on a panel, and separately shows the percentage of runtime occupied by the preconditioned Cholesky QR and Householder restoration. The bottom row represents \code{BQRRP\_GPU} with Householder QR on a panel.
  The results are captured on an NVIDIA H100 GPU (see \cref{table:gpu_config}).
  In the algorithm with Cholesky QR on a panel, observe that our naive approach for \code{ORHR\_COL\_GPU} appears to be too costly.
  In the algorithm with Householder QR on a panel, observe that the main bottleneck is the \code{apply\_trans\_q} function, similar to the CPU results. 
  }
  \label{fig:gpu_runtime_breakdown}
  \endminipage
\end{figure}

\paragraph{GPU performance breakdown.}
In \cref{fig:gpu_runtime_breakdown}, we present the runtime breakdown of the two implicit versions of \code{BQRRP\_GPU} to see any bottlenecks in their respective subroutines.
Any given implementation of \code{BQRRP\_GPU} uses a single GPU stream, and hence timing its computational kernels does not involve any complications.
We depict the percentage of runtime that is occupied by a given subcomponent of \code{BQRRP\_GPU} on the $y$-axis.
We use a square test matrix with $32{,}768$ rows and columns and the block size parameter $b$ varying as powers of two from $32$ to $2048$ ($x$-axis).

The glaring issue with the top plot can be seen as the block size increases: the fact that our simple implementation of \code{ORHR\_COL} dominates the runtime.
This suggests that the simplest approach to \code{ORHR\_COL} is simply not viable in practice. 
As such, the version of \code{BQRRP\_GPU} with Cholesky QR on a panel is expected to have worse overall performance than that with Householder QR.
As we can see from the bottom plot in \cref{fig:gpu_runtime_breakdown}, whenever a naive implementation of \code{ORHR\_COL} is not an issue, the main algorithm bottleneck is \code{ORMQR}, just like in the CPU version of \code{BQRRP}.
Both plots in \cref{fig:gpu_runtime_breakdown} show that when smaller block sizes are in use, permuting columns in the matrix $\mtx{M}$ is rather costly.

\section{Pivot quality on the Kahan matrix}
\label{sec:piv_qual}

This section gives experimental comparisons of pivot quality using
LAPACK's default QRCP subroutine \code{GEQP3}, compared to
those produced by \code{BQRRP}, configured to use \cref{alg:qrcp_practical} in Step \ref{bqrrp:qrcp}.
For a QR factorization with column pivoting, the pivot quality directly coincides with the \textit{reconstruction quality} of the factors that a given algorithm produces.

We use two pivot quality metrics, following our prior work 
 \cite[Section 4]{MBM2024}.
The first is the Frobenius norms of the trailing submatrix of the output $\mtx{R}$-factor, $\mtx{R}(\tslice{i}, \tslice{i})$ for $0 \le i < n$.
This has the natural interpretation as the norm of the residual from a rank-$i$ approximation of $\mtx{M}$ as $\mtx{Q}(\fslice{}, \lslice{i})\mtx{R}(\lslice{i}, \fslice{})$.
We plot this metric as ratios \[
\|\mtx{R}^{\text{geqp3}}(\tslice{i}, \tslice{i})\|_{\mathrm{F}} / \|\mtx{R}^{\text{bqrrp}}(\tslice{i}, \tslice{i})\|_{\mathrm{F}}.
\]
The second pivot quality metric involves the ratios of $|\mtx{R}(i,i)|$
to the singular values of $\mtx{M}$.
If $\mtx{R}$ comes from \code{GEQP3}, then this ratio can be quite bad in the worst case.
Letting $\sigma_i$ denote the $i^{\text{th}}$ singular value of $\mtx{M}$, this only guarantees that $|\mtx{R}(i,i)| / \sigma_i$ is between $(n(n + 1)/2)^{-1/2}$ and $2^{n-1}$ \cite{Higham:blog:rrf}.
Since there is a chance for large deviations, we plot $|\mtx{R}(i,i)| / \sigma_i$ for \code{BQRRP} and \code{GEQP3} separately (rather than plotting the ratio $|\mtx{R}^{\text{geqp3}}(i,i)| / |\mtx{R}^{\text{bqrrp}}(i,i)|$).

The quality of pivots produced by \code{BQRRP} naturally depends on the type of matrix used as input. Rather than exhaustively testing \code{BQRRP} with various commonly used matrices in QRCP verification schemes,  we focused directly on a challenging case: the Kahan matrix, which is notoriously difficult for QRCP to handle.
The Kahan matrix is known for having small differences in column norms, hence potentially causing pivoting to fail, which results in inaccurate factorizations.
From the description in \cite{kahan_matrix_generator}, we can parameterize the Kahan matrix by $n, p, \theta$, by taking $\alpha = \sin(\theta)$, $\beta = -\cos(\theta)$, and 
\begin{equation}\label{alg:kahan_generator}
    \mathsmaller{
    \mtx{M} = 
    \underbrace{\begin{bmatrix}
     1 &          &         &   \\
       & \alpha^1 &        &    \\
       &          & \ddots &     \\
       &           &        &   \alpha^{n-1} 
    \end{bmatrix}}_{\text{diagonal}}
    \underbrace{\begin{bmatrix}
\beta & 1       & \cdots     & 1          \\
      & \beta   & \ddots     &   \vdots   \\
      &         & \ddots     &  1         \\
      &         &            &  \beta     \\
    \end{bmatrix}}_{\text{upper-triangular}}
    +~\epsilon_{\mathrm{mach}} \cdot p 
    \underbrace{\begin{bmatrix}
    n &     &        & \\
      & n-1 &        & \\
      &      & \ddots & \\
      &     &        & 1    
    \end{bmatrix}}_{\text{diagonal}}}
\end{equation}

\begin{figure}[htp]
\minipage{0.4\textwidth}
\centering
    \begin{overpic}[width=0.75\textwidth]{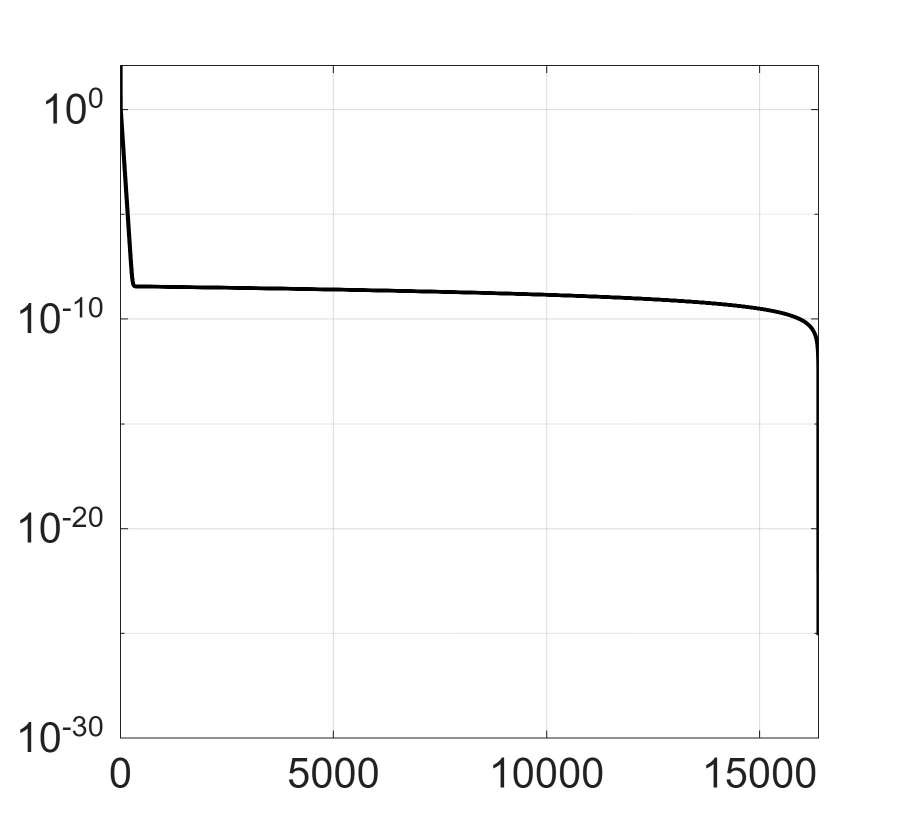}
    \put (-3, 45) {\rotatebox[origin=c]{90}{\textbf{$\mathbf{\sigma_{i}}$}}}
    \put (47, -1) {$\scalemath{0.9}{\mathbf{i}}$}
    \end{overpic}
  \endminipage
  \minipage{0.6\textwidth}
  \captionof{figure}{\footnotesize Spectrum of Kahan matrix of order $16{,}384$, generated as described in \cref{alg:kahan_generator}, using $p = 1000$ and $\theta = 1.2$. 
  The spectrum was obtained using LAPACK's most accurate SVD function, \code{GESVD}.
  Note that the trailing singular values fall below the double-precision machine $\epsilon$.
  }
  \label{fig:kahan_spectrum}
  \endminipage
\end{figure}

\begin{figure}[htb]
\minipage{0.4\textwidth}
    \captionof{figure}{\footnotesize Pivot quality results for \code{BQRRP} with block sizes $64$ and $4096$ show that block size has limited impact on pivot quality. The residual-norm ratio is generally similar to that of \code{GEQP3}, diverging mainly near sharp singular value drops, especially for $b = 4096$. The second metric shows near-identical behavior, except at the final singular value.
    }\label{fig:piv_qual}
\endminipage
\minipage{0.6\textwidth}
    \centering
    \begin{overpic}[width=1\textwidth,trim={0 0 2cm 0},clip]{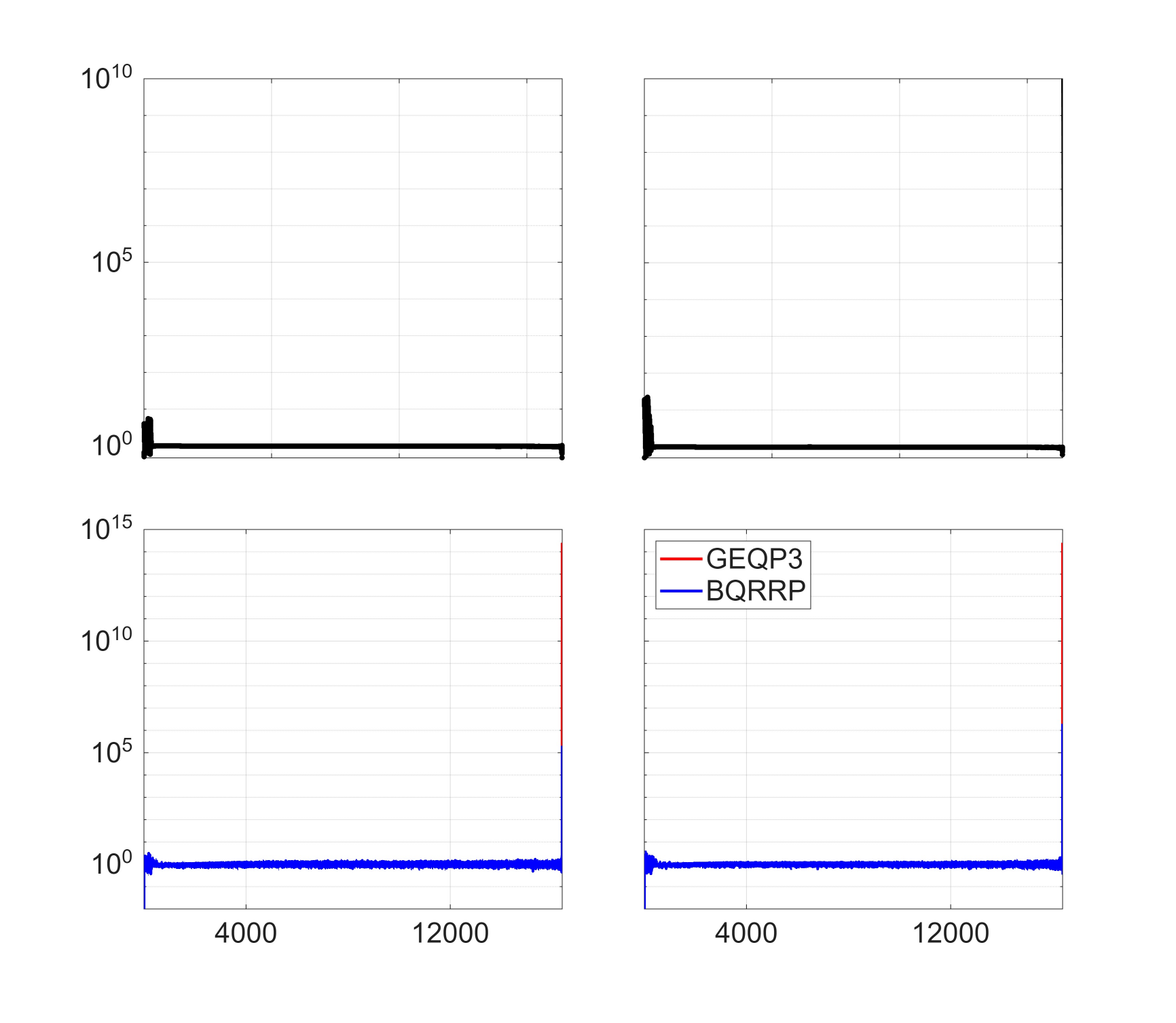}
    \put (20, 64) {$\scalemath{0.8}{\displaystyle\frac{\|\mtx{R}^{\text{geqp3}}(\tslice{i}, \tslice{i})\|_{\mathrm{F}}}{ \|\mtx{R}^{\text{bqrrp}}(\tslice{i}, \tslice{i})\|_{\mathrm{F}}}}$}
    \put (32, 4) {$\scalemath{0.9}{\mathbf{i}}$}
    \put (25, 25) {$\scalemath{0.8}{\displaystyle\frac{|\mtx{R}(i,i)|}{\sigma_{i}}}$}
    \put (75, 4) {$\scalemath{0.9}{\mathbf{i}}$}
    \put (28, 88) {$\scalemath{0.9}{b = 64}$}
    \put (68, 88) {$\scalemath{0.9}{b = 4096}$}
    \end{overpic}
\endminipage
\end{figure}

\cref{fig:kahan_spectrum} shows
the spectrum of a Kahan matrix (with default choices for the values of the tuning parameters), obtained by the Jacobi SVD function, \code{GESVD}.
\cref{fig:piv_qual} shows \code{BQRRP} pivot quality compared against \code{GEQP3} pivot quality in the two aforementioned metrics.
The experiment was conducted using two different \code{BQRRP} block sizes.

\FloatBarrier
\section{Performance results}
\label{sec:performance}

The experiments that we conduct in this section compare the performance of the following QR and QRCP algorithms:
\begin{itemize}
    \item \code{BQRRP\_CQR} -- a version of \code{BQRRP\_CPU} that uses Cholesky QR (and its dependencies) on a panel, \code{ORMQR} for the updating step, and \cref{alg:qrcp_practical} in Step \cref{bqrrp:qrcp}.
    \item \code{BQRRP\_HQR} -- a version of \code{BQRRP\_CPU} that uses Householder QR on a panel, \code{ORMQR} for the updating step, and \cref{alg:qrcp_practical} as a black-box \code{qrcp\_wide}.
    \item \code{GEQRF} -- standard unpivoted Householder QR.
\end{itemize}

This section concentrates on square matrices; tall and wide experiments are found in \cref{app:more_aspect_ratios}.
The sampling factor, $\gamma$, is set to the default value of $1.0$ in all experiments (since this is the only reasonable choice if \cref{alg:qrcp_practical} is in use at step \ref{bqrrp:qrcp}).
In all experiments, the performance is measured via \textit{canonical} FLOP rate, relying on the FLOP count of \code{GEQRF}.

\subsection{CPU algorithms performance}
\label{subsec:performance_cpu}
In addition to the algorithms listed above, CPU experiments also involve benchmarking \code{GEQP3} -- standard pivoted QR, and \code{HQRRP} -- randomized pivoted QR algorithm from \cite{MOHvdG:2017:QR}. We ported an LAPACK-compatible implementation \cite{hqrrp_lapack_sources} of HQRRP into \RandLAPACK{} (found in \code{/RandLAPACK/drivers/rl\_hqrrp.hh}) for easier benchmarking. 

\begin{figure}[htb!]
  \centering
  \hspace*{-0.2cm}
  \begin{tikzpicture}
    \node[inner sep=0pt] at (0,0) {\includegraphics[width=1.0\linewidth]{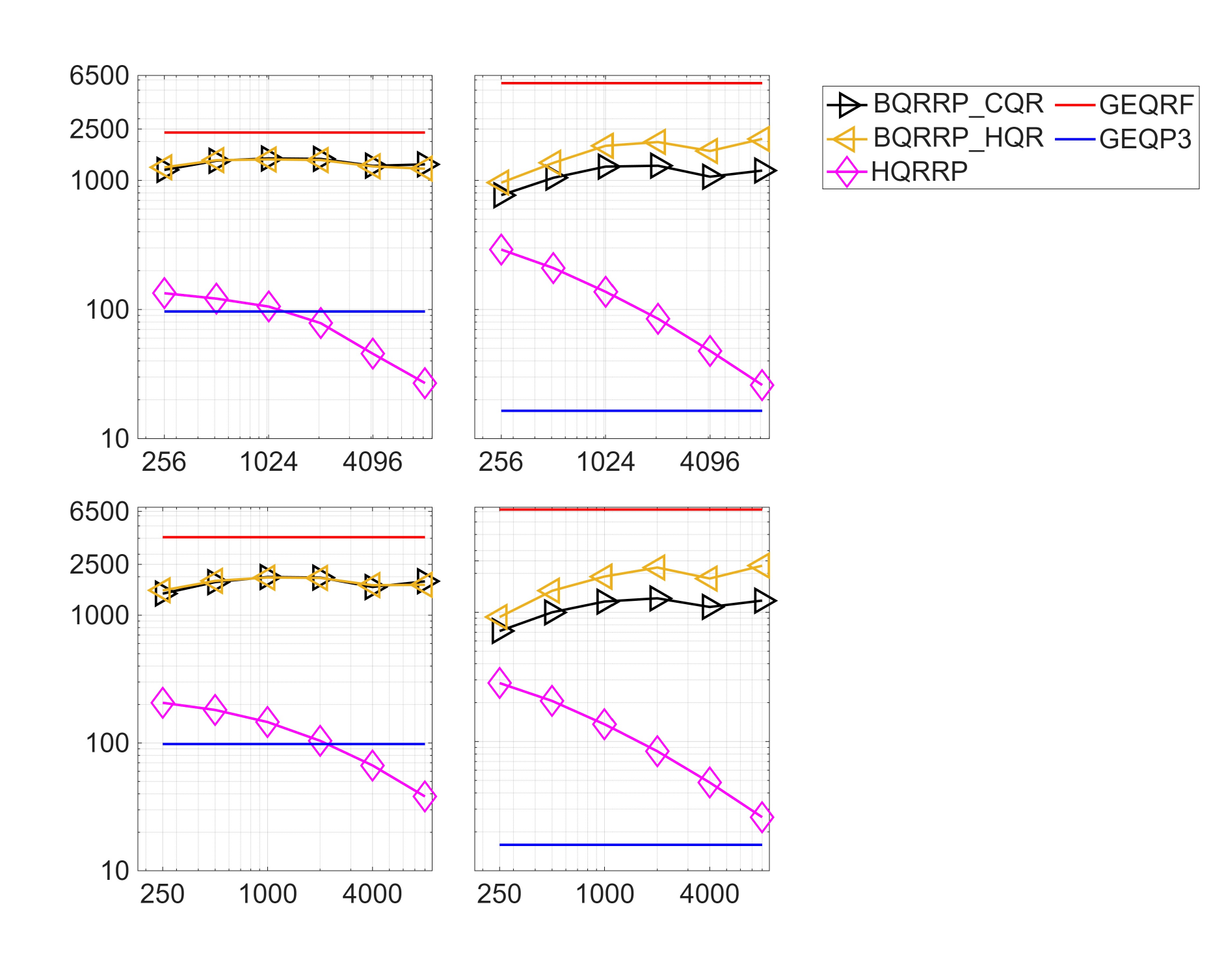}};

    \node[anchor=north west, text width=4.9cm, align=justify] at (1.9, 2.5) {%
    \caption{\footnotesize{Performance of various QR and QRCP methods, captured on an Intel and AMD systems (see \cref{table:cpu_config}). 
    The execution of \code{GEQRF} and \code{GEQP3} does not depend on the block size parameter $b$, and hence the performance is depicted as constant. 
    Observe that on an Intel system, both versions of \code{BQRRP\_CPU} exhibit comparable performance.
    On an AMD system, \code{BQRRP\_CPU} with Householder QR is the fastest method.
    }}
    \label{fig:qr_performance_block_size}
    };

    \node[anchor=north west, font=\bfseries] at (-0.35\linewidth, 5.2) {Intel CPU};
    \node[anchor=north west, font=\bfseries] at (-0.08\linewidth, 5.2) {AMD CPU};

    \node[anchor=north west, rotate=90, font=\bfseries] at (-0.51\linewidth, 0.1\linewidth) {\textbf{GigaFLOP/s}};
    \node[anchor=north west, rotate=90, font=\bfseries] at (-0.48\linewidth, 0.11\linewidth) {\textbf{$\mathbf{m_1 = 65{,}536}$}};

    \node[anchor=north west, rotate=90, font=\bfseries] at (-0.51\linewidth, -0.26\linewidth) {\textbf{GigaFLOP/s}};
    \node[anchor=north west, rotate=90, font=\bfseries] at (-0.48\linewidth, -0.25\linewidth) {\textbf{$\mathbf{m_1 = 64{,}000}$}};

    \node at (-0.27\linewidth, -5) {$\mathbf{b}$};
    \node at (0.005\linewidth, -5) {$\mathbf{b}$};
  \end{tikzpicture}
\end{figure}
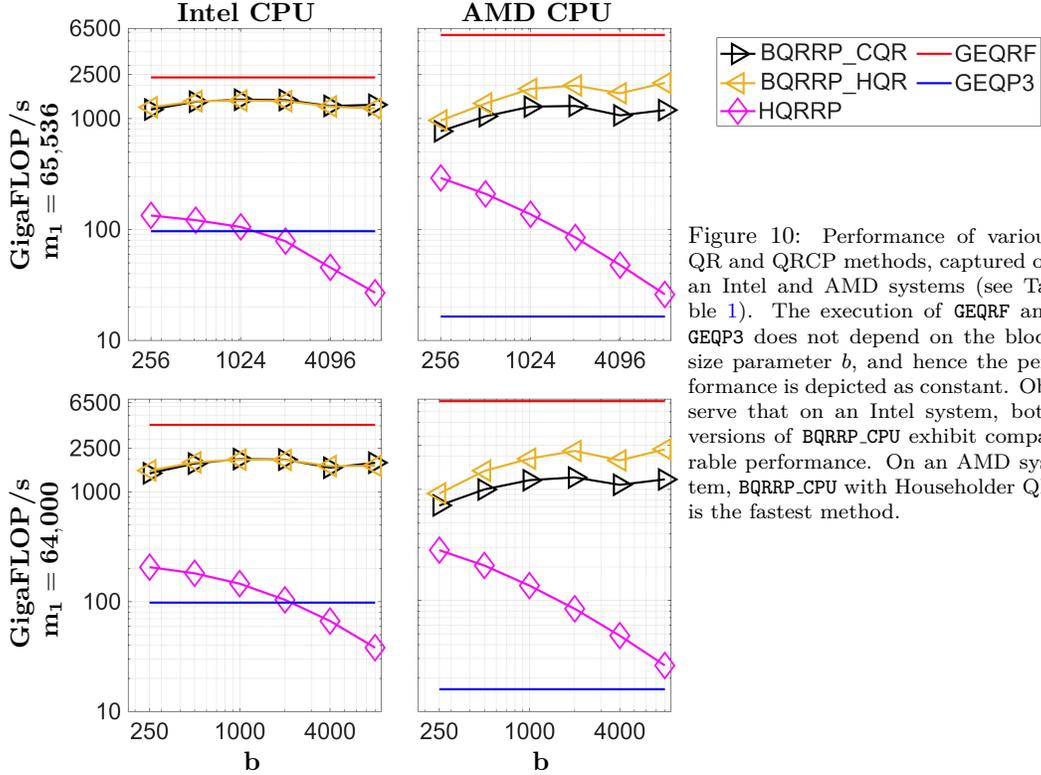

The first set of experimental results is shown in \cref{fig:qr_performance_block_size}.
Algorithms were run on $m_{1,2} \times m_{1,2}$ matrices using block sizes $b_{1,2}$, with $m_{1} = 65{,}536$ (\cref{fig:qr_performance_block_size}, row one) and $m_{2} = 64{,}000$ (\cref{fig:qr_performance_block_size}, row two), with block size in \code{BQRRP\_CPU} and \code{HQRRP} varying as the powers of two 
of multiples of ten as $b_{1} = 256 \cdot \{1, 2, 4, \dots, 32\}$ and $b_{2} = 250 \cdot \{1, 2, 4, \dots, 32\}$.

We set the \code{HQRRP} block sizes to the same values as \code{BQRRP} block sizes, given that the internal logic of \code{HQRRP} is extremely similar to that of \code{BQRRP}. 
This allows us to explore how similar block size decisions impact both algorithms.
A detailed exploration of the optimal \code{HQRRP} block size is provided in \cref{app:hqrrp_var_block_size}.

\cref{fig:qr_performance_block_size} shows that on an Intel system, \code{BQRRP\_CPU} with Cholesky QR on a panel has near-identical performance to that with Householder QR on a panel. 
\code{BQRRP\_CQR} and \code{BQRRP\_HQR} are up to $19\times$ faster than the standard pivoted QR, \code{GEQP3} (as seen in the bottom left plot in \cref{fig:qr_performance_block_size}), and they achieve up to $60\%$ of performance of the unpivoted QR, \code{GEQRF} (as seen in the top left plot in \cref{fig:qr_performance_block_size}).
Furthermore, \code{BQRRP\_CPU} algorithms are $7 \times$--$20 \times$ faster than \code{HQRRP}, depending on the block size used.

On an AMD system, \code{BQRRP\_CPU} with Householder QR on a panel is the fastest QRCP method. \code{BQRRP\_HQR} is up to $148 \times$ faster than the standard pivoted QR, \code{GEQP3} (as seen in the bottom right plot in \cref{fig:qr_performance_block_size}), and it achieves up to $35\%$ of performance of the unpivoted QR, \code{GEQRF} (as seen in the top right plot in \cref{fig:qr_performance_block_size}).
Furthermore, it is $3 \times$--$148 \times$ faster than \code{HQRRP}, depending on the block size used.

\begin{figure}[htb!]
  \centering
  \hspace*{-0.5cm}
  \begin{tikzpicture}
    \node[inner sep=0pt] at (0,0) {\includegraphics[width=1.0\linewidth]{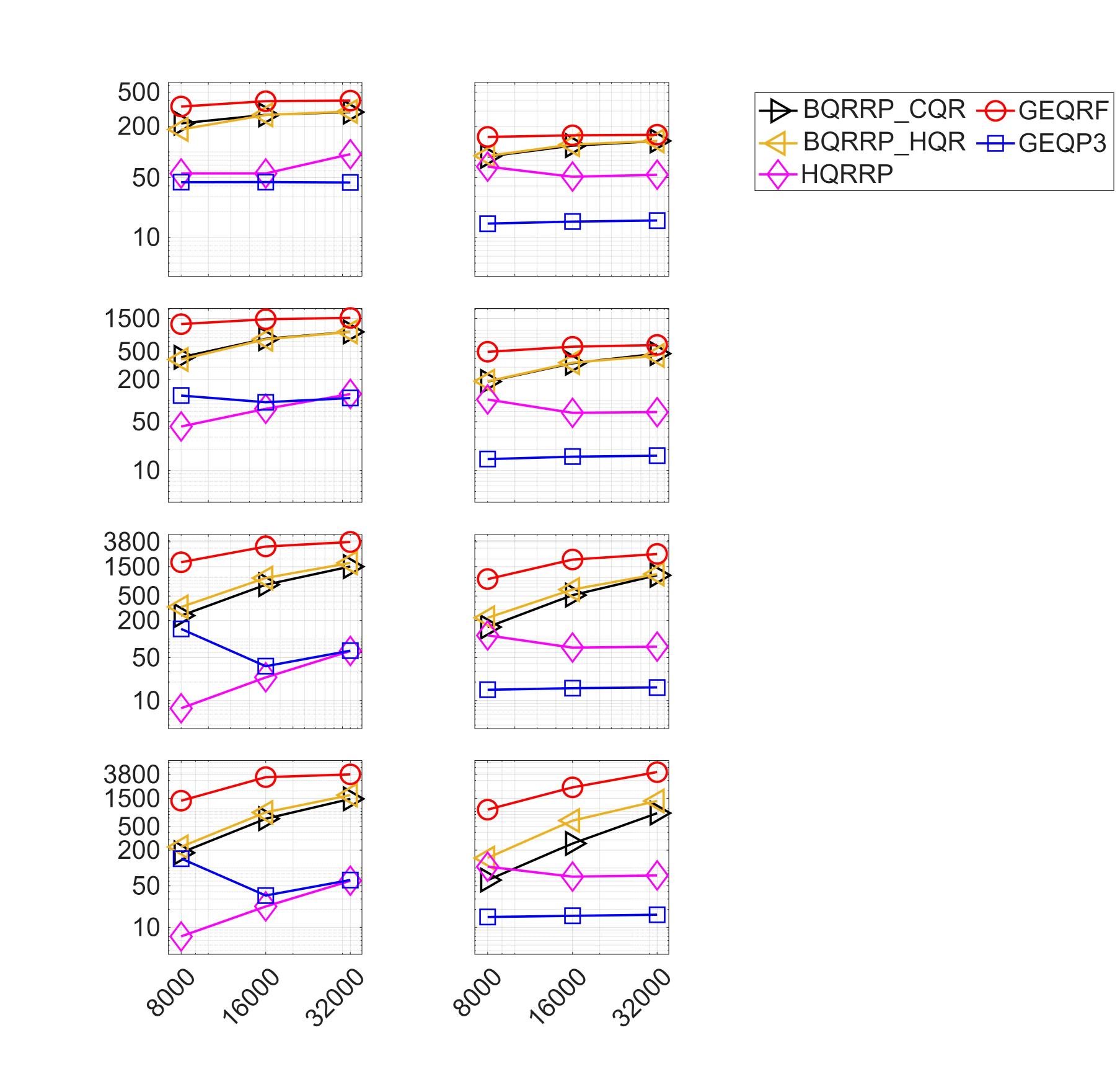}};
    
    \node[anchor=north west, text width=5.5cm, align=justify] at (1.7, 3.5) {\caption{\footnotesize Effects of varying the numbers of threads on the FLOP rates in QR and QRCP methods, captured on an Intel and AMD systems (see \cref{table:cpu_config}).
    In these experiments, $m \times m$ matrices are used, where $m \in \{8{,}000,~16{,}000,~32{,}000\}$; \code{HQRRP}, \code{BQRRP\_CQR}, and \code{BQRRP\_HQR} block size is set to $b = m / 32$.
    Observe that both versions of \code{BQRRP\_CPU} depict excellent thread scaling on par with \code{GEQRF} for the larger matrix sizes.
    Simultaneously, both \code{HQRRP} and \code{GEQP3} exhibit poor thread scaling (especially on the AMD system).}
    \label{fig:qr_performance_mat_size}
    };

     (manually adjusted coordinates to match image)
    \node[anchor=north west, font=\bfseries] at (-0.34\linewidth, 6.3) {Intel CPU};
    \node[anchor=north west, font=\bfseries] at (-0.067\linewidth, 6.3) {AMD CPU};

    \node[anchor=north west, rotate=90, font=\bfseries] at (-0.48\linewidth, 0.23\linewidth) {Threads=$4$};
    \node[anchor=north west, rotate=90, font=\bfseries] at (-0.45\linewidth, 0.23\linewidth) {GigaFLOP/s};

    \node[anchor=north west, rotate=90, font=\bfseries] at (-0.48\linewidth, 0.03\linewidth) {Threads=$16$};
    \node[anchor=north west, rotate=90, font=\bfseries] at (-0.45\linewidth, 0.03\linewidth) {GigaFLOP/s};

    \node[anchor=north west, rotate=90, font=\bfseries] at (-0.48\linewidth, -0.17\linewidth) {Threads=$64$};
    \node[anchor=north west, rotate=90, font=\bfseries] at (-0.45\linewidth, -0.17\linewidth) {GigaFLOP/s};

    \node[anchor=north west, rotate=90, font=\bfseries] at (-0.48\linewidth, -0.38\linewidth) {Threads=$128$};
    \node[anchor=north west, rotate=90, font=\bfseries] at (-0.45\linewidth, -0.38\linewidth) {GigaFLOP/s};

    \node at (-0.27\linewidth, -6.5) {$\mathbf{m=n}$};
    \node at (0.005\linewidth, -6.5) {$\mathbf{m=n}$};
  \end{tikzpicture}
\end{figure}
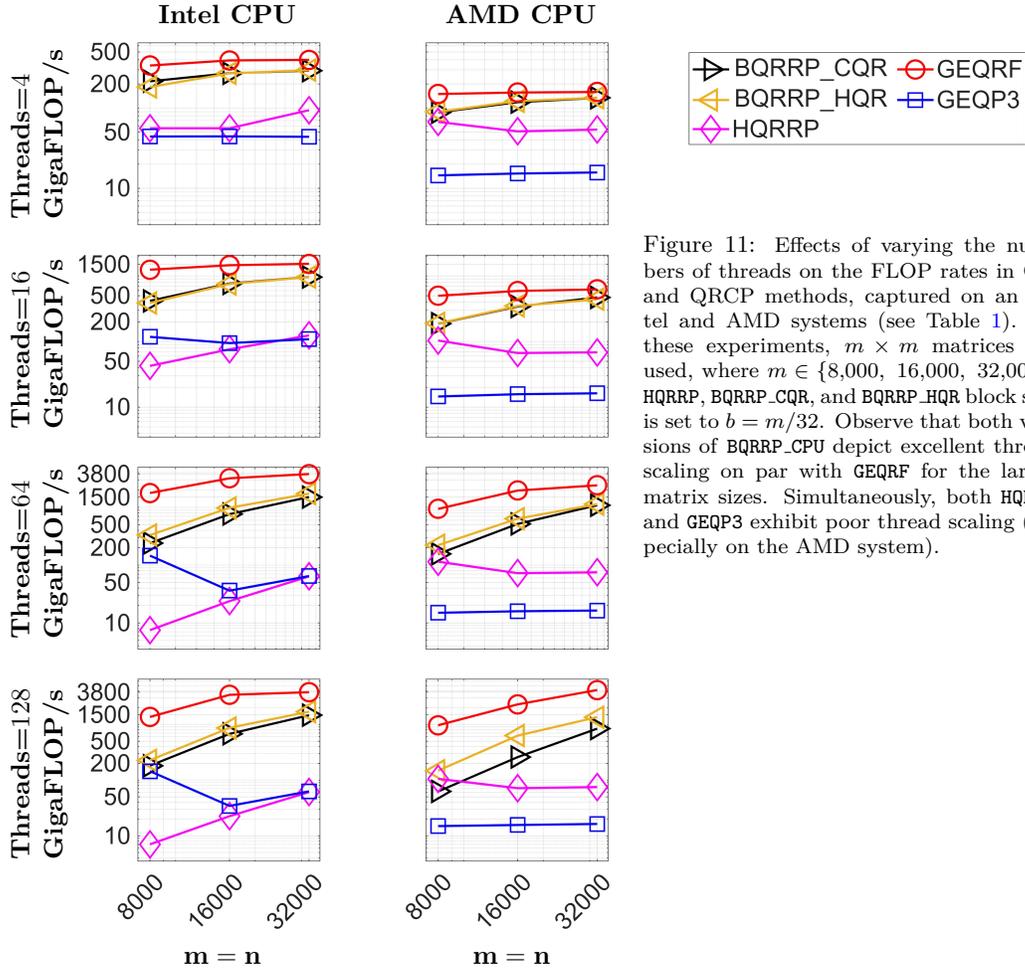

\paragraph{Thread scaling results. }\cref{fig:qr_performance_mat_size} depicts thread scaling results for the QR and QRCP schemes run on $m \times m$ matrices with $m \in \{8{,}000,~16{,}000,~32{,}000\}$.
In these plots, we set the block size parameter in \code{BQRRP\_CQR} and \code{BQRRP\_HQR} to $b = m / 32$.
This is because \cref{fig:qr_performance_block_size} shows that this block size setting generally yields the best $\code{BQRRP}$ performance across all experiments.

\cref{fig:qr_performance_mat_size} shows that \code{BQRRP} exhibits good thread scaling. For a $32{,}000 \times 32{,}000$ matrix, \code{BQRRP}'s performance improves by a factor of $20\times$ on the Intel system and $37\times$ on the AMD system when scaling from $1$ to $128$ threads.
In comparison, \code{GEQP3} achieves only a $6\times$ speedup on Intel and exhibits negligible thread scaling on AMD.

The standout performer is, as expected, \code{GEQRF}, with speedups of $43\times$ on Intel and $99\times$ on AMD.
Observe also that the performance of \code{GEQRF} on the AMD system begins to exceed that on the Intel system at $128$ threads for the matrix of size $32{,}000 \times 32{,}000$.
This is in line with what we saw in \cref{fig:qr_performance_block_size}, where \code{GEQRF} and \code{BQRRP} were showing much better overall performance on an AMD system.

We also observe that the performance of \code{HQRRP} begins to stagnate at $64$ threads used on an Intel system. 
We present a thorough investigation of \code{HQRRP} thread scaling results for various block sizes in \cref{app:hqrrp_var_block_size}.

In the bottom-left plot of \cref{fig:qr_performance_mat_size}, the performance of \code{BQRRP} approaches that of \code{GEQP3} for a matrix of size $8{,}000 \times 8{,}000$, with only a $1.5\times$ difference. 
This suggests that a \code{BQRRP} block size of $m/32$ may be suboptimal for smaller input matrices.
We further explore the performance of the discussed QR and QRCP schemes on smaller matrices in \cref{app:small_mat_performance}.

\FloatBarrier

\subsection{Performance of GPU Implementations}

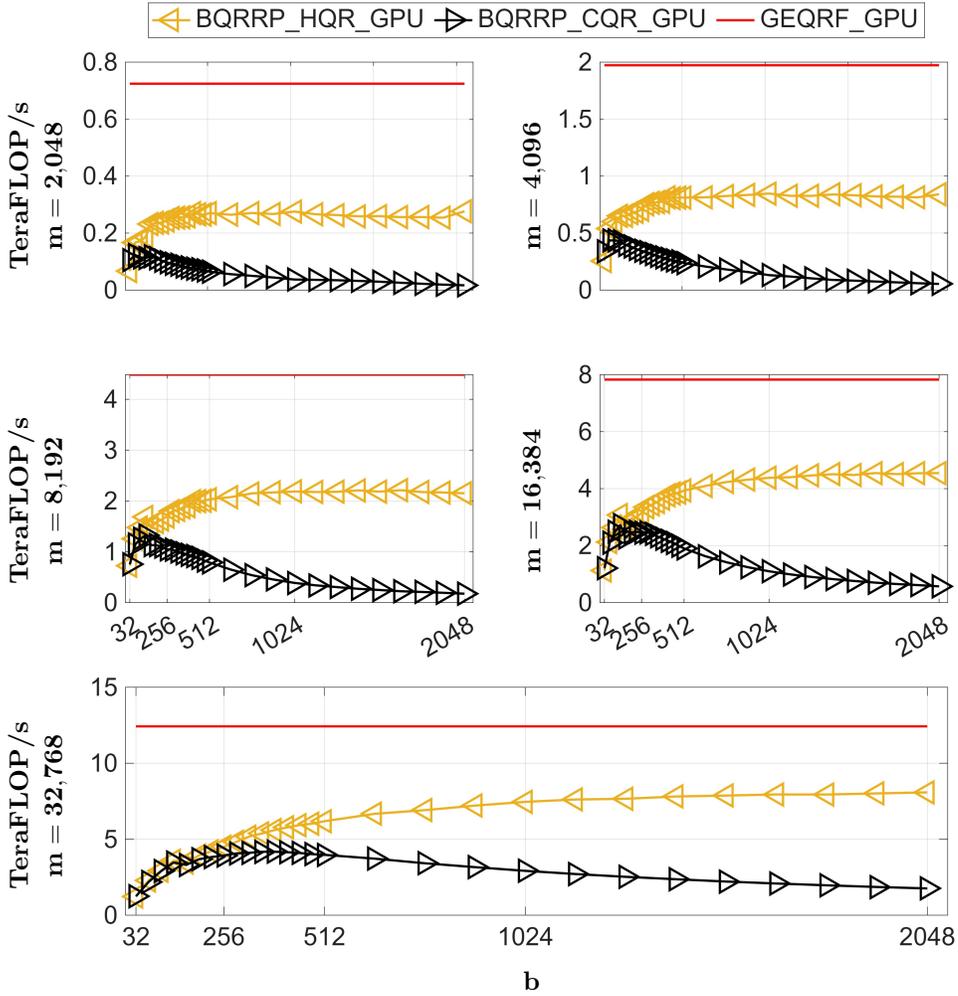
\begin{figure}[htp]
\centering
\begin{overpic}[width=1.0\textwidth]{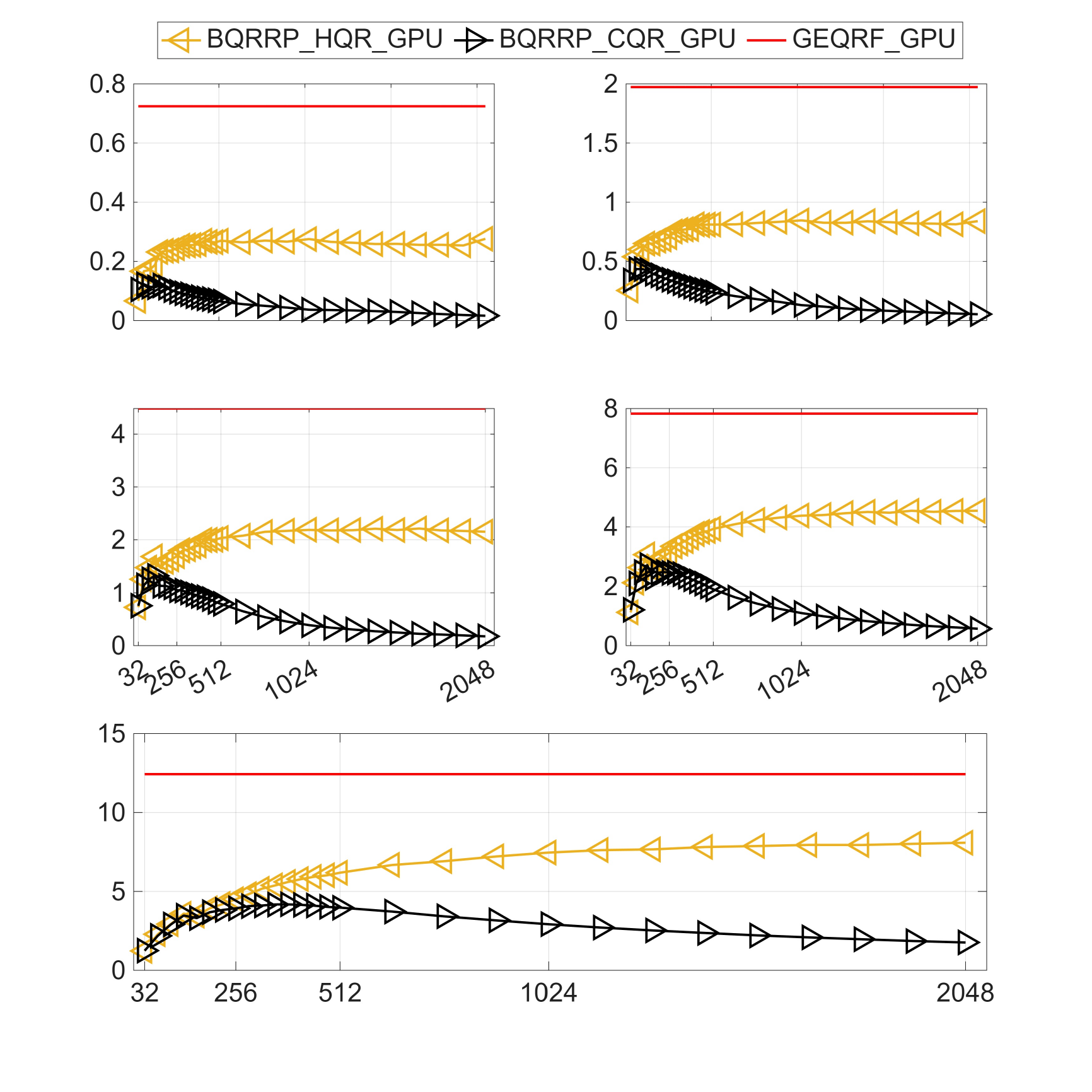}
\put (4.5, 80) {\rotatebox[origin=c]{90}{\textbf{$\mathbf{m = 2{,}048}$}}}
\put (1, 80) {\rotatebox[origin=c]{90}{\textbf{TeraFLOP/s}}}
\put (50, 80) {\rotatebox[origin=c]{90}{\textbf{$\mathbf{m = 4{,}096}$}}}
\put (4.5, 50) {\rotatebox[origin=c]{90}{\textbf{$\mathbf{m = 8{,}192}$}}}
\put (1, 50) {\rotatebox[origin=c]{90}{\textbf{TeraFLOP/s}}}
\put (50, 50) {\rotatebox[origin=c]{90}{\textbf{$\mathbf{m = 16{,}384}$}}}
\put (4.5, 21) {\rotatebox[origin=c]{90}{\textbf{$\mathbf{m = 32{,}768}$}}}
\put (1, 21) {\rotatebox[origin=c]{90}{\textbf{TeraFLOP/s}}}
\put (50, 4) {\textbf{b}}
\end{overpic}
\captionof{figure}{\footnotesize  Performance of standard QR and two versions of \code{BQRRP\_GPU}, captured on an NVIDIA GPU (for the details on system configuration, refer to \cref{table:gpu_config}). 
cuSOLVER’s \code{GEQRF} performance is constant, as it is independent of the block size $b$. \code{BQRRP} with Householder QR outperforms its Cholesky-based variant, aligning with our profiling results in \cref{fig:gpu_runtime_breakdown}. The relative performance of both \code{BQRRP\_GPU} versions to cuSOLVER remains stable across input sizes. Notably, smaller matrices reach peak performance with smaller block sizes.
}
\label{fig:qr_performance_gpu}
\end{figure}

As noted in \cref{sec:bqrrp_gpu}, a major challenge in designing GPU versions of algorithms is the lack of GPU implementations for many LAPACK-level functions. Unlike the CPU experiments, the only readily available GPU algorithm to compare with \code{BQRRP\_GPU} is cuSOLVER’s \code{GEQRF}\footnote{A GPU QRCP exists in MAGMA \cite{magma_zgeqp3_gpu}, \cite{tomov2010magma}, but is not widely used.}. We therefore compare cuSOLVER's \code{GEQRF} with two \code{BQRRP\_GPU} variants: one using Cholesky QR, and another using Householder QR for panel factorization. Experiments are run on $m \times m$ matrices with $m \in 256 \cdot {8, 16, 32, \dots, 128}$ (\cref{fig:qr_performance_gpu}), and block sizes from $32$ to $2048$.

\cref{fig:qr_performance_gpu} shows that the implementation of \code{BQRRP\_GPU} with Householder QR on a panel is able to achieve up to $60\%$ of the performance of the unpivoted QR method offered by cuSOLVER.
The performance of \code{BQRRP\_GPU} relying on Cholesky QR is far from acceptable due to the fact that a slow implementation of \code{ORHR\_COL} was used.
\cref{fig:qr_performance_gpu} also shows that the relative performance of the two \code{BQRRP\_GPU} schemes to cuSOLVER's \code{GEQRF} scales with the change in the input matrix size.

\section{Conclusion}
\label{sec:conclusion}

We have introduced BQRRP, a powerful algorithmic framework for QR factorization with column pivoting (QRCP) for general matrices. The framework enables the design of practical QRCP algorithms by allowing users to control key subroutine choices. We provide a detailed analysis of how these choices can be navigated for modern hardware, presenting formulations of \code{BQRRP\_CPU} and \code{BQRRP\_GPU}. The CPU version is designed for maximally in-place computation, while the GPU version prioritizes performance at the cost of additional storage. Both implementations produce output in the same format as \code{GEQP3}.

Looking ahead, the relative performance of core subroutines within the BQRRP framework may evolve due to advancements in computational techniques. Nevertheless, our work remains relevant, as it offers a plug-and-play algorithmic framework and a structured approach for analyzing the efficiency of individual subroutines in modern QRCP methods.

Our \RandLAPACK{} implementation of \code{BQRRP\_CPU} achieves up to $140\times$ the speed of GEQP3 and up to $60\%$ of the performance of GEQRF. Similarly, \code{BQRRP\_GPU}, also implemented in \RandLAPACK, reaches up to $60\%$ of \code{GEQRF} performance. Given these results and the flexibility of the BQRRP framework, it presents a strong case for inclusion as an alternative to \code{GEQP3} in LAPACK.

Our CPU benchmarks were conducted on Intel Sapphire Rapids and AMD Zen4c, while GPU results were obtained using an NVIDIA H100. The core subroutine choices were tuned to optimize performance on these architectures. For large matrices, we recommend setting the \code{BQRRP} block size to $1/32$ of the number of columns; while for smaller matrices, it should be set to its maximum feasible value. While this tuning strategy is not rigorously derived, it serves as a general guideline. For more specialized cases, where computations are performed on specific hardware and vendor-optimized libraries, subroutine choices within \code{BQRRP} should be re-evaluated accordingly.

This adaptability opens the door for future research and practical implementations by engineers looking to integrate \code{BQRRP} into their software. In particular, it would be valuable to assess \code{BQRRP}'s performance on ARM-based architectures and consumer-grade hardware such as Apple M-series silicon. Additionally, broader GPU testing is of interest, although current evaluations are limited by \RandLAPACK's CUDA-specific GPU support.


\FloatBarrier

\section*{Acknowledgements}

This work was partially funded by an NSF Collaborative Research Framework: Basic ALgebra LIbraries for Sustainable Technology with Interdisciplinary Collaboration (BALLISTIC), a project of the International Computer Science Institute, the University of Tennessee’s ICL, the University of California at Berkeley, and the University of Colorado at Denver (NSF Grant Nos. 2004235, 2004541, 2004763, 2004850, respectively).
MWM would also like to acknowledge the NSF, DOE, and ONR Basic Research Challenge on RLA for providing partial support for this work.

RM was partially supported by Laboratory Directed Research and Development (LDRD) funding from Sandia National Laboratories; Sandia is a multimission laboratory managed and operated by National Technology \& Engineering Solutions of Sandia, LLC, a wholly-owned subsidiary of Honeywell International Inc., for the U.S. Department of Energy’s National Nuclear Security Administration under contract DENA0003525. 

PL was supported in part by the Department of the Air Force Artificial Intelligence Accelerator and was accomplished under Cooperative Agreement Number FA8750-19-2-1000.

The views and conclusions contained
in this document are those of the authors and should not be
interpreted as presenting the official policies, either expressed or
implied, of the Department of the Air Force, the Department of Energy, or the U.S.\ Government.
The U.S.\ Government is authorized to reproduce and distribute reprints
for Government purposes notwithstanding any copyright notation herein.

\bibliography{references/refs.bib}
\bibliographystyle{alpha}

\normalsize

\appendix

\section{Investigating HQRRP performance}
\label{app:hqrrp_performance}

Recall from \cref{subsec:contribution} that we discussed prior work on developing a modern QRCP approach, highlighting \code{HQRRP} \cite{MOHvdG:2017:QR} as a particularly promising scheme. 
Nevertheless, comparing the results in \cref{fig:hqrrp_plot_remake} with those reported in \cite[Fig. 1]{MOHvdG:2017:QR} and \cite[Fig. 5]{MOHvdG:2017:QR}, we observe that the performance gap between \code{GEQRF} and \code{GEQP3} has increased from under $10\times$ to around $100\times$ on a modern AMD CPU, while the speedup of \code{HQRRP} over \code{GEQP3} has not scaled proportionately, remaining at approximately $13\times$ at best.
Moreover, from column one in \cref{fig:hqrrp_plot_remake}, we see that the performance of \code{HQRRP} may fall below that of \code{GEQP3} as the number of OpenMP threads in use increases. 

This overall difference in performance is, of course, largely attributable to the drastic differences in hardware platforms: our experiments were conducted on modern CPUs, described in \cref{table:cpu_config}, whereas the experiments from \cite{MOHvdG:2017:QR} were performed on an Intel Xeon E5-2695 v3 (the Haswell platform) processor featuring ``only'' $14$
cores in a single socket, a platform that is now over a decade old.
Nonetheless, to thoroughly assess the \code{HQRRP} algorithm, we shall investigate how tuning its \textit{block size} parameter affects performance on modern systems (although \cite[Sec. 4.1]{MOHvdG:2017:QR} suggests that using the block size of $64$ or $128$ should yield near-best performance, this may not hold true on our hardware systems).
Furthermore, we shall perform the subroutine performance profiling (similar to \cref{sec:performance_profiling}) in \code{HQRRP} to identify any computational bottlenecks.
Finally, it is worth analyzing the performance of \textit{each individual} algorithm involved in \cref{fig:hqrrp_plot_remake} in order to detect any performance instabilities of each given scheme (of particular interest is the relationship between the performance of \code{GEQRF} and \code{GEQP3} on an Intel system).
The following subsections address each of these tasks.

\subsection{Alternative view of \cref{fig:hqrrp_plot_remake}}
\label{app:subsec:hqrrp_remake_alternative}

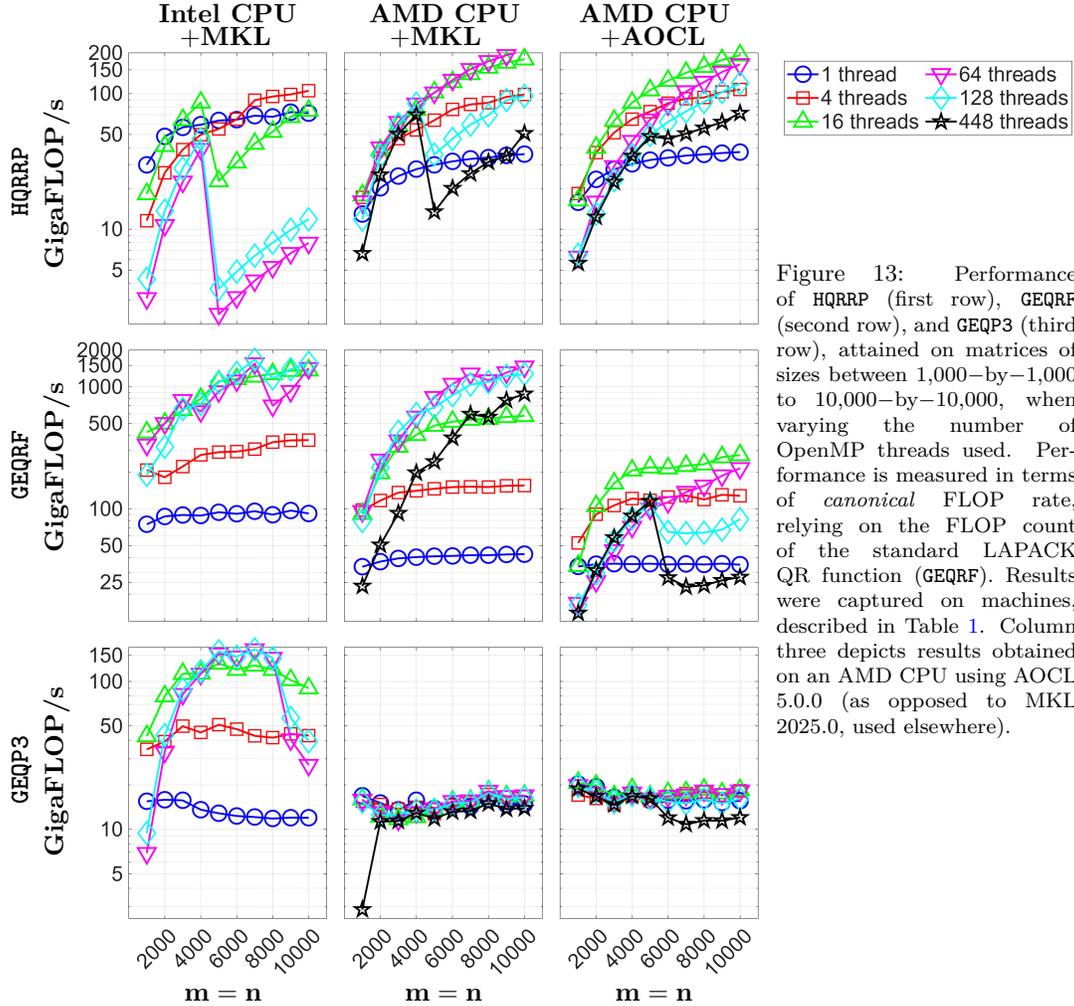
\begin{figure}[htb!]
  \centering
  \hspace*{-0.3cm}
  \begin{tikzpicture}
    \node[inner sep=0pt] at (0,0) {\includegraphics[width=1.0\linewidth]{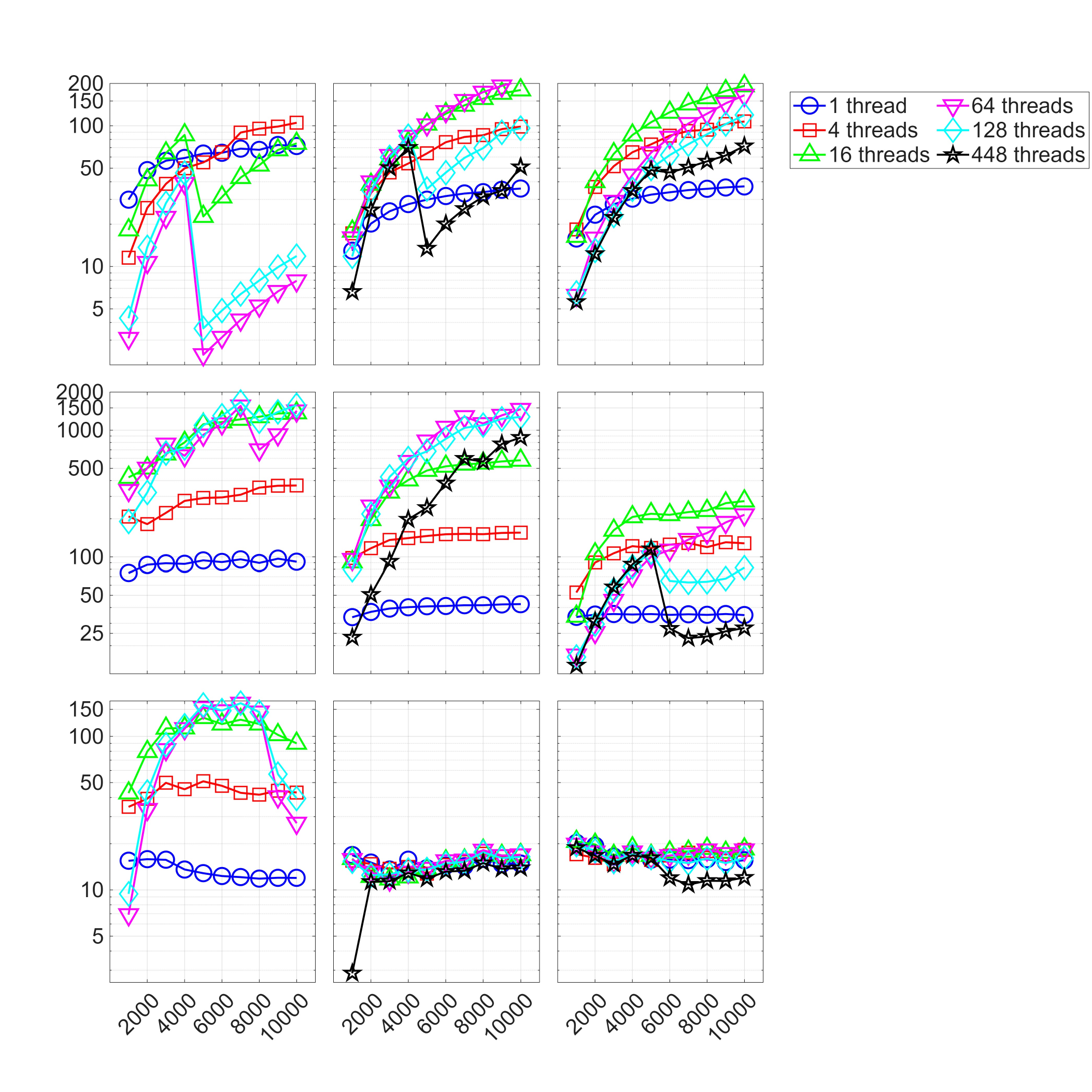}};

    \node[anchor=north west, text width=4cm, align=justify] at (2.9, 3.5) {%
    \caption{\footnotesize{Performance of \code{HQRRP} (first row), \code{GEQRF} (second row), and \code{GEQP3} (third row), attained on matrices of sizes between $1{,}000-$by$-1{,}000$ to $10{,}000-$by$-10{,}000$, when varying the number of OpenMP threads used.
    Performance is measured in terms of \textit{canonical} FLOP rate, relying on the FLOP
    count of the standard LAPACK QR function (\code{GEQRF}).
    Results were captured on machines, described in \cref{table:cpu_config}. Column three depicts results obtained on an AMD CPU using AOCL 5.0.0 (as opposed to MKL 2025.0, used elsewhere).
    }}
    \label{fig:hqrrp_plot_remake_flops}
    };

    \node[anchor=north west, font=\bfseries] at (-0.38\linewidth, 6.7) {Intel CPU};
    \node[anchor=north west, font=\bfseries] at (-0.36\linewidth, 6.4) {$+$MKL};
    
    \node[anchor=north west, font=\bfseries] at (-0.18\linewidth, 6.7) {AMD CPU};
    \node[anchor=north west, font=\bfseries] at (-0.16\linewidth, 6.4) {$+$MKL};

    \node[anchor=north west, font=\bfseries] at (0.02\linewidth, 6.7) {AMD CPU};
    \node[anchor=north west, font=\bfseries] at (0.04\linewidth, 6.4) {$+$AOCL};

    \node[anchor=north west, rotate=90, font=\bfseries] at (-0.49\linewidth, 0.21\linewidth) {\textbf{GigaFLOP/s}};
    \node[anchor=north west, rotate=90, font=\bfseries] at (-0.52\linewidth, 0.26\linewidth) {\textbf{\code{HQRRP}}};

    \node[anchor=north west, rotate=90, font=\bfseries] at (-0.49\linewidth, -0.06\linewidth) {\textbf{GigaFLOP/s}};
    \node[anchor=north west, rotate=90, font=\bfseries] at (-0.52\linewidth, -0.01\linewidth) {\textbf{\code{GEQRF}}};

    \node[anchor=north west, rotate=90, font=\bfseries] at (-0.49\linewidth, -0.35\linewidth) {\textbf{GigaFLOP/s}};
    \node[anchor=north west, rotate=90, font=\bfseries] at (-0.52\linewidth, -0.3\linewidth) {\textbf{\code{GEQP3}}};

    \node at (-0.31\linewidth, -6.6) {$\mathbf{m = n}$};
    \node at (-0.1\linewidth, -6.6) {$\mathbf{m = n}$};
    \node at (0.1\linewidth, -6.6) {$\mathbf{m = n}$};
  \end{tikzpicture}
\end{figure}

\cref{fig:hqrrp_plot_remake_flops} presents the performance results of running \code{HQRRP}, \code{GEQRF}, and \code{GEQP3}, measured in terms of the canonical FLOP rate, as opposed to relative speedup (\cref{fig:hqrrp_plot_remake}).
As such, we are able to assess any performance instabilities of each individual algorithm.
We observe that the algorithms perform rather unstably on an Intel CPU across the board, even though, as stated in \cref{subsec:exp_setup}, we perform $20$ runs of each algorithm per given matrix size to address the potential instabilities.
Of particular interest is the fact that increasing the number of OpenMP threads used only occasionally has a positive effect on the performance of \code{HQRRP} on Intel hardware. 
Additionally, it is worth noting that the performance of \code{GEQP3} on Intel CPU is far superior to that on AMD CPU, regardless of which vendor library the algorithm is sourced from.

Additional observations can be made regarding the relative performance of algorithms run on AMD hardware when using AOCL versus MKL.
As seen from the two rightmost columns in \cref{fig:hqrrp_plot_remake_flops}, while the \code{HQRRP} performs roughly the same at its peak, the performance of \code{GEQRF} and \code{GEQP3} sourced from MKL is far superior to that from AOCL.
This observation is the basis for our decision to report AMD system results using MKL instead of AOCL.

\subsection{Varying \code{HQRRP} block size}
\label{app:hqrrp_var_block_size}
\begin{figure}[htb!]
  \centering
  \hspace*{-0.3cm}
  \begin{tikzpicture}
    \node[inner sep=0pt] at (0,0) {\includegraphics[width=1.0\linewidth]{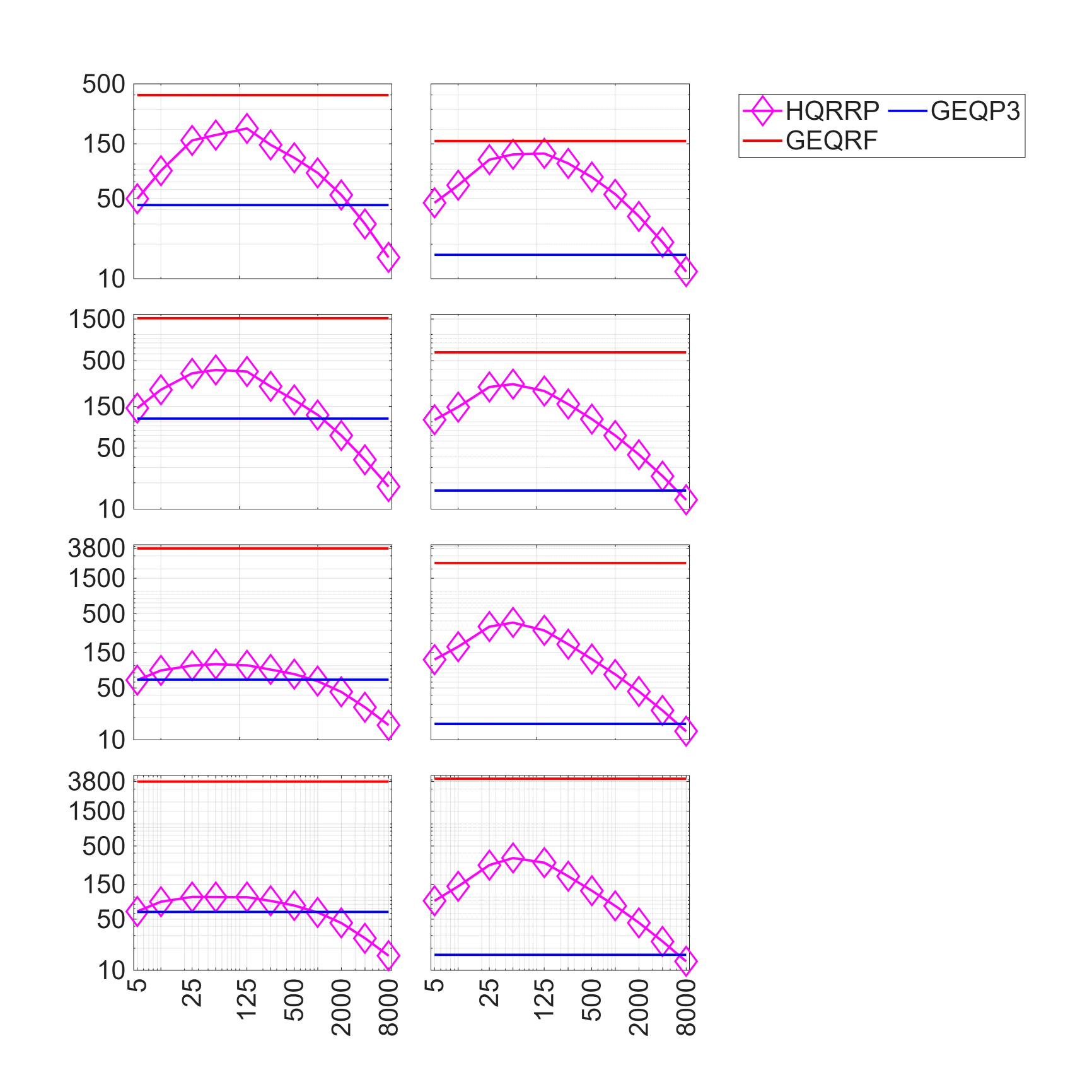}};
    
    \node[anchor=north west, text width=4.8cm, align=justify] at (1.9, 3.5) {\caption{\footnotesize Performance comparison of \code{HQRRP}, \code{GEQRF}, and \code{GEQP3}, captured on Intel and AMD systems (see \cref{table:cpu_config}). 
    The experiments are conducted on matrices with $32{,}000$ rows and columns and the \code{HQRRP} block size parameter $b \in \{5,~10,~25,~50,~125,~250,~500, \\ ~1000,~2000,~4000,~8000\}$ ($x$-axis).}
    \label{fig:hqrrp_performance_varying_block_size}
    };

     (manually adjusted coordinates to match image)
    \node[anchor=north west, font=\bfseries] at (-0.33\linewidth, 6.4) {Intel CPU};
    \node[anchor=north west, font=\bfseries] at (-0.067\linewidth, 6.4) {AMD CPU};

    \node[anchor=north west, rotate=90, font=\bfseries] at (-0.51\linewidth, 0.24\linewidth) {Threads=$4$};
    \node[anchor=north west, rotate=90, font=\bfseries] at (-0.48\linewidth, 0.23\linewidth) {GigaFLOP/s};

    \node[anchor=north west, rotate=90, font=\bfseries] at (-0.51\linewidth, 0.04\linewidth) {Threads=$16$};
    \node[anchor=north west, rotate=90, font=\bfseries] at (-0.48\linewidth, 0.03\linewidth) {GigaFLOP/s};

    \node[anchor=north west, rotate=90, font=\bfseries] at (-0.51\linewidth, -0.17\linewidth) {Threads=$64$};
    \node[anchor=north west, rotate=90, font=\bfseries] at (-0.48\linewidth, -0.17\linewidth) {GigaFLOP/s};

    \node[anchor=north west, rotate=90, font=\bfseries] at (-0.51\linewidth, -0.38\linewidth) {Threads=$128$};
    \node[anchor=north west, rotate=90, font=\bfseries] at (-0.48\linewidth, -0.38\linewidth) {GigaFLOP/s};

    \node at (-0.27\linewidth, -6.5) {$\mathbf{b}$};
    \node at (0.005\linewidth, -6.5) {$\mathbf{b}$};
  \end{tikzpicture}
\end{figure}

As stated previously, in the experiments presented in \cref{fig:hqrrp_plot_remake} and \cref{fig:hqrrp_plot_remake_flops}, the \code{HQRRP} block size parameter is set to $128$, per the suggestion in \cite[Sec. 4.1]{MOHvdG:2017:QR}.
\cref{fig:hqrrp_performance_varying_block_size} shows how varying the block size in \code{HQRRP} affects its performance for the various numbers of OpenMP threads used.

Results in \cref{fig:qr_performance_mat_size} show that the performance of \code{HQRRP} indeed peaks at around block size $64-128$, as \cite{MOHvdG:2017:QR} suggests.
However, comparing \cref{fig:qr_performance_mat_size} with \cref{fig:hqrrp_performance_varying_block_size}, we see that even at its best, \code{HQRRP} does not reach the performance of either version of \code{BQRRP} (except, of course, in a single-threaded case).
Furthermore, as seen previously, the performance of \code{HQRRP} begins to stagnate at $64$ OpenMP threads, suggesting that some suboptimal (possibly, Level-2 BLAS) subroutines may have been used in \code{HQRRP}.

\subsection{\code{HQRRP} subroutines profiling}
\label{sec:hqrrp_profiling}
\begin{figure}[htp]
\centering
    \begin{overpic}[width=1.0\textwidth]{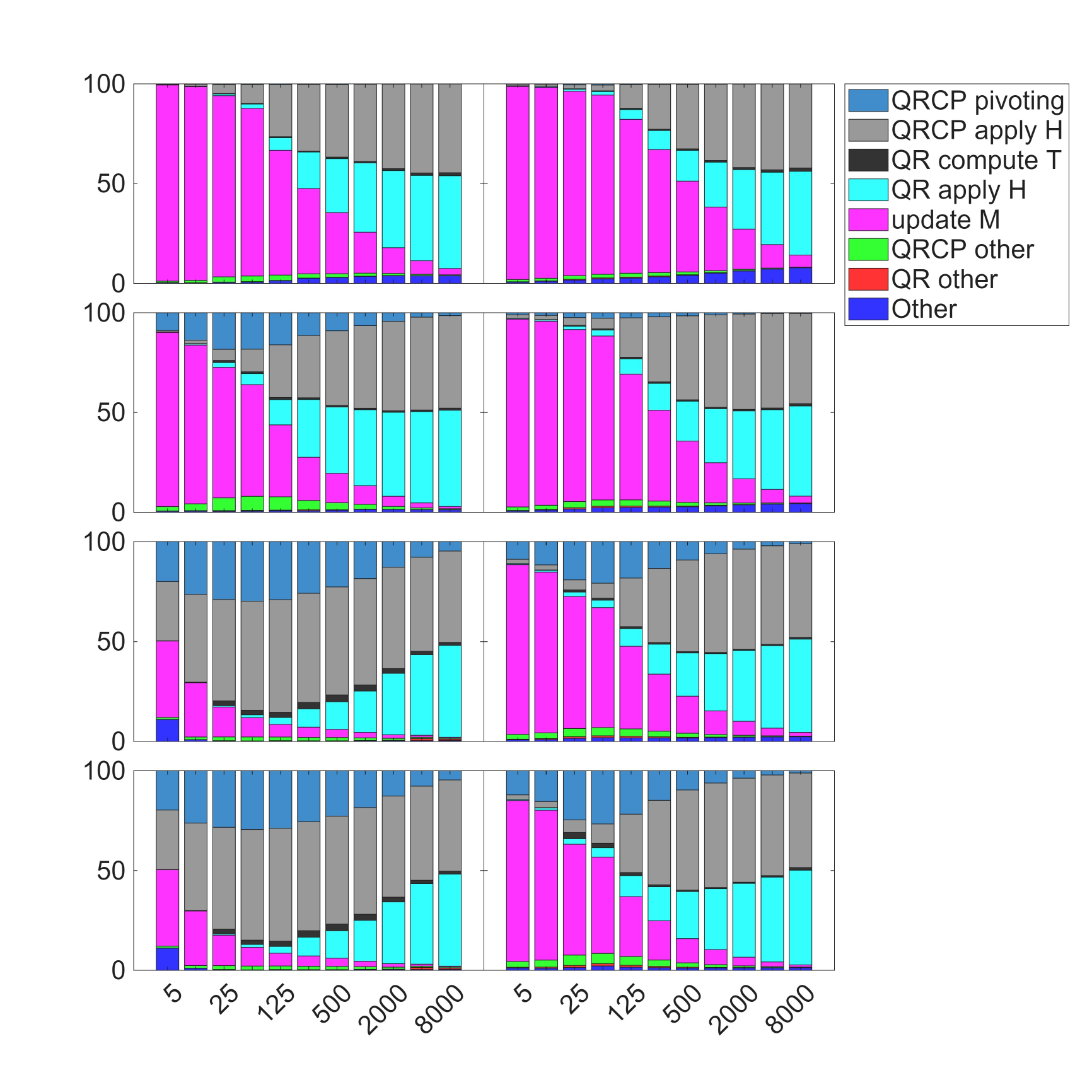}
    \put (23, 93) {\textbf{Intel CPU}}
    \put (53, 93) {\textbf{AMD CPU}}
    \put (3, 82) {\rotatebox[origin=c]{90}{\textbf{Threads=$4$}}}
    \put (5, 82) {\rotatebox[origin=c]{90}{\textbf{GigaFLOP/s}}}
    \put (3, 61) {\rotatebox[origin=c]{90}{\textbf{Threads=$16$}}}
    \put (5, 61) {\rotatebox[origin=c]{90}{\textbf{GigaFLOP/s}}}
    \put (3, 40) {\rotatebox[origin=c]{90}{\textbf{Threads=$64$}}}
    \put (5, 40) {\rotatebox[origin=c]{90}{\textbf{GigaFLOP/s}}}
    \put (3, 19) {\rotatebox[origin=c]{90}{\textbf{Threads=$128$}}}
    \put (5, 19) {\rotatebox[origin=c]{90}{\textbf{GigaFLOP/s}}}
    \put (28, 2) {$\mathbf{b}$}
    \put (60, 2) {$\mathbf{b}$}
    \end{overpic}
  \captionof{figure}{\footnotesize Percentages of \code{HQRRP} runtime, occupied by its respective subroutines.
  Experminets were conducted on square matrices of size $32{,}000\times 32{,}000$ with the \code{HQRRP} block size $b$ taking values $\{5,~10,~25,~50,~125,~250,~500,~1000,~2000,~4000,~8000\}$.
  The results are captured on Intel and AMD systems (see \cref{table:cpu_config}).
  }
  \label{fig:hqrrp_runtime_breakdown}
\end{figure}

In an effort to better understand the performance gap between \code{BQRRP} and \code{HQRRP} seen in \cref{fig:qr_performance_mat_size} and \cref{fig:hqrrp_performance_varying_block_size} (as well as the reason for poor thread scaling in \code{HQRRP}), we present the subroutines performance breakdown of \code{HQRRP} below (similar to how we did it for \code{BQRRP} in \cref{sec:performance_profiling}). 
\cref{fig:hqrrp_runtime_breakdown} depicts the percentage of runtime that is occupied by a given subcomponent of \code{HQRRP} on the $y$-axis.
We use square test matrices with $32{,}000$ rows and columns and the block size parameter $b \in \{5,~10,~25,~50,~125,~250,~500,~1000,~2000,~4000,~8000\}$ ($x$-axis).

As seen in \cref{fig:hqrrp_runtime_breakdown}, the function that is responsible for updating the input matrix tends to occupy most of the \code{HQRRP} runtime when smaller block sizes are in use (and especially for the lower numbers of OpenMP threads used).
In the LAPACK-compatible \code{HQRRP} implementation, this function is called ``\code{NoFLA\_Apply\_Q\_WY\_lhfc\_blk\_var4}.'' 
It relies on a single LAPACK subroutine, \code{LARFB}, which applies a block reflector to a given rectangular matrix.
Other notable subroutines come from the QR and QRCP functions within \code{HQRRP}, which are performed via calling ``\code{NoFLA\_QRPmod\_WY\_unb\_var4}'' in pivoted and unpivoted modes, respectively. There, the most costly functions are \code{LARF} (applies a single reflector to a given rectangular matrix).
Finally, the column permutation strategy in \code{HQRRP}, implemented in ``\code{NoFLA\_QRP\_pivot\_G\_B\_C}'' becomes costly when larger numbers of OpenMP threads are used (particularly at $64$ threads, where we start seeing the stagnation in \code{HQRRP} performance).

Overall, we conclude that using \code{LARF} within \code{HQRRP} is suboptimal, since this function operates on a single reflector at a time, and consequently is largely cast in terms of level 2 BLAS.
Furthermore, it could be the case that depending on the system and the input problem, \code{ORMQR} used in \code{BQRRP} is superior to \code{LARFB} used in \code{HQRRP}.

\section{Cholesky QR background}
\label{app:cholqr}

This section is intended to provide background information on \textit{Cholesky QR} that can be used in the context of \cref{subsec:qr_tall}.

\paragraph{Cholesky QR.} Given an $m \times n$ matrix $\mtx{M}$, with $m \geq n$, Cholesky QR computes the Gram matrix $\mtx{G} = \mtx{M}^{\trans}\mtx{M}$, factors $\mtx{G} = \mtx{R}^{\trans}\mtx{R}$, computing a non-singular upper-triangular matrix $\mtx{R}$, and obtains an orthonormal factor $\mtx{Q} = \mtx{M}\mtx{R}^{-1}$.
Note that this procedure works only if $\mtx{M}$ has rank $n$.
The FLOP count in Cholesky QR is close to that of standard LAPACK unpivoted QR, \code{GEQRF}, as long as $m \gg n$.
In a practical implementation, Cholesky QR can significantly outperform \code{GEQRF} even when the input matrices are not conventionally considered ``very tall,'' with the ratio $m / n$ being on the order of $10$. 
Despite its simplicity and speed, Cholesky QR is rarely used in practice, as it fails\footnote{\code{POTRF} always outputs a factorization of the whole input or a leading principal submatrix thereof.} to provide accurate output when the numerical rank of the matrix $\mtx{G}$ falls below $n$. 
This phenomenon can be mitigated with a variety of preconditioning and truncation strategies.

\paragraph{Preconditioned Cholesky QR.}
The use of Cholesky QR in the context of step \ref{bqrrp:qr_tall} is motivated by the fact that the suitable preconditioner in the form of $\sk{\mtx{R}}(\lslice{k},\lslice{k})$ is acquired ``for free'' in step \ref{bqrrp:qrcp} (as this step is necessary for acquiring the permutation vector).
The preconditioning would be performed by applying $(\sk{\mtx{R}}(\lslice{k},\lslice{k}))^{-1}$ to a portion of the permuted matrix $\mtx{M}$ from the right.
The effectiveness of the preconditioning of this flavor has been thoroughly analyzed in Section 2.1, Appendix A1 of \cite{MBM2024}.
This idea was first introduced by Fan, Guo, and Lin
\cite{FGL:2021:CholeskyQR} and studied in detail by others \cite{Balabanov:2022:cholQR,HSBY:2023:rand_chol_qr}.
In the context of Cholesky QR, the numerical rank computation (step \ref{bqrrp:rank_est}, described in \cref{subsec:rank_est}) is not strictly necessary;
this step is solely used to ensure that the preconditioning is performed safely (meaning that no infinite or not-a-number values should appear when inverting $\sk{\mtx{R}}(\lslice{k},\lslice{k})$).
As such, naive rank estimation suffices. 

\section{Additional CPU performance experiments}
\subsection{Performance results on smaller inputs}
\label{app:small_mat_performance}

As stated in \cref{subsec:performance_cpu}, although the results from \cref{fig:qr_performance_block_size} show that \code{BQRRP} performs best when the block size $b$ is set to $b = n/32$ (given an $m \times n$ input matrix), this may only hold for larger input matrix sizes (as seen in \cref{fig:qr_performance_mat_size}, the performance of \code{BQRRP} gets suspiciously close to that of \code{GEQRF} when the input is of size $8{,}000\times 8{,}000$).
As such, we investigate what block size setting works best for the smaller input matrices.

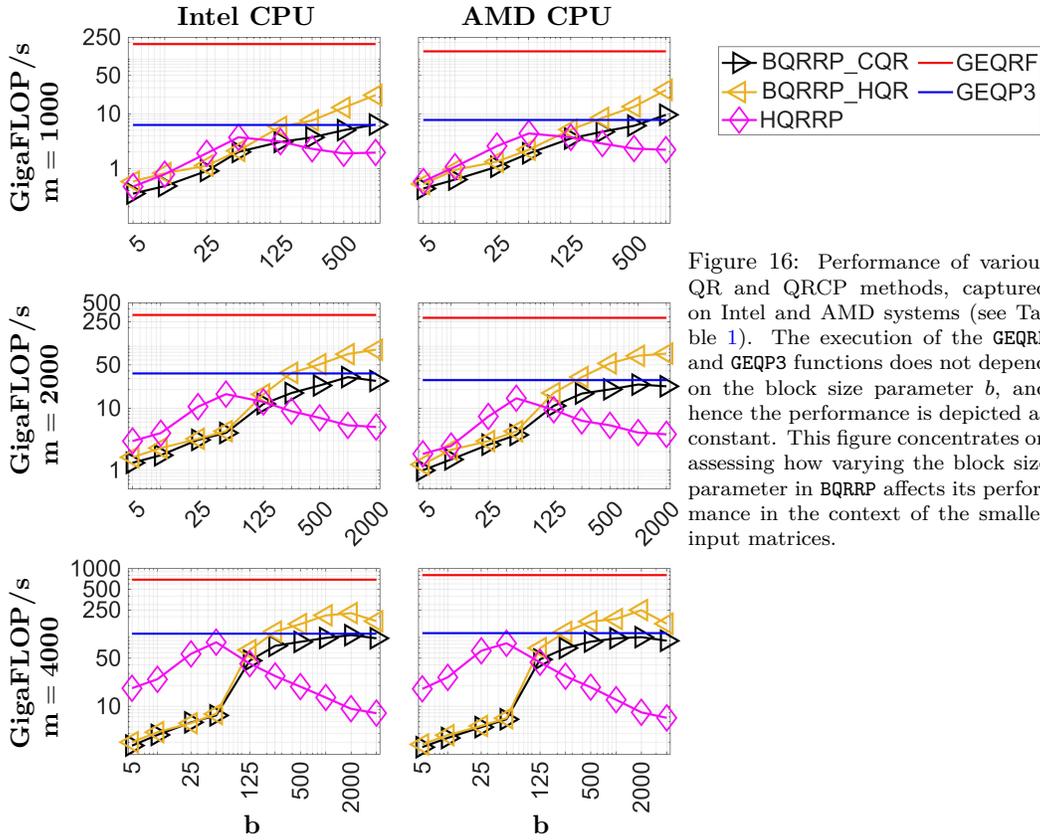
\begin{figure}[htb!]
  \centering
  \hspace*{-0.2cm}
  \begin{tikzpicture}
    \node[inner sep=0pt] at (0,0) {\includegraphics[width=1.0\linewidth]{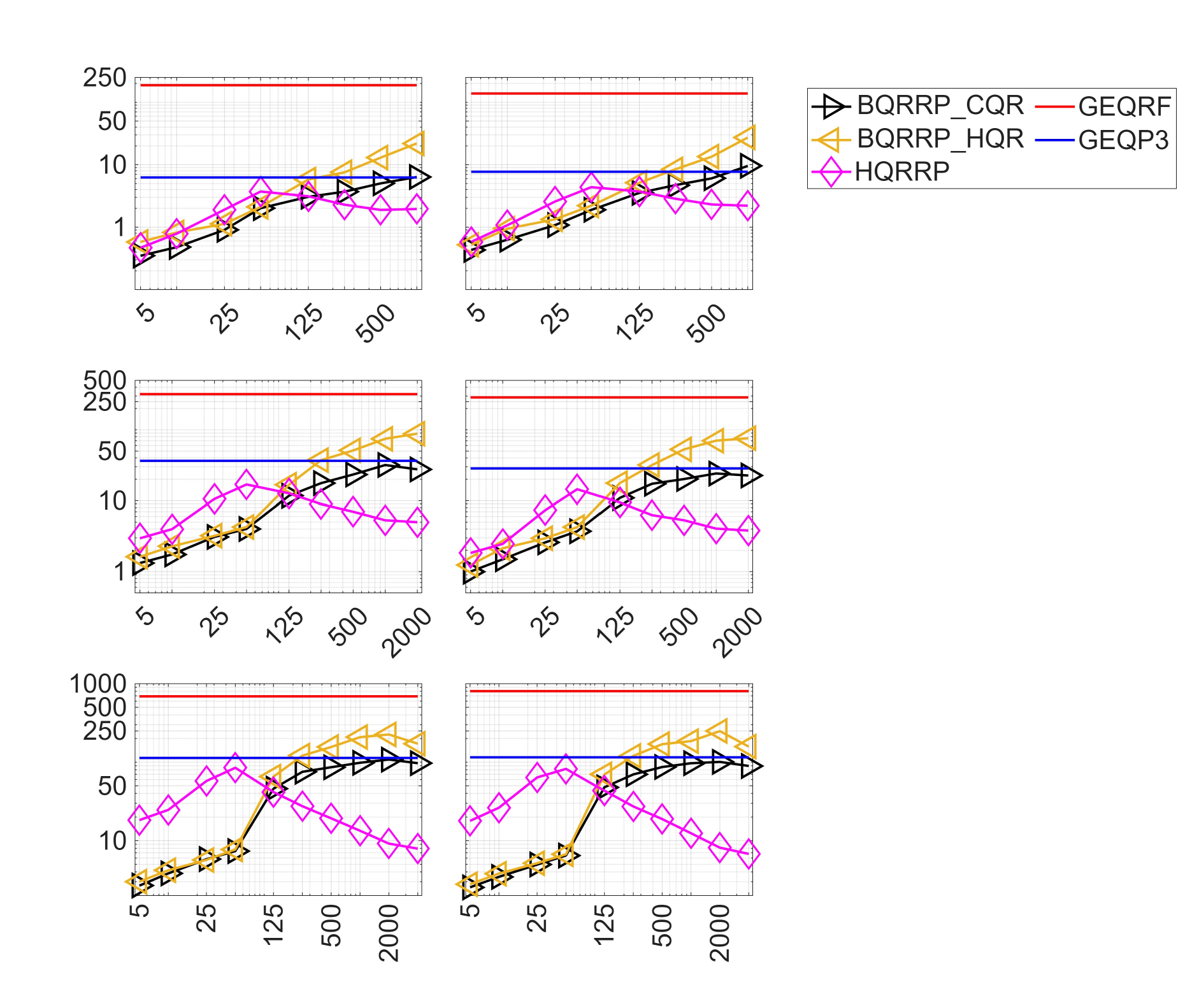}};

    \node[anchor=north west, text width=4.8cm, align=justify] at (1.9, 2.5) {%
    \caption{\footnotesize{Performance of various QR and QRCP methods, captured on Intel and AMD systems (see \cref{table:cpu_config}). 
    The execution of the \code{GEQRF} and \code{GEQP3} functions does not depend on the block size parameter $b$, and hence the performance is depicted as constant. 
    This figure concentrates on assessing how varying the block size parameter in \code{BQRRP} affects its performance in the context of the smaller input matrices.
    }}
    \label{fig:qr_performance_block_size_small_mat}
    };

    \node[anchor=north west, font=\bfseries] at (-0.35\linewidth, 5.5) {Intel CPU};
    \node[anchor=north west, font=\bfseries] at (-0.08\linewidth, 5.5) {AMD CPU};

    \node[anchor=north west, rotate=90, font=\bfseries] at (-0.51\linewidth, 0.18\linewidth) {\textbf{GigaFLOP/s}};
    \node[anchor=north west, rotate=90, font=\bfseries] at (-0.48\linewidth, 0.19\linewidth) {\textbf{$\mathbf{m = 1000}$}};

    \node[anchor=north west, rotate=90, font=\bfseries] at (-0.51\linewidth, -0.07\linewidth) {\textbf{GigaFLOP/s}};
    \node[anchor=north west, rotate=90, font=\bfseries] at (-0.48\linewidth, -0.06\linewidth) {\textbf{$\mathbf{m = 2000}$}};

    \node[anchor=north west, rotate=90, font=\bfseries] at (-0.51\linewidth, -0.33\linewidth) {\textbf{GigaFLOP/s}};
    \node[anchor=north west, rotate=90, font=\bfseries] at (-0.48\linewidth, -0.32\linewidth) {\textbf{$\mathbf{m = 4000}$}};

    \node at (-0.27\linewidth, -5.5) {$\mathbf{b}$};
    \node at (0.005\linewidth, -5.5) {$\mathbf{b}$};
  \end{tikzpicture}
\end{figure}

\cref{fig:qr_performance_block_size_small_mat} shows that in all the smaller matrices tested across both systems, using block sizes that are either twice smaller or as large as $m = n$ is the best choice. 
\cref{fig:qr_performance_block_size_small_mat} also suggests that using \code{BQRRP\_HQR} over \code{BQRRP\_CQR} is always by far the best option and that at no point does the performance of \code{HQRRP} exceeds that of \code{GEQP3}.

\subsection{Varying the aspect ratio in the test matrices}
\label{app:more_aspect_ratios}
\cref{subsec:exp_setup} briefly explains our choice of test matrix types and sizes. As noted there and in \cref{sec:performance}, we primarily use square matrices to evaluate QR and QRCP schemes. However, since \code{BQRRP} applies to any aspect ratio, we also present results for tall and wide matrices. We do not, however, explore the \textit{extremely} tall or wide cases, where specialized algorithms may be more suitable \cite{MBM2024, Fukaya2024CholeskyQR, Armstrong2025}.

\begin{figure}[htb!]
  \centering
  \hspace*{-0.2cm}
  \begin{tikzpicture}
    \node[inner sep=0pt] at (0,0) {\includegraphics[width=1.0\linewidth]{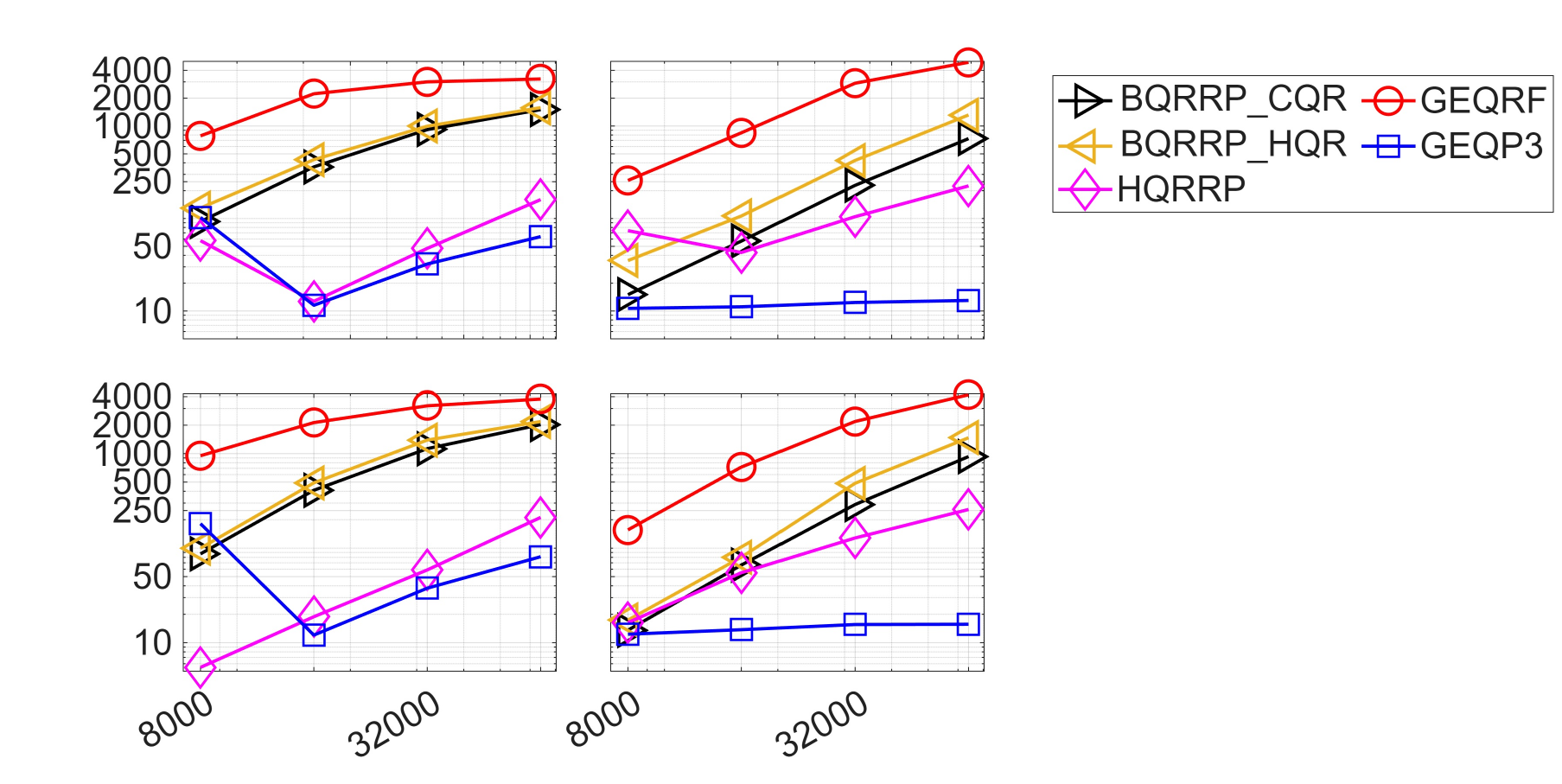}};

    \node[anchor=north west, text width=4.8cm, align=justify] at (1.9, 1.5) {%
    \caption{\footnotesize{Performance of various QR and QRCP methods, captured on Intel and AMD systems (see \cref{table:cpu_config}). The first row depicts tall input matrices, such that $m = 2n$. 
    The second row depicts wide input matrices, such that $n = 2m$. \code{BQRRP} block size is set to $n / 32$, \code{HQRRP} block size is set to $128$.
    }}
    \label{fig:qr_performance_mat_size_rectangle}
    };

    \node[anchor=north west, font=\bfseries] at (-0.35\linewidth, 3.5) {Intel CPU};
    \node[anchor=north west, font=\bfseries] at (-0.07\linewidth, 3.5) {AMD CPU};

    \node[anchor=north west, rotate=90, font=\bfseries] at (-0.48\linewidth, 0.03\linewidth) {\textbf{GigaFLOP/s}};

    \node[anchor=north west, rotate=90, font=\bfseries] at (-0.48\linewidth, -0.17\linewidth) {\textbf{GigaFLOP/s}};

    \node at (-0.27\linewidth, 0.13) {$\mathbf{m}$};
    \node at (0.005\linewidth, 0.13) {$\mathbf{m}$};

    \node at (-0.27\linewidth, -3.5) {$\mathbf{n}$};
    \node at (0.005\linewidth, -3.5) {$\mathbf{n}$};
  \end{tikzpicture}
\end{figure}
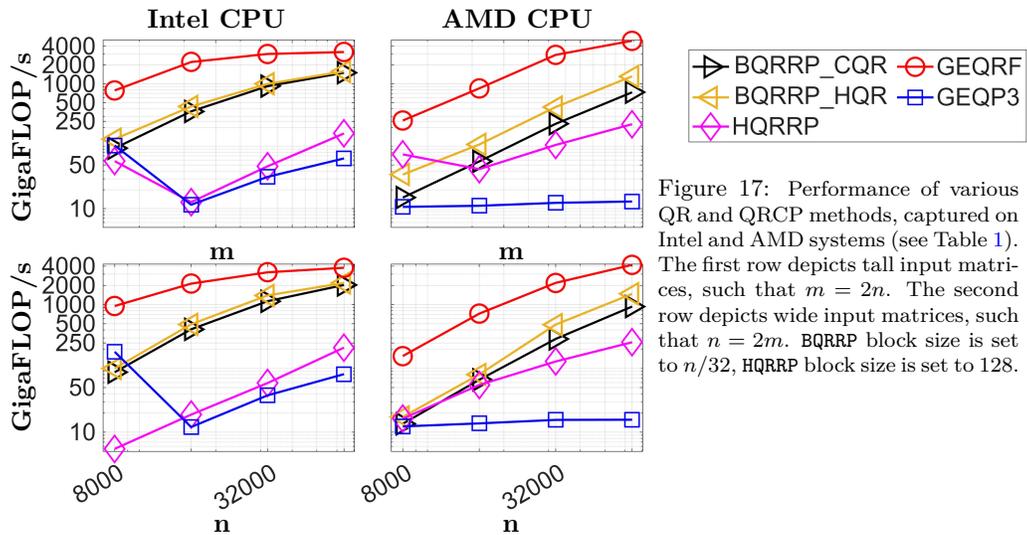

\cref{fig:qr_performance_mat_size_rectangle} depicts the expected results of the performance in the two \code{BQRRP} versions, reaching closer to that of \code{GEQRF} as the input matrix size increases.
Additionally, as seen previously in \cref{app:small_mat_performance}, we observe that using the \code{BQRRP} block size $b = n / 32$ is suboptimal when testing smaller matrices. 


\end{document}